\documentclass[12pt,times]{article}

\title{Multi-fidelity surrogates for mechanics of composites: from co-kriging to multi-fidelity neural networks}

\author{Haizhou Wen$^1$, Elham Kiyani$^2$, Gang Li$^1$, Srikanth Pilla$^{3,4,5,6,7}$, \\George Em Karniadakis$^2$, and Zhen Li$^1$\footnote{Corresponding author: \href{mailto:zli7@clemson.edu}{zli7@clemson.edu}} \\
\small{$^1$ Department of Mechanical Engineering, Clemson University, Clemson, SC 29634, USA}\\
\small{$^2$ Division of Applied Mathematics, Brown University, Providence, RI 02912, USA}\\
\small{$^3$ Center for Composite Materials, University of Delaware, Newark, DL 19716, USA}\\
\small{$^4$ Department of Mechanical Engineering, University of Delaware, Newark, DL 19716, USA}\\
\small{$^5$ Department of Chemical and Biomolecular Engineering, University of Delaware, Newark, DL 19716, USA}\\
\small{$^6$ Department of Materials Science and Engineering, University of Delaware, Newark, DL 19716, USA}\\
\small{$^7$ Department of Computer and Information Sciences, University of Delaware, Newark, DL 19716, USA}}

\date{}

\RequirePackage{lineno}
\usepackage[colorlinks=true]{hyperref}
\usepackage{graphicx}
\usepackage{multirow,array,color,mathrsfs,subcaption}
\usepackage{epstopdf} 

\usepackage{caption}
\DeclareGraphicsExtensions{.eps,.mps,.pdf,.jpg,.png}
\graphicspath{{figures/}{../figures/}}

\usepackage{fancyhdr}
\usepackage{xcolor}
\usepackage{amsmath}
\usepackage{bm}
\usepackage{mathtools}
\numberwithin{equation}{subsection}

\usepackage{amsfonts}

\usepackage{dsfont}

\usepackage[top=1.5cm,bottom=2.5cm,left=1.5cm,right=1.5cm]{geometry}

\usepackage[figuresright]{rotating}
\usepackage{extarrows}

\usepackage{fixltx2e}
\usepackage{xspace}

\usepackage{amsmath,amssymb}

\usepackage{graphicx} 
\usepackage{longtable}
\usepackage{array}
\usepackage{amssymb} 
\newcommand{\yes}{\checkmark}
\newcommand{\no}{}
\usepackage{makecell}

\begin{document}
\maketitle

\begin{abstract}
Composite materials exhibit strongly hierarchical and anisotropic properties governed by coupled mechanisms spanning constituents, plies, laminates, structures, and manufacturing history. This intrinsic complexity makes predictive modeling of composites expensive, because repeated experiments and high-fidelity simulations are needed to cover large design spaces of material, structure, and manufacturing. Multi-fidelity surrogate modeling addresses this challenge by combining abundant, less expensive data with limited high-accuracy data to recover reliable high-fidelity predictions. This review presents a structured overview of multi-fidelity modeling for composite mechanics, covering Gaussian-process or Kriging-based methods, including co-Kriging, coregionalization models, autoregressive formulations, nonlinear autoregressive Gaussian processes, multi-fidelity deep Gaussian processes, and multi-fidelity neural networks. Their distinctions are examined in terms of cross-fidelity correlation, discrepancy representation, uncertainty quantification, and scalability. Selected examples of their applications to composites are introduced according to the roles that multi-fidelity surrogates play in engineering problems, including forward prediction for rapid exploration of material design spaces, inverse optimization for composite parameter identification and design search under limited high-fidelity access, and workflow integration, where heterogeneous data sources, constraints, and validation requirements determine model utility. Open question discussions highlight recurring challenges specific to composites, such as regime-dependent fidelity gaps associated with nonlinear damage and manufacturing history, mismatches between simulations and experiments, and uncertainty propagation across multi-fidelity models. 
\end{abstract}

\newpage
\tableofcontents

\newpage
\section*{Abbreviations \& Symbols}

\begin{table}[htbp]
\centering
\renewcommand{\arraystretch}{1.1}
\caption{List of abbreviations used in this review}
{\fontsize{10}{12}\selectfont
\label{tab:abb}
\begin{tabular}{ll|ll}
\hline
\multicolumn{2}{c|}{\textbf{Methodology}} & \multicolumn{2}{c}{\textbf{Applications}} \\
\hline
\textbf{Abb.} & \textbf{Definition} & \textbf{Abb.} & \textbf{Definition} \\
\hline
MF & Multi-fidelity & DFT & Density functional theory \\
LF & Low-fidelity & FE & Finite element \\
HF & High-fidelity & CFD & Computational fluid dynamics \\
MFM & Multi-fidelity modeling & PID & Process-induced deformation \\
MFSM & Multi-fidelity surrogate modeling & CFRP & Carbon fiber reinforced polymer \\
SM & Surrogate model & SPR & Self-piercing riveting \\
GP & Gaussian process & AE & Acoustic emission \\
SOGP & Single-output Gaussian process & TOF & Time of flight \\
MOGP & Multi-output Gaussian process & SFRC & Short fiber reinforced composite \\
ICM & Intrinsic coregionalization model & CLT & Classical lamination theory \\
LMC & Linear model of coregionalization & ESL & Equivalent single layer \\
AR & Auto-regressive & LW & Layer-wise \\
NARGP & Nonlinear auto-regressive Gaussian process & CUF & Carrera unified formulation \\
MF-DGP & Multi-fidelity deep Gaussian process & UMAT & User material subroutine \\
NLML & Negative log marginal likelihood & FFT & Fast Fourier transform \\
UQ & Uncertainty quantification &  &  \\
BLUP & Best linear unbiased predictor &  &  \\
MLE & Maximum likelihood estimation &  &  \\
REML & Restricted maximum likelihood &  &  \\
ANN & Artificial neural network &  &  \\
FNN & Feedforward neural network &  &  \\
PINN & Physics-informed neural network &  &  \\
MFPINN & Multi-fidelity physics-informed neural network &  &  \\
RNN & Recurrent neural network &  &  \\
LSTM & Long short-term memory &  &  \\
GRU & Gated recurrent unit &  &  \\
MFNN & Multi-fidelity neural network &  &  \\
BO & Bayesian optimization &  &  \\
EGO & Efficient global optimization &  &  \\
RBDO & Reliability-based design optimization &  &  \\
MRGP & Multi-response Gaussian process &  &  \\
SWGPR & Spatially weighted Gaussian process regression &  &  \\
MORNN & Multi-output regression neural network &  &  \\
SK & Simple Kriging &  &  \\
OK & Ordinary Kriging &  &  \\
UK & Universal Kriging &  &  \\
\hline
\end{tabular}
}\end{table}

\begin{table}[htbp]
\centering
\renewcommand{\arraystretch}{1.05}
{\fontsize{9}{12}\selectfont
\caption{List of mathematical symbols used in this review}
\label{tab:symbols}
\begin{tabular}{ll|ll}
\hline
\multicolumn{4}{c}{\textbf{Gaussian Process / Kriging}} \\
\hline
\textbf{Symbol} & \textbf{Definition} & \textbf{Symbol} & \textbf{Definition} \\
\hline
$\mathbf{x}$ & Input vector & $y$ & Scalar output \\
$\mathbf{y}$ & Output vector & $f(\mathbf{x})$ & Latent function \\
$f_q(\mathbf{x})$ & Output at task/fidelity $q$ & $\mathbf{X}$ & Training input set \\
$\mathbf{Y}$ & Training output set & $\mathbf{X}^*$ & Test input set \\
$\mathbf{Y}^*$ & Predicted output set & $\mathbf{F}$ & Latent function values \\
$\mathbf{F}_{\mathbf{X}}$ & Latent values at $\mathbf{X}$ & $\mathbf{F}_{\mathbf{X}^*}$ & Latent values at $\mathbf{X}^*$ \\
$m_\theta(\mathbf{x})$ & Mean function in GP & $\mu(\mathbf{x})$ & Trend function in Kriging \\
$k_\theta(\mathbf{x},\mathbf{x}')$ & Kernel function & $\psi(\mathbf{x},\mathbf{x}')$ & Correlation function in Kriging \\
$\mathbf{K}_{\mathbf{X}\mathbf{X}}$ & Kernel matrix on training inputs & $\mathbf{K}_{*\mathbf{X}}$ & Test--train cross-covariance \\
$\mathbf{K}_{\mathbf{X}*}$ & Train--test cross-covariance & $\mathbf{K}_{**}$ & Kernel matrix on test inputs \\
$\mathbf{\Psi}$ & Correlation matrix in Kriging & $\mathbf{R}$ & Covariance matrix in Co-kriging \\
$\mathbf{B}$ & Task covariance matrix & $K_{qq'}$ & Cross-task covariance \\
$\mathbf{c}_*$ & Cross-covariance vector in Co-kriging & $\mathbf{w}$ & Co-kriging weight vector \\
$\mathbf{f}_*$ & Trend basis at test input & $\boldsymbol{\lambda}$ & Lagrange multipliers \\
$\boldsymbol{\mu}_*$ & Predictive mean & $\boldsymbol{\Sigma}_*$ & Predictive covariance \\
$\hat{y}(\mathbf{x}^*)$ & BLUP at test input $\mathbf{x}^*$ & $y_{\text{Co-kri}}(\mathbf{x}^*)$ & Co-kriging predictor \\
$\sigma_f^2$ & Kernel signal variance & $\sigma_s^2$ & Observation noise variance \\
$\sigma_{s,q}^2$ & Noise variance for task $q$ & $\sigma^2$ & Process variance in Kriging \\
$\sigma^2_{\text{Kri}}$ & Kriging prediction variance & $\sigma^2_{\text{Co-kri}}(\mathbf{x}^*)$ & Co-kriging prediction variance \\
$\epsilon$ & Observation noise & $Z(\mathbf{x})$ & Random field in Kriging \\
$\delta_q(\mathbf{x})$ & Discrepancy function & $\rho_q$ & Scaling coefficient in AR model \\
$z_q(\cdot)$ & Nonlinear mapping in NARGP & $g_q(\cdot)$ & GP mapping function \\
$u_r(\mathbf{x})$ & Latent GP in LMC & $a_{q,r}$ & Mixing coefficient in LMC \\
$\mathbf{Z}_q$ & Inducing inputs & $\mathbf{u}_q$ & Inducing variables \\
$\mathbf{m}_q$ & Variational mean & $\mathbf{S}_q$ & Variational covariance \\
$A_q(\mathbf{x})$ & Interpolation matrix & $\xi_q(\mathbf{x})$ & Conditional residual \\
$\theta$ & Kernel hyperparameters & $\Theta$ & Hyperparameters in MOGP \\
$\eta$ & Hyperparameter set in SOGP & $\zeta$ & Hyperparameter set in MOGP / Co-kriging \\
$l_j$ & Length scale & $N$ & Number of samples \\
$N_q$ & Number of samples at task/fidelity $q$ & $N_{\text{tot}}$ & Total number of stacked samples \\
$Q$ & Number of tasks/fidelities & $M$ & Number of test points \\
\hline

\multicolumn{4}{c}{\textbf{Neural Networks}} \\
\hline
\textbf{Symbol} & \textbf{Definition} & \textbf{Symbol} & \textbf{Definition} \\
\hline
$\mathbf{x}$ & Input vector & $\mathbf{y}$ & Output vector \\
$\mathbf{W}$ & Weight matrix & $\mathbf{b}$ & Bias vector \\
$\mathbf{b}_{\text{in}}$ & Bias from input to hidden layer & $\mathbf{b}_{\text{out}}$ & Bias from hidden to output layer \\
$\mathbf{h}^{(\ell)}$ & Hidden-state vector at layer $\ell$ & $A(\cdot)$ & Activation function \\
$A^{(\ell)}(\cdot)$ & Activation at layer $\ell$ & $\Theta$ & Trainable network parameters \\
$\mathcal{L}$ & Loss function & $l_r$ & Learning rate \\
$k$ & Training iteration index & $L$ & Number of layers \\
$y_L$ & Low-fidelity output & $y_H$ & High-fidelity output \\
$\mathcal{F}_l$ & Linear LF--HF mapping & $\mathcal{F}_{nl}$ & Nonlinear LF--HF mapping \\
$\beta_i$ & Weight in regularization term & $\lambda_r$ & Regularization coefficient \\
$u(t,\mathbf{x})$ & Solution field in PINN & $f(t,\mathbf{x})$ & PDE residual in PINN \\
$\mathcal{N}[u;\boldsymbol{\lambda_u}]$ & Differential operator & $\boldsymbol{\lambda_u}$ & Physical parameters in PDE operator \\
$h_t^{(\ell)}$ & Hidden state at time step $t$, layer $\ell$ & $c_t^{(\ell)}$ & Cell state in LSTM \\
$i_t^{(\ell)}$ & Input gate in LSTM & $f_t^{(\ell)}$ & Forget gate in LSTM \\
$o_t^{(\ell)}$ & Output gate in LSTM & $\hat{c}_t^{(\ell)}$ & Candidate cell state in LSTM \\
$x_t$ & Input at time step $t$ & $T$ & Sequence length / final time \\
\hline
\end{tabular}
}\end{table}

\newpage
\section{Introduction}
Composite materials are engineered solids formed by combining two or more distinct constituents, most commonly high-stiffness fibers embedded in a compliant polymer matrix, so that the resulting laminate or structure exhibits tailored performance that a single homogeneous material cannot provide~\cite{clyne2019introduction,ABRAMOVICH20171}. Their heterogeneity naturally organizes composite mechanics into a micro–meso–macro hierarchy. At the microscale, constituent properties and local architecture such as fiber packing, matrix regions, and fiber–matrix interfaces govern load transfer~\cite{MICHEL1999109,hale1976physical}. At the mesoscale, a lamina or ply behaves as an anisotropic medium whose effective response reflects the underlying microstructure and defects, and multiple plies interact through stacking sequence and interlaminar behavior~\cite{doi:10.1177/002199837100500106}. At the macroscale, the laminate or full component response is further shaped by geometry, boundary conditions, and service loading, often requiring structural-level models for realistic prediction~\cite{doi:10.1061/(ASCE)1090-0268(2002)6:2(73)}. In practice, analysis therefore moves across these levels using different representations, from the micromechanics-based homogenization at micro or meso scales~\cite{10.1115/1.4001911}, to the classical lamination theory (CLT)~\cite{doi:10.1177/002199837500900110} as well as the equivalent single-layer plate/shell formulations~\cite{10.1007/BF02736224}, and full-scale finite element (FE) simulations~\cite{Ochoa1992} at the macro scale while experiments are commonly used for calibration and validation. At the same time, the very features that make composites valuable in engineering practice also create the central computational difficulty. Strong anisotropy~\cite{daniel1994engineering}, layerwise discontinuities~\cite{PAGANO1978385}, damage initiation and evolution such as matrix cracking~\cite{10.1098/rspa.1985.0055}, fiber breakage~\cite{doi:10.1177/002199839803202204}, delamination~\cite{HUTCHINSON199163}, and manufacturing-induced variability~\cite{jcs3020056} introduce nonlinearities and history dependence; consequently, reliable high-accuracy prediction often demands expensive multiscale or multiphysics simulations and extensive calibration. In design and certification workflows, these solvers need to be executed repeatedly across many stacking sequences, geometries, defect scenarios, and uncertainty realizations, which quickly becomes prohibitive. One practical way to alleviate this cost is to replace repeated high-cost analyses with surrogate models (SMs), which learn an efficient input–output map for quantities of interest in composites~\cite{DEY2017227,sharma2022advances}, for example from layup, geometry, and loading to stiffness, strength, or failure indicators using Gaussian process (GP) regression~\cite{doi:10.2514/1.6386,HAERI201626,mukhopadhyay2017critical,carrico20193d,WANG2020112821}, reduced-order models~\cite{aguado2017simulation,Liang02012021}, and neural networks~\cite{lo1995artificial,VASSILOPOULOS200720,KARNIK20081768,MCCRORY2015424,HAMDIA2015304,kiyani2025probabilisticpredictionsprocessinduceddeformation}. These approaches can significantly accelerate repeated evaluations once trained. However, most of these surrogate approaches especially neural networks are strongly data-driven, so their accuracy hinges on the availability of high-quality labeled data. Generating such high-quality datasets from detailed simulations or experiments is expensive, while training on cheaper low-detail data is easier to scale but typically limits accuracy due to inherent model simplifications and bias. This gap motivates a strategy that can integrate information across the micro–meso–macro ladder while explicitly correcting discrepancies between levels.

Building on this perspective, the notion of fidelity provides a useful language for describing and organizing these information sources within a unified framework. In computational modeling, fidelity refers to how closely a model or dataset represents the underlying physical response of interest~\cite{10.1093/biomet/87.1.1,doi:10.1098/rspa.2007.1900}, as determined by the governing assumptions, numerical resolution, and the degree of experimental grounding. Under this view, a low-fidelity (LF) source is typically inexpensive and enables broad exploration, but often carries structured errors due to idealizations or omitted mechanisms. A high-fidelity (HF) source is generally more trusted and more predictive, yet more expensive to obtain, whether through refined multiscale or multiphysics simulations or through experimental testing. Multi-fidelity (MF) modeling formalizes the joint use of these sources by leveraging abundant LF information to capture global trends while using limited HF evidence to correct bias and anchor predictions. A representative early example is the design optimization of a cracked stiffened composite panel by Vitali et al.~\cite{vitali2002multi}, where a coarse FE model combined with a closed-form stress-intensity solution was treated as the LF source and a refined crack-tip FE model served as the HF source, with the two sources linked through correction response surfaces to approximate the crack-propagation constraint efficiently. Multi-scale modeling naturally fits within this framework because micro-, meso-, and macro-scale descriptions often differ in both cost and reliability and can therefore be interpreted as distinct fidelity levels rather than unrelated modeling choices. This interpretation is especially natural for composites, whose response is governed by coupled mechanisms across multiple levels and whose practical analysis already moves among constituent behavior, ply and laminate theories, and structural simulations. Seen more broadly, this interpretation is consistent with the modern MF literature, which treats fidelity as a relative, task-dependent notion and organizes MF methods around information fusion across models and data sources rather than around scale alone~\cite{peherstorfer2018survey,fernandez2016review}. Correspondingly, later MF formulations expanded from efficient recursive co-kriging strategies~\cite{perdikaris2015multi} to nonlinear information-fusion frameworks capable of learning more complex cross-fidelity relations~\cite{perdikaris2017nonlinear}.

More broadly, the MF concept is not restricted to multi-scale settings, since the relative meaning of LF and HF is always comparative and problem-dependent rather than tied to a specific length scale. In many physical domains, LF may correspond to an idealized but abundant numerical source, while HF corresponds to a smaller set of measurements that better reflect real operating conditions, as commonly seen when computational fluid dynamics (CFD) is contrasted with expensive experiments in fluid dynamics~\cite{doi:10.2514/2.456,OBERKAMPF2002209}. Even within CFD itself, fidelity can be defined through solver choices and discretization, where a lower-order method or a coarser resolution provides inexpensive trend information and a higher-order method or a finer resolution provides more accurate but costly supervision~\cite{https://doi.org/10.1002/fld.3767}. Similar distinctions arise in computational biology, where efficient coarse-grained particle models can serve as LF surrogates for high-cost molecular descriptions that resolve finer interactions~\cite{annurev:/content/journals/10.1146/annurev-biophys-083012-130348,Li2016}. In materials mechanics, fidelity can also be defined through the degree of novelty or uncertainty in the system under evaluation, for instance using tests on well-characterized single-constituent materials as a broad and relatively cheap baseline while reserving expensive testing for newly designed composites whose graded or heterogeneous architectures introduce additional mechanisms and uncertainty~\cite{SABA20191}. Under all these interpretations, MF modeling provides a transferable principle, namely to couple a large amount of inexpensive information that captures the global structure with a limited amount of trusted evidence that corrects bias and stabilizes prediction.
In this context, the emphasis here is on surrogate modeling under the MF framework rather than on developing new multi-scale solvers, because the dominant bottleneck in composite design and qualification is often not the existence of physics-based models but the need to evaluate them repeatedly across large design spaces and uncertainty scenarios. MF surrogates target this repeated-use regime by reusing established solvers and experiments as information sources, then learning an efficient predictor that preserves the high-accuracy behavior with far fewer expensive evaluations.
 
This review surveys multi-fidelity surrogate modeling (MFSM) for composite mechanics through several sections. Section~\ref{Section Classic MFSM} summarizes classical MFSM through the GP and Kriging families, with emphasis on the mathematical foundations of the correlation assumptions and fusion strategies that couple information across fidelity. Section~\ref{Section NN-based MFSM} first introduces the fundamentals of neural network modeling and its variants commonly employed in the MFSM of composite mechanics. Then it continues to review NN-based MFSM as it appears in recent composite-mechanics studies, focusing on two dominant constructions, namely multi-fidelity fused networks that learn cross-fidelity mappings within a single architecture, and transfer learning frameworks that reuse features learned from abundant inexpensive sources and recalibrate them using limited trusted data. Section~\ref{Section Apps} collects representative applications of MFSM in composite mechanics and arranges the literature using two complementary viewpoints. One viewpoint follows the surrogate model family, separating GP and Kriging-based approaches from neural-network-based approaches. The other viewpoint follows how the surrogate is used in practice, distinguishing prediction-oriented tasks from inverse tasks such as design, calibration, or parameter identification, and further highlighting cases where the surrogate is embedded as a component in a broader engineering workflow. Within the neural network stream, applications are additionally organized by the dominant construction, contrasting fused network architectures with transfer-learning-based multi-fidelity strategies. Then, section~\ref{Section OCRO} discusses key observations and open issues that emerge from these threads and highlights potential directions of MFSM in composite mechanics. Finally, section~\ref{Section summary} ends the review with discussions and a summary.

\section{Classical Multi-Fidelity Surrogate Modeling} \label{Section Classic MFSM}
The objective of multi-fidelity modeling (MFM) is to bridge the gap between fast computation and high accuracy by combining different fidelities and leveraging their strengths~\cite{MGiselle2023}. In the context of MFM, Peherstorfer~\emph{et al.}~\cite{doi:10.1137/16M1082469,BREVAULT2020106339} classified the specific MF techniques into three categories: adaptation, filtering, and fusion. The adaptation models focus on enhancing the performance of LF models with the results from HF ones. For instance, the hierarchy between different fidelity outputs captured by an autoregressive process is used for model correction~\cite{10.1093/biomet/87.1.1}. The filtering model enables the LF model to act as a gate to decide when to call the HF model. An example is incorporating a multi-level estimator into the process of uncertainty quantification (UQ), where the estimator evaluates the cheap coarse posterior or LF model to determine whether to use a fine likelihood or HF model for further correction~\cite{doi:10.1137/130915005}. Therefore, the LF model acts as a stochastic filtering gate, delaying the acceptance of the HF model. The fusion models aim to integrate low and high-fidelity model outputs or source data of corresponding fidelity levels. The purposes of such integration utilizing fusion techniques are many, including cutting variance and computational cost~\cite{doi:10.1007/s00791-011-0160-x}, boosting robustness by combining complementary models~\cite{10.1214/ss/1009212519}, and improving state estimation and UQ~\cite{doi:10.1137/15M100955X}. Based on the different types of the assumed relations between two fidelities, Park~\emph{et al.}~\cite{10.1007/s00158-016-1550-y-park} further divided fusion into three kinds. The first is the discrepancy-based fusion~\cite{Qian01052008,doi:10.2514/6.2015-1375}, which assumes the HF is the scaled LF plus a discrepancy correction term. The second is the calibration-based fusion~\cite{doi:10.2514/1.C031808}, which assumes the LF model has tunable parameters that can be calibrated to make the LF model identical to the HF model. The third one is comprehensive fusion~\cite{kennedy2001supplementary}, which combines both discrepancy-based and calibration-based approaches.

The integration of LF and HF information is often converted into a single, queryable model, referred to as SM or surrogate~\cite{doi:10.1007/PL00007198}, to predict the HF quantity. SMs are essentially mathematical approximations employed to mimic a system’s behavior, and built primarily to cut down on prediction time and costs~\cite{MGiselle2023}, especially in the numerous and expansive simulations involving optimization~\cite{doi:10.1098/rspa.2007.1900} or UQ~\cite{doi:10.1098/rspa.2015.0018}. A surrogate model may be deterministic or probabilistic, and can be purely data-driven, physics-informed, or hybrid. One example is the surrogate modeling of CFD~\cite{doi:10.1007/s10409-021-01148-1}, where the surrogate physics informed neural networks (PINNs) are the mathematical function, which are optimized by CFD-generated data and physical constraints, mapping from spatial-temporal coordinates to specific physical quantities of interest. In the conceptualization of the majority, when MFM with the fusion technique is pursued with the explicit goal of building such a surrogate for the HF quantity, it is referred to as MFSM. Within MFSM, fusion approaches fall broadly into several families including GP or Co-kriging formulations~\cite{doi:10.1098/rspa.2007.1900}, neural-network-based models~\cite{MENG2020109020}, polynomial or sparse-grid surrogates~\cite{doi:10.1137/130929461}, low-rank multifidelity projection methods~\cite{hampton2017parametricstochasticmodelreductionlowrank}, and other probabilistic schemes~\cite{NIPS2010_fc8001f8}. Among these, GP and Kriging are widely adopted due to their sample efficiency and interpretability. Therefore, in this particular section, the specific classical fusion techniques GP and Kriging for surrogate models within MFSM will be reviewed. Additionally, the authors distinguish the fusion techniques like GP and Kriging from surrogate models to avoid ambiguity since many contexts directly refer to GP or Kriging as the surrogate model. In this review section, the fusion technique like GP or Kriging are referred to as a methodology explaining how LF and HF data are combined, and the surrogate model is referred to as a surrogate artifact or predictor after training.

\subsection{Single-output \& Multi-output Gaussian Process}
From a statistical perspective, a process refers to a family of random variables. A GP describes a distribution over a collection of random variables, any finite subset of which follows a joint Gaussian distribution~\cite{NIPS1995_7cce53cf}. From a functional perspective, GP is used to approximate any unknown mapping function $f(\cdot)$ in a supervised learning problem. According to the output dimension of the function $f$, GP is generally divided into single-output GP (SOGP) and multi-output GP (MOGP). SOGP gives a non-deterministic prediction of $f$, where each input $\mathbf{x}$ (scalar or vector) has a scalar output $y$, while in the approximated $f$ by MOGP, each input $\mathbf{x}$ (scalar or vector) has a vector output $\mathbf{y}$.

\subsubsection{SOGP}
Assuming an observed sample dataset of finite size $N$, with input samples $\{\mathbf{x}^1,\mathbf{x}^2,\ldots,\mathbf{x}^N\}$, where $\mathbf{x}\in\mathbb{R}^d$, and output samples $\{y^1=f(\mathbf{x}^1),y^2=f(\mathbf{x}^2),\ldots,y^N=f(\mathbf{x}^N)\}$, where $y\in\mathbb{R}$, SOGP regression~\cite{SCHULZ20181,LIU2018102} can be used to provide a probabilistic, rather than exact deterministic, prediction of the function $f$ at any new unobserved input location based on the observed dataset. The first step in GP regression is to place a GP prior on $f$, assuming that $f$ can be modeled as a Gaussian process, expressed as
\begin{equation}
    f \sim \mathcal{GP}\!\left(m_\theta(\mathbf{x}),\,k_\theta(\mathbf{x},\mathbf{x}')\right),
    \label{GP-prior}
\end{equation}
where $m_\theta(\mathbf{x})$ denotes the mean function and $k_\theta(\mathbf{x},\mathbf{x}')$ denotes the covariance function between two input locations $\mathbf{x}$ and $\mathbf{x}'$, with $\theta$ representing the corresponding hyperparameters. The mean function is commonly taken as zero, i.e., $m_\theta(\mathbf{x})=0$, without loss of generality~\cite{LIU2018102}. A typical example of the covariance or kernel function is the squared exponential covariance function~\cite{Rasmussen2004}, which is expressed as
\begin{equation}
    k^{\mathrm{SE}}_\theta(\mathbf{x},\mathbf{x}')= \sigma_f^{2} \exp\left(-\frac{1}{2}\sum_{j=1}^{d}\frac{(x_j-x_j')^{2}}{l_j^{2}}\right),
    \label{KE-variance}
\end{equation}
where the specific kernel signal variance $\sigma_f^\text{2}$ and scale length $l_\text{j}$ are considered as hyperparameters $\theta$. In many realistic scenarios, the sample dataset can be seen as the observations of the exact latent function $f$ value that are perturbed from the latent function values due to factors such as sensor noise and calibration drift during measurement~\cite{PAN2013140,NASAJPOURESFAHANI2025115416}, approximation error from ideal modeling~\cite{yu2016unified}, and unmodeled covariates~\cite{MAJEWSKA2021113263}. In GP, a Gaussian noise model is used to model the discrepancy, which can be expressed as
\begin{equation}
    {y} = f(\mathbf{x}) + \epsilon, \label{noise}
\end{equation}
where $\epsilon\sim \mathcal{N}(0,\sigma_s^\text{2})$ is the independent and identically distributed noise with zero mean and observation noise variance $\sigma_s^\text{2}$ from samples. The likelihood specifies the probability distribution of the observed outputs conditioned on the latent function values at the input locations. Let the vectorized datasets be defined as $\mathbf{X}=\{\mathbf{x}^1,\mathbf{x}^2,\ldots,\mathbf{x}^N\}^\mathsf{T}$, $\mathbf{Y}=\{y^1,y^2,\ldots,y^N\}^\mathsf{T}$, and $\mathbf{F}_{\mathbf{X}}=\{f(\mathbf{x}^1),f(\mathbf{x}^2),\ldots,f(\mathbf{x}^N)\}^\mathsf{T}$. The likelihood in GP regression is then expressed as
\begin{equation}
    p({\mathbf{Y}}\mid \mathbf{F}_\mathbf{X},\sigma_s^\text{2}) = \mathcal{N}(\mathbf{Y}\mid\mathbf{F}_\mathbf{X},\sigma_s^\text{2}\mathbf{I}_\text{N}), \label{likelihood}
\end{equation}

where $\mathcal{N}(\mathbf{Y}\mid\mathbf{F}_\mathbf{X},\sigma_s^\text{2}\mathbf{I}_\text{N})$ denotes a multivariate Gaussian distribution ($\mathbf{I}$ is an identity matrix) with mean vector $\mathbf{F}_\mathbf{X}$ and covariance matrix $\sigma_s^\text{2}\mathbf{I}_\text{N}$. With Eqn.~\ref{GP-prior}, a GP prior with zero-mean is expressed as

\begin{equation}
    \mathbf{F}_\mathbf{X}\sim\mathcal{N}(\mathbf{0},\mathbf{K}_{\mathbf{X}\mathbf{X}}), \label{GP-prior2}
\end{equation}
where $\mathbf{K}_{\mathbf{X}\mathbf{X}}\in\mathbb{R}^{N\times N}$ with entries $[\mathbf{K}_{\mathbf{X}\mathbf{X}}]_\text{ij}=k_\theta(\mathbf{x}^\text{i},\mathbf{x}^\text{j})$ with $\text{i, j}\in[\text{1},\text{N}]$ is the covariance matrix or kernel matrix on the observation dataset. GP prior gives a prior estimation of the unknown latent functions before utilizing the knowledge of observation data. The prior estimation under Eqn.~\ref{GP-prior2} can be expressed as
\begin{equation}
    p(\mathbf{F}_{\mathbf{X}}\mid \mathbf{X},\theta)
    =
    \mathcal{N}\!\left(\mathbf{F}_{\mathbf{X}}\mid \mathbf{0},\mathbf{K}_{\mathbf{X}\mathbf{X}}\right).
    \label{GP-prior3}
\end{equation}
Since the latent function values are unknown, they are integrated out using Eqns.~\ref{likelihood} and \ref{GP-prior3}, yielding the marginal likelihood
\begin{equation}
\begin{aligned}
   p(\mathbf{Y}\mid\mathbf{X},\{\theta,\sigma_s^2\})
   &=
   \int
   p(\mathbf{Y}\mid \mathbf{F}_{\mathbf{X}},\sigma_s^2)\,
   p(\mathbf{F}_{\mathbf{X}}\mid \mathbf{X},\theta)\,
   d\mathbf{F}_{\mathbf{X}} \\
   &=
   \mathcal{N}\!\left(
   \mathbf{Y}\mid
   \mathbf{0},
   \mathbf{K}_{\mathbf{X}\mathbf{X}}+\sigma_s^2\mathbf{I}_N
   \right).
\end{aligned}
\end{equation}
The integrated outcome $p(\mathbf{Y}\mid\mathbf{X},\eta)$ is the marginal likelihood, representing the probability of observing $\mathbf{Y}$ at inputs $\mathbf{X}$ under a GP with the whole hyperparameters $\eta = \{\theta, \sigma_s^{2}\}$. It evaluates how well the GP, with a given prior and noise specification, explains the observed dataset, and is therefore often referred to as the model evidence. By measuring the overall plausibility of the data under the GP model, the marginal likelihood provides a principled objective for hyperparameter learning and model selection through its maximization with respect to $\eta$. One conventional approach for hyperparameter learning is to minimize the negative log marginal likelihood (NLML)~\cite{10.1162/neco.1992.4.3.415}, which is expressed as
\begin{equation}
\begin{aligned}
   \hat{\eta}
   &=
   \arg\min_{\eta}
   \left(
   -\log p(\mathbf{Y}\mid\mathbf{X},\eta)
   \right)\\
   &=
   \frac{1}{2}\mathbf{Y}^\mathsf{T}
   \left(\mathbf{K}_{\mathbf{X}\mathbf{X}}+\sigma_s^2\mathbf{I}_N\right)^{-1}
   \mathbf{Y}
   +\frac{1}{2}\log\left|
   \mathbf{K}_{\mathbf{X}\mathbf{X}}+\sigma_s^2\mathbf{I}_N
   \right|
   +\frac{N}{2}\log(2\pi).
\end{aligned}
\end{equation}
After the optimal hyperparameters $\eta$ are obtained by maximizing the marginal likelihood, posterior prediction at new input locations can be performed. Let the set of $M$ prediction inputs be defined as
\[
\mathbf{X}^*=\{ {\mathbf{x}^{*}}^{1}, {\mathbf{x}^{*}}^{2}, \ldots, {\mathbf{x}^{*}}^{M} \}^\mathsf{T}.
\]
Define the kernel matrices
\begin{equation}
\begin{aligned}
    \mathbf{K}_{*\mathbf{X}} &=
    \left[
    k_\theta({\mathbf{x}^{*}}^{\text{i}},\mathbf{x}^{\text{j}})
    \right], \quad \text{i}=1,\ldots,M,\; \text{j}=1,\ldots,N,\\
    \mathbf{K}_{\mathbf{X}*} &= \mathbf{K}_{*\mathbf{X}}^\mathsf{T},\\
    \mathbf{K}_{**} &=
    \left[
    k_\theta({\mathbf{x}^{*}}^{\text{i}},{\mathbf{x}^{*}}^{\text{j}})
    \right], \quad \text{i},\text{j}=1,\ldots,M.
\end{aligned}
\end{equation}

The posterior distribution over the latent function values
\[
\mathbf{F}_{\mathbf{X}^*}=\{f({\mathbf{x}^{*}}^{1}),f({\mathbf{x}^{*}}^{2}),\ldots,f({\mathbf{x}^{*}}^{M})\}^\mathsf{T}
\]
is Gaussian:
\begin{equation}
\begin{aligned}
    \mathbf{F}_{\mathbf{X}^*}\mid\mathbf{X},\mathbf{Y},\eta
    &\sim\mathcal{N}(\boldsymbol{\mu}_*,\boldsymbol{\Sigma}_*),\\
    \boldsymbol{\mu}_* &=
    \mathbf{K}_{*\mathbf{X}}
    \left(\mathbf{K}_{\mathbf{X}\mathbf{X}}+\sigma_s^2\mathbf{I}_N\right)^{-1}
    \mathbf{Y},\\
    \boldsymbol{\Sigma}_* &=
    \mathbf{K}_{**}
    -
    \mathbf{K}_{*\mathbf{X}}
    \left(\mathbf{K}_{\mathbf{X}\mathbf{X}}+\sigma_s^2\mathbf{I}_N\right)^{-1}
    \mathbf{K}_{\mathbf{X}*}.
\end{aligned}
\end{equation}

With the observation noise model in Eqn.~\ref{noise}, the posterior over the predictive outputs
\[
\mathbf{Y}^*=\{ {y^{*}}^{1}, {y^{*}}^{2}, \ldots, {y^{*}}^{M} \}^\mathsf{T}
\]
is given by
\begin{equation}
\mathbf{Y}^*\mid\mathbf{X},\mathbf{Y},\eta
\sim
\mathcal{N}\!\left(
\boldsymbol{\mu}_*,
\boldsymbol{\Sigma}_*+\sigma_s^2\mathbf{I}_M
\right).
\end{equation}

\begin{figure}[hbt!]
\centering
\includegraphics[width=0.8\linewidth]{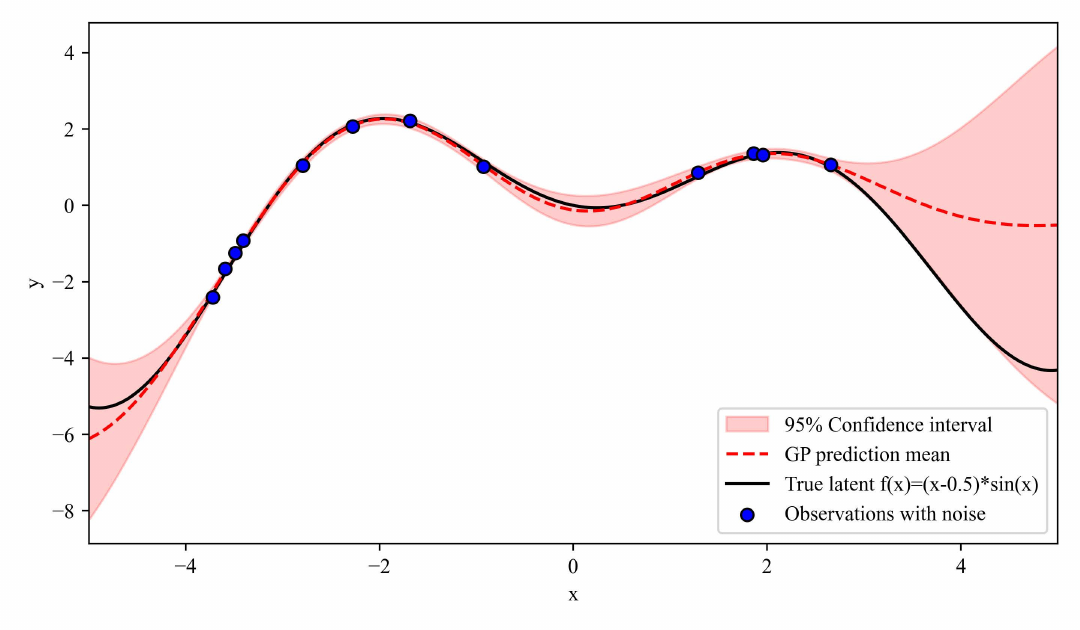}
\caption{Example of Gaussian process prediction and associated 95\% confidence interval. The latent function $f(x)=(x-0.5)\sin x$, the observation samples are biased with 10\% standard deviation (1\% variance). Example uses a GP prior of zero-mean and squared exponential covariance function (Eqn.~\ref{KE-variance}), and the hyperparameters are learnt by maximizing NLML using Cholesky factorization numerics~\cite{https://doi.org/10.1002/wics.18}.}
\label{fig:GP_demo}
\end{figure}

The posterior over $\mathbf{Y}^*$ represents the non-deterministic prediction of the latent function $f$ with consideration of the noise at any new unmapped location $\mathbf{X}^*$ informed by observation
samples ($\mathbf{X}$,$\mathbf{Y})$ through a typical GP regression. An example using simple SOGP regression to predict the non-deterministic distribution of a simple latent function $f(x)=(x-0.5)\sin x$ with several noisy observation points is shown in Fig.~\ref{fig:GP_demo}.

\subsubsection{MOGP}
\begin{figure}[hbt!]
\centering
\includegraphics[width=0.95\linewidth]{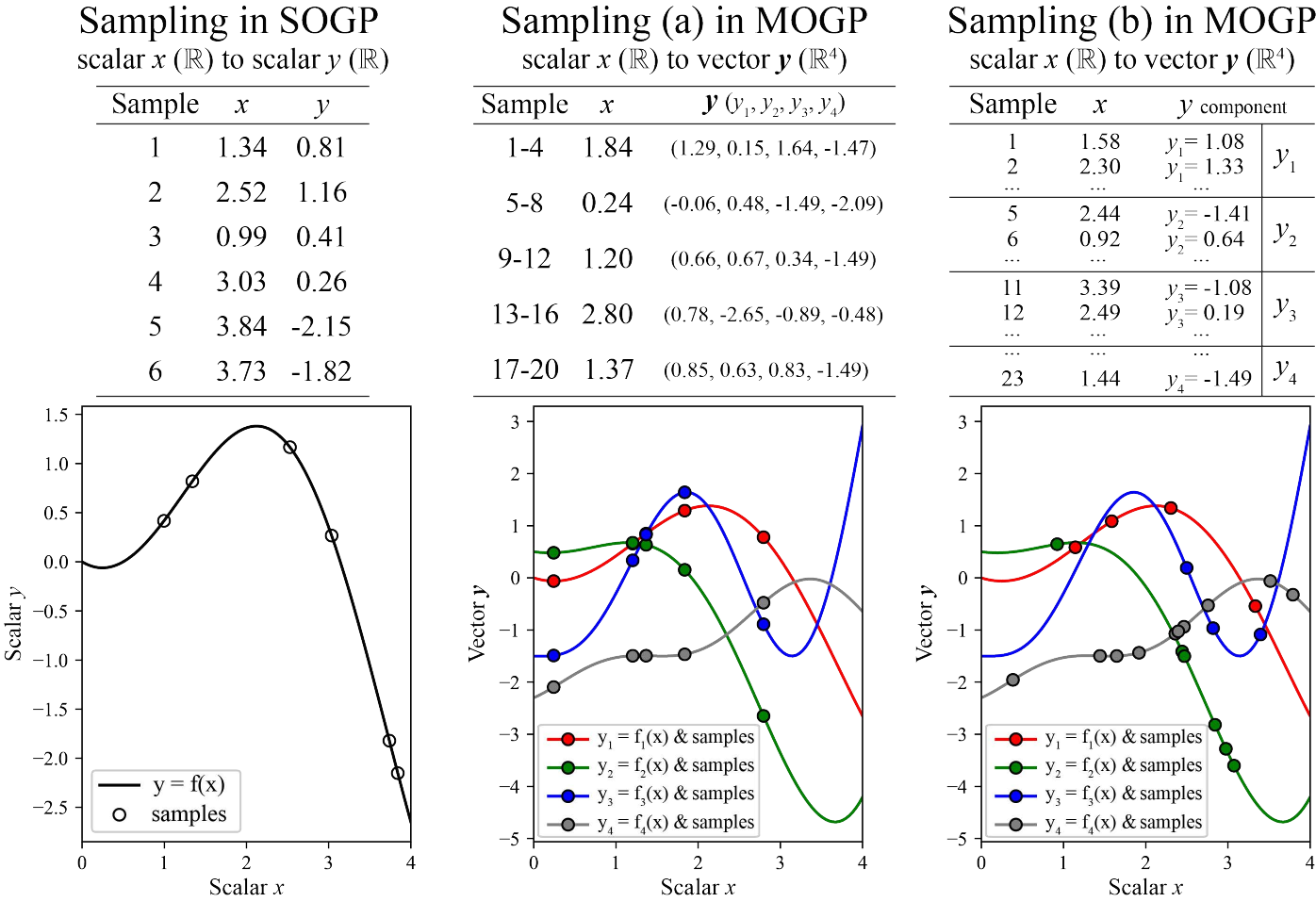}
\caption{Demonstration of varying dataset types in GP: scalar to scalar sampling in SOGP (left), scalar to vector sampling~(a) in MOGP (middle), and scalar to vector sampling~(b) in MOGP (right). Sampling~(a) indicates the data samples at every $\mathcal{x}$ are fully informed, and sampling~(b) indicates the data samples at every $\mathcal{x}$ are not fully informed.}
\label{fig:GP_sample}
\end{figure}

MOGP can handle output vectors of higher dimension compared to single-dimensional output scalars of SOGP. Assuming an observed sample set with a finite size of N, its vectorized input dataset is $\mathbf{X}=\{\mathbf{x}^\text{1},\mathbf{x}^\text{2},...,\mathbf{x}^\text{N}\}^\mathsf{T}$ ($\mathbf{x}\in \mathbb{R}^d$) and its vectorized output dataset $\mathbf{Y}=\{\mathbf{y}^\text{1},\mathbf{y}^\text{2},...,\mathbf{y}^\text{N}\}^\mathsf{T}$ ($\mathbf{y}\in \mathbb{R}^Q$). It is important to notice that some sample vectors in $\mathbf{X}$ may be overlapped, and some sample vectors in $\mathbf{Y}$ may not be fully informed. To distinguish this feature in the MOGP scenario from SOGP, a comparison demonstration is shown in Fig.~\ref{fig:GP_sample}. Sampling~(b) shown in Fig.~\ref{fig:GP_sample} is more universal in a typical MOGP regression problem, in which the sampling quantity for each $\mathbf{y}$ component may not be unified. Therefore, MOGP initially recognizes the different components or dimensions of $\mathbf{y}$ output as different tasks, and each task can be seen as a SOGP problem. Let the output dimension or the task number of $\mathbf{Y}$ be Q, $\mathbf{X}$ and $\mathbf{Y}$ is further expressed as
\begin{equation}
    \begin{aligned}
        \mathbf{X}=&\{\underset{\text{N}_\text{1}}{\underline{\mathbf{x}^\text{1,1},..,\mathbf{x}^{\text{N}_\text{1}\text{,1}}}} ,...,\underset{\text{N}_\text{q}}{\underline{\mathbf{x}^\text{i,q},..,\mathbf{x}^{\text{i}+\text{N}_\text{q}\text{,q}}}},...,\underset{\text{N}_\text{Q}}{\underline{\mathbf{x}^{\text{N}_\text{tot}-\text{N}_\text{Q}\text{,Q}},..,\mathbf{x}^{\text{N}_\text{tot}\text{,Q}}}}\}^\mathsf{T}=\{\underset{\text{N}_\text{1}}{\underline{\mathbf{X}^\text{(1)}}},\underset{\text{N}_\text{2}}{\underline{\mathbf{X}^\text{(2)}}},...,\underset{\text{N}_\text{Q}}{\underline{\mathbf{X}^\text{(Q)}}}\}=\{\mathbf{x}^\text{i,q}\}^\mathsf{T},\\
        \mathbf{Y}=&\{\underset{\text{N}_\text{1}}{\underline{y^\text{1,1},..,y^{\text{N}_\text{1}\text{,1}}}} ,...,\underset{\text{N}_\text{q}}{\underline{y^\text{i,q},..,y^{\text{i}+\text{N}_\text{q}\text{,q}}}},...,\underset{\text{N}_\text{Q}}{\underline{y^{\text{N}_\text{tot}-\text{N}_\text{Q}\text{,Q}},..,y^{\text{N}_\text{tot}\text{,Q}}}}\}^\mathsf{T}=\{\underset{\text{N}_\text{1}}{\underline{\mathbf{y}^\text{(1)}}},\underset{\text{N}_\text{2}}{\underline{\mathbf{y}^\text{(2)}}},...,\underset{\text{N}_\text{Q}}{\underline{\mathbf{y}^\text{(Q)}}}\}=\{y^\text{i,q}\}^\mathsf{T},\\&(\text{i}\in[\text{1},\text{N}_\text{tot}],\text{q}\in[\text{1},\text{Q}]),
    \end{aligned}\label{mogp-obdata}
\end{equation}
where $\mathbf{Y}$ is the output long vector consisting of the decomposed scalar components of length $\text{N}_\text{tot}=\sum_\text{q=1}^\text{Q}\text{N}_\text{q}$, and $\mathbf{X}$ is the input matrix whose elements may be overlapped. 

For each output task, a corresponding task function $f_\text{q}$ (where $\text{q}\in[\text{1},\text{Q}]$) exists to map any input $\mathbf{x}$ to that output task component $y^\text{i,q}$. With identical consideration of observation noise of Eqn.~\ref{noise}, each output is modelled as
\begin{equation}
    \begin{aligned}
        y^\text{i,q}&=f_\text{q}(\mathbf{x}^\text{i})+\epsilon^\text{q,i},\\ \epsilon^\text{q,i}&\sim\mathcal{N}(\text{0},\sigma_{s,\text{q}}^\text{2}), \text{i} \in [\text{1},\text{N}_\text{tot}]\text{ and }\text{q}\in[\text{1},\text{Q}]. \label{mogp-likelihood}
    \end{aligned}
\end{equation}
Let $\mathbf{F}^{(\text{q})} = \big[f_\text{q}(\mathbf{x}^{\text{i},\text{q}})\big]_{\text{i}=1}^{N_\text{q}} \in \mathbb{R}^{N_\text{q}}$ and $\mathbf{F}=({\mathbf{F}^\text{(1)}}^\mathsf{T}, {\mathbf{F}^\text{(2)}}^\mathsf{T},..., {\mathbf{F}^\text{(Q)}}^\mathsf{T})^\mathsf{T}$. As SOGP places a GP prior in Eqn.~\ref{GP-prior}, MOGP places a joint Gaussian prior over all task functions, and it is expressed as

\begin{equation}
    (f_\text{1},f_\text{2},...,f_\text{Q})\sim\mathcal{GP}(\mathbf{m}_\Theta(\mathbf{x}),\mathbf{K}_\Theta(\mathbf{x},\mathbf{x}')),
\end{equation}

where all hyperparameters of covariance matrices are collected in $\Theta$, $\mathbf{m}_\Theta(\mathbf{x})\in \mathbb{R}^\text{Q}$ is the mean function, and $\mathbf{K}_\Theta(\mathbf{x},\mathbf{x}^\prime)\in \mathbb{R}^{\text{Q}\times\text{Q}}$ is the matrix-valued kernel, whose (q,$\text{q}^\prime$)-th scalar entry is the cross-covariance
\begin{equation}
    K_{\text{q}\text{q}^\prime}(\mathbf{x},\mathbf{x}^\prime)=\text{Cov}(f_\text{q}(\mathbf{x}),f_{\text{q}^\prime}(\mathbf{x}^\prime)).
\end{equation}
In the multi-output setting of MOGP, the matrix-valued kernel encodes both (i) the input-space correlation as in SOGP (Eqn.~\ref{GP-prior} \& \ref{KE-variance}) and (ii) the inter-task or cross-output correlation. The latter specifies how different task outputs co-vary, allowing the model to borrow statistical strength across tasks. Therefore, MOGP formulations naturally support multi-fidelity problems, and different approaches to model inter-task correlation will be presented in a later context. A simple demonstration example is the intrinsic coregionalization model (ICM)~\cite{Jeffrey1994}, which assumes all tasks share the uniform input-space kernel 
\begin{equation}
    k_{\text{q}\text{q}^\prime}(\mathbf{x},\mathbf{x}^\prime) = k(\mathbf{x},\mathbf{x}^\prime),
\end{equation}
and differ only through a positive semidefinite task–task covariance matrix $\mathbf{B}$ by

\begin{equation}
\begin{aligned}
    K_{\text{q}\text{q}^\prime}(\mathbf{x},\mathbf{x}^\prime)&=B_{\text{q}\text{q}^\prime}k_{\text{q}\text{q}^\prime}(\mathbf{x},\mathbf{x}^\prime) \label{mogp-icmkernel}
    \\ &= B_{\text{q}\text{q}^\prime}k(\mathbf{x},\mathbf{x}^\prime)
\end{aligned}
\end{equation}
where $B_{\text{q}\text{q}^\prime}$ is the $(\text{q},\text{q}^\prime)$-th scalar entry of $\mathbf{B}$, which quantifies the covariance strength between tasks $\text{q}$ and $\text{q}^\prime$, and $k_{\text{q}\text{q}^\prime}(\mathbf{x},\mathbf{x}^\prime)$ is the addressed covariance function in Eqn.~\ref{GP-prior} for one specific $(\text{q},\text{q}^\prime)$ entry. Since the selection of $k(\mathbf{x},\mathbf{x}^\prime)$ is uniform in ICM across all entries, a zero-mean GP prior of all stacked tasks is given by
\begin{equation}
\begin{aligned}
    &\mathbf{F}\sim\mathcal{N}(\mathbf{0},\mathbf{K}_{\mathbf{X}\mathbf{X}}),\\
    &\mathbf{K}_{\mathbf{X}\mathbf{X}} =
    \begin{bmatrix}
    B_\text{11}\,\mathbf{K}_{\mathbf{X}^{(\text{1})}\mathbf{X}^{(\text{1})}} & \cdots & B_\text{1Q}\,\mathbf{K}_{\mathbf{X}^{(\text{1})}\mathbf{X}^{(\text{Q})}} \\
    \vdots & \ddots & \vdots \\
    B_\text{Q1}\,\mathbf{K}_{\mathbf{X}^{(\text{Q})}\mathbf{X}^{(\text{1})}} & \cdots & B_\text{QQ}\,\mathbf{K}_{\mathbf{X}^{(\text{Q})}\mathbf{X}^{(\text{Q})}}
    \end{bmatrix},
\end{aligned}
\end{equation}
where $\mathbf{K}_{\mathbf{X}^{(\text{q})}\mathbf{X}^{(\text{q}^\prime)}} \in \mathbb{R}^{N_\text{q}\times N_{\text{q}^\prime}}$ is the covariance matrix for one $(\text{q},\text{q}^\prime)$ block, with entries
\[
\left[\mathbf{K}_{\mathbf{X}^{(\text{q})}\mathbf{X}^{(\text{q}^\prime)}}\right]_{\text{i}\text{j}}
=
k_\theta(\mathbf{x}^{\text{i},\text{q}},\mathbf{x}^{\text{j},\text{q}^\prime}),
\quad \text{i}\in[1,N_\text{q}],\;\text{j}\in[1,N_{\text{q}^\prime}].
\]

The MOGP prior is therefore written as
\begin{equation}
    p(\mathbf{F}\mid\mathbf{X},\{\Theta,\mathbf{B}\})
    =
    \mathcal{N}(\mathbf{F}\mid\mathbf{0},\mathbf{K}_{\mathbf{X}\mathbf{X}}).
\end{equation}

With Eqn.~\ref{mogp-likelihood}, the stacked likelihood is
\begin{equation}
    p(\mathbf{Y}\mid\mathbf{F},\mathbf{\Sigma}_s)
    =
    \mathcal{N}(\mathbf{Y}\mid \mathbf{F},\mathbf{\Sigma}_s),
\end{equation}
where
\[
\mathbf{\Sigma}_s=
\mathrm{blkdiag}\!\left(
\sigma_{s,\text{1}}^2\mathbf{I}_{N_\text{1}},
\sigma_{s,\text{2}}^2\mathbf{I}_{N_\text{2}},
\ldots,
\sigma_{s,\text{Q}}^2\mathbf{I}_{N_\text{Q}}
\right)
\]
is the block-diagonal noise covariance matrix.

By integrating out the latent function values $\mathbf{F}$, the marginal likelihood is
\begin{equation}
\begin{aligned}
   p(\mathbf{Y}\mid\mathbf{X},\zeta)
   &=
   \int
   p(\mathbf{Y}\mid \mathbf{F},\mathbf{\Sigma}_s)\,
   p(\mathbf{F}\mid \mathbf{X},\{\Theta,\mathbf{B}\})
   \,d\mathbf{F}\\
   &=
   \mathcal{N}(\mathbf{Y}\mid\mathbf{0},\mathbf{K}_{\mathbf{X}\mathbf{X}}+\mathbf{\Sigma}_s),
\end{aligned}
\end{equation}
where $\zeta=\{\Theta,\mathbf{B},\sigma_s^2\}$ denotes all hyperparameters.

The hyperparameters are learned by minimizing NLML:
\begin{equation}
\begin{aligned}
   \hat{\zeta}
   &=
   \arg\min_{\zeta}
   \left(-\log p(\mathbf{Y}\mid\mathbf{X},\zeta)\right)\\
   &=
   \frac{1}{2}\mathbf{Y}^\mathsf{T}
   (\mathbf{K}_{\mathbf{X}\mathbf{X}}+\mathbf{\Sigma}_s)^{-1}
   \mathbf{Y}
   +
   \frac{1}{2}\log\left|
   \mathbf{K}_{\mathbf{X}\mathbf{X}}+\mathbf{\Sigma}_s
   \right|
   +
   \frac{N_\text{tot}}{2}\log(2\pi).
\end{aligned}
\end{equation}

Then, posterior prediction can be performed using identical test input sets for all tasks,
\[
\mathbf{X}^*=\{ {\mathbf{x}^{*}}^\text{1}, {\mathbf{x}^{*}}^\text{2}, \ldots, {\mathbf{x}^{*}}^\text{M} \}^\mathsf{T}.
\]
The MOGP cross-covariance blocks are constructed as
\begin{equation}
    \begin{aligned}
        \mathbf{K}_{\mathbf{*}\mathbf{X}} &=
        \begin{bmatrix}
        B_\text{11}\,\mathbf{K}_{\mathbf{X}^{*}\mathbf{X}^\text{(1)}} & \cdots & B_\text{1Q}\,\mathbf{K}_{\mathbf{X}^{*}\mathbf{X}^\text{(Q)}} \\
        \vdots & \ddots & \vdots \\
        B_\text{Q1}\,\mathbf{K}_{\mathbf{X}^{*}\mathbf{X}^\text{(1)}} & \cdots & B_\text{QQ}\,\mathbf{K}_{\mathbf{X}^{*}\mathbf{X}^\text{(Q)}}
        \end{bmatrix},\\
        \mathbf{K}_{\mathbf{X}\mathbf{*}} &= \mathbf{K}_{\mathbf{*}\mathbf{X}}^\mathsf{T},\\
        \mathbf{K}_{\mathbf{*}\mathbf{*}} &=
        \begin{bmatrix}
        B_\text{11}\,\mathbf{K}_{\mathbf{X}^{*}\mathbf{X}^{*}} & \cdots & B_\text{1Q}\,\mathbf{K}_{\mathbf{X}^{*}\mathbf{X}^{*}} \\
        \vdots & \ddots & \vdots \\
        B_\text{Q1}\,\mathbf{K}_{\mathbf{X}^{*}\mathbf{X}^{*}} & \cdots & B_\text{QQ}\,\mathbf{K}_{\mathbf{X}^{*}\mathbf{X}^{*}}
        \end{bmatrix}.
    \end{aligned}
\end{equation}

The posterior distribution over latent outputs at $\mathbf{X}^*$ is
\begin{equation}
    \begin{aligned}
        \mathbf{F}_{*}\mid\mathbf{X},\mathbf{Y},\zeta
        &\sim
        \mathcal{N}(\bm{\mu}_*,\mathbf{\Sigma}_*),\\
        \bm{\mu}_*&=
        \mathbf{K}_{\mathbf{*}\mathbf{X}}
        (\mathbf{K}_{\mathbf{X}\mathbf{X}}+\mathbf{\Sigma}_s)^{-1}
        \mathbf{Y},\\
        \mathbf{\Sigma}_*&=
        \mathbf{K}_{\mathbf{*}\mathbf{*}}
        -
        \mathbf{K}_{\mathbf{*}\mathbf{X}}
        (\mathbf{K}_{\mathbf{X}\mathbf{X}}+\mathbf{\Sigma}_s)^{-1}
        \mathbf{K}_{\mathbf{X}\mathbf{*}}.
    \end{aligned}
\end{equation}

With Eqn.~\ref{mogp-likelihood}, the predictive posterior is
\begin{equation}
    \mathbf{Y}_{*}\mid\mathbf{X},\mathbf{Y},\zeta
    \sim
    \mathcal{N}\!\left(
    \bm{\mu}_*,
    \mathbf{\Sigma}_*
    +
    \mathrm{blkdiag}(
    \sigma_{s,\text{1}}^2\mathbf{I}_M,
    \sigma_{s,\text{2}}^2\mathbf{I}_M,
    \ldots,
    \sigma_{s,\text{Q}}^2\mathbf{I}_M)
    \right).
\end{equation}

The overall methodology of MOGP parallels that of SOGP, except that MOGP considers the additional cross-output correlation among the different tasks or output dimensions, which provides the fundamental basis for its strong performance in multi-fidelity settings where each fidelity is treated as a separate task. The demonstrated simple ICM model for handling cross-output correlation assumes that multiple tasks share the same spatial kernel (Eqn.~\ref{mogp-icmkernel}), which employs a unified kernel covariance function for every task-task pair and only a single covariance matrix $\mathbf{B}$ to represent the weights of each task-task component. In addition to the simple ICM, many other classic models, which introduce a higher level of manipulation on task-task correlation, will be introduced in later content.

\subsection{Kriging \& Co-kriging}
The concept of Kriging arose in geostatistics during the 1950s to 60s as a method to optimally interpolate a spatial field from sparse measurements~\cite{cressie1990origins}. The naming of this method originates from D.G. Krige, who pioneered the methodology by introducing several fundamental flaws in this technique~\cite{doi:10.10520/AJA0038223X4792,georges1963principles}. In the following decades, the method has been applied and developed, and many extended models within the Kriging family, including Co-kriging~\cite{doi:10.1098/rspa.2007.1900}, are widely utilized as reliable approximation techniques in the field of geospatial statistics~\cite{OLIVER01071990,10.1093/biomet/87.1.1}. Kriging and GP share the same primary goal, which is to predict an unknown function at new inputs and quantify uncertainty using a model of correlation between observations. For comparison, GP uses posterior mean and posterior variance, while Kriging uses best linear unbiased predictor (BLUP) and kriging variance. Despite different terminologies due to the discrepancy in machine learning and geospatial statistics, the mathematical essence of GP and Kriging is identical. The process of a single Kriging approximation exactly resembles the GP regression. 

\subsubsection{Kriging}
Given observations $\{(\mathbf{x}^\text{i},y^\text{i})\}_{\text{i}=1}^{\text{N}}$ with $\mathbf{x}^\text{i}\in\mathbb{R}^\text{d}$ and $y^\text{i}\in\mathbb{R}$, a prediction $y(\mathbf{x}^*)$ at a new input $\mathbf{x}^*$ with principled uncertainty quantification can be obtained via the Kriging process. As GP places a Gaussian prior, Kriging models an unknown field $f(\mathbf{x})$ as a trend or mean base term $\mu(\mathbf{x})$ plus a zero-mean Gaussian random field $Z(\mathbf{x})$, expressed as
\begin{equation}
    f(\mathbf{x})=\mu(\mathbf{x})+Z(\mathbf{x}). \label{kg-1}
\end{equation}
Here, $\mu(\mathbf{x})$ can be seen as the mean function in the GP prior. Kriging is generally divided into three types according to the modeling of $\mu$, including simple Kriging (SK, where $\mu$ is known and constant), ordinary Kriging (OK, where $\mu$ is unknown but constant), and universal Kriging (UK, where $\mu$ is a known linear basis with unknown coefficients).

The covariance of $Z(\mathbf{x})$ is defined as
\begin{equation}
    \text{Cov}(Z(\mathbf{x}), Z(\mathbf{x}^\text{i}))=\sigma^\text{2}\psi(\mathbf{x},\mathbf{x}^\text{i}),
\end{equation}
where $\psi(\mathbf{x},\mathbf{x}^\text{i})$ denotes the correlation between $Z(\mathbf{x})$ and $Z(\mathbf{x}^\text{i})$, and $\sigma^\text{2}$ is the process variance representing the amplitude scale. A commonly used anisotropic power exponential form for the correlation function is~\cite{doi:10.1098/rspa.2007.1900}
\begin{equation}
    \psi(\mathbf{x},\mathbf{x}^\text{i})=\exp\left( -\sum^\text{d}_{\text{j}=1}q_\text{j}|x_\text{j}-x_\text{j}^\text{i}|^{p_\text{j}} \right),
\end{equation}
where $x_\text{j}$ denotes the $\text{j}$-th component of $\mathbf{x}$, $q_\text{j}>0$ is a scale parameter, and $p_\text{j}\in(0,2]$ controls smoothness in the $\text{j}$-th dimension.

Accordingly, the correlation matrix $\mathbf{\Psi}$ is constructed with entries
\begin{equation}
    \mathbf{\Psi}_{\text{i}\text{j}}=\psi(\mathbf{x}^\text{i},\mathbf{x}^\text{j}).
\end{equation}

With Eqn.~\ref{kg-1}, the likelihood under ordinary Kriging (OK), which corresponds to the marginal likelihood in the GP framework, is given by
\begin{equation}
    \mathcal{N}(\mathbf{y}\mid\mu\mathbf{1},\sigma^\text{2}\mathbf{\Psi}).
\end{equation}

Then, similar to GP, maximum likelihood estimation (MLE)~\cite{MYUNG200390} is performed to obtain optimal hyperparameters $\hat{\eta}=\{\hat{\mu},\hat{\sigma}^\text{2},\hat{q},\hat{p}\}$. Using $\hat{\eta}$, the BLUP is given by
\begin{equation}
    y(\mathbf{x}^*)=\hat{\mu}+{\boldsymbol{\psi}^*}^\mathsf{T}\mathbf{\Psi}^{-1}(\mathbf{y}-\hat{\mu}\mathbf{1}),
\end{equation}
where $\boldsymbol{\psi}^*=\big[\psi(\mathbf{x}^*,\mathbf{x}^\text{i})\big]_{\text{i}=1}^{\text{N}}$, and $\hat{\mu}=\mathbf{1}^\mathsf{T}\mathbf{\Psi}^{-1}\mathbf{y}/(\mathbf{1}^\mathsf{T}\mathbf{\Psi}^{-1}\mathbf{1})$.

The mean-squared prediction error is given by
\begin{equation}
    \sigma^\text{2}_\text{Kri}=\sigma^\text{2}\left(
    1-{\boldsymbol{\psi}^*}^\mathsf{T}\mathbf{\Psi}^{-1}\boldsymbol{\psi}^*
    +\frac{(1-\mathbf{1}^\mathsf{T}\mathbf{\Psi}^{-1}\boldsymbol{\psi}^*)^2}{\mathbf{1}^\mathsf{T}\mathbf{\Psi}^{-1}\mathbf{1}}
    \right).
\end{equation}

\subsubsection{Co-kriging}
As MOGP handles a multi-dimensional output $\mathbf{y}$ compared to SOGP, Co-kriging extends the simple Kriging method to multiple, related random fields observed on different input sets. Co-kriging with ICM is equivalent to the MOGP described in the previous context. Consider Q-related regionalized variables, which are essentially the tasks in GP (Eqn.~\ref{mogp-obdata}),
\begin{equation}
    \{y_\text{q}=f_\text{q}(\mathbf{x})\}_{\text{q}=1}^{\text{Q}},\quad\mathbf{x}\in\mathbb{R}^\text{d},\;y_\text{q}\in\mathbb{R},
\end{equation}
each of which can be decomposed into a trend and a zero-mean fluctuation as in Eqn.~\ref{kg-1}, expressed as
\begin{equation}
    f_\text{q}(\mathbf{x})= \mu_\text{q}(\mathbf{x})+Z_\text{q}(\mathbf{x}).
\end{equation}
Here, the trend term $\mu_\text{q}(\mathbf{x})$ follows the usual SK, OK, or UK choices, and $Z_\text{q}(\mathbf{x})$ is encoded by the cross-covariance
\begin{equation}
    C_{\text{q}\text{q}^\prime}(\mathbf{x},\mathbf{x}^\prime)
    =
    \text{Cov}\!\big(Z_\text{q}(\mathbf{x}),Z_{\text{q}^\prime}(\mathbf{x}^\prime)\big),
    \quad \text{q},\text{q}^\prime=1,\ldots,Q.
\end{equation}
Under the ICM, a valid cross-covariance is specified as
\begin{equation}
    C_{\text{q}\text{q}^\prime}(\mathbf{x},\mathbf{x}^\prime)
    =
    B_{\text{q}\text{q}^\prime}\,C(\mathbf{x},\mathbf{x}^\prime), 
    \quad \mathbf{B}\succeq 0,
\end{equation}
where $\mathbf{B}$ is a positive semidefinite task--task covariance matrix.

Let $\mathbf{X}_\text{q}=\{\mathbf{x}^{\text{i},\text{q}}\}_{\text{i}=1}^{\text{N}_\text{q}}$ be the sampling set for variable $\text{q}$, and stack the observations as
\begin{equation}
    \mathbf{Y}
    =
    \big(
    \mathbf{y}_\text{1}(\mathbf{X}_\text{1})^\mathsf{T},
    \ldots,
    \mathbf{y}_\text{Q}(\mathbf{X}_\text{Q})^\mathsf{T}
    \big)^\mathsf{T}
    \in \mathbb{R}^{\text{N}_{\text{tot}}},
    \quad
    \text{N}_{\text{tot}}=\sum_{\text{q}=1}^{\text{Q}} \text{N}_\text{q}.
\end{equation}

The block covariance among all sampled locations is
\begin{equation}
    \mathbf{C}
    =
    \begin{bmatrix}
        \mathbf{C}_{\text{11}}(\mathbf{X}_\text{1},\mathbf{X}_\text{1}) & \cdots & \mathbf{C}_{\text{1Q}}(\mathbf{X}_\text{1},\mathbf{X}_\text{Q})\\
        \vdots & \ddots & \vdots\\
        \mathbf{C}_{\text{Q1}}(\mathbf{X}_\text{Q},\mathbf{X}_\text{1}) & \cdots & \mathbf{C}_{\text{QQ}}(\mathbf{X}_\text{Q},\mathbf{X}_\text{Q})
    \end{bmatrix}.
\end{equation}

Considering measurement noise as indicated in Eqn.~\ref{noise}, a block-diagonal nugget is included:
\begin{equation}
    \boldsymbol{\Sigma}_\varepsilon
    =
    \mathrm{blkdiag}\!\left(
    \sigma_{\varepsilon,\text{1}}^2 \mathbf{I}_{\text{N}_\text{1}},
    \ldots,
    \sigma_{\varepsilon,\text{Q}}^2 \mathbf{I}_{\text{N}_\text{Q}}
    \right).
\end{equation}

The observations’ covariance matrix is therefore
\begin{equation}
    \mathbf{R}=\mathbf{C}+\boldsymbol{\Sigma}_\varepsilon.
\end{equation}

The trend bases are included through a block design matrix. For UK,
\begin{equation}
    \mathbf{F}
    =
    \mathrm{blkdiag}(H_\text{1},\ldots,H_\text{Q}),
    \quad
    H_\text{q}
    =
    \big[
    h_\text{q}(\mathbf{x}^{\text{1},\text{q}}),
    \ldots,
    h_\text{q}(\mathbf{x}^{\text{N}_\text{q},\text{q}})
    \big]^\mathsf{T},
\end{equation}
where $h_\text{q}$ is a specified basis vector with unknown coefficients. For OK,
\begin{equation}
    \mathbf{F}
    =
    \mathrm{blkdiag}(\mathbf{1}_{\text{N}_\text{1}},\ldots,\mathbf{1}_{\text{N}_\text{Q}}).
\end{equation}
For SK, the trend is known and $\mathbf{F}$ is omitted.

Select a target variable $\text{t}\in\{1,\ldots,Q\}$ at a new location $\mathbf{x}^*$ and define the cross-covariance vector
\begin{equation}
    \mathbf{c}_*
    =
    \big(
    \mathbf{C}_{\text{t1}}(\mathbf{x}^*,\mathbf{X}_\text{1})^\mathsf{T},
    \ldots,
    \mathbf{C}_{\text{tQ}}(\mathbf{x}^*,\mathbf{X}_\text{Q})^\mathsf{T}
    \big)^\mathsf{T},
\end{equation}
and the corresponding trend basis
\begin{equation}
    \mathbf{f}_*
    =
    \begin{cases}
        h_\text{t}(\mathbf{x}^*) & \text{(UK)},\\
        1 & \text{(OK)},\\
        \text{(omitted)} & \text{(SK)}.
    \end{cases}
\end{equation}

The co-kriging weights $\mathbf{w}$ and Lagrange multipliers $\boldsymbol{\lambda}$ (for OK and UK) solve the bordered system
\begin{equation}
    \begin{bmatrix}
        \mathbf{R} & \mathbf{F}\\
        \mathbf{F}^\mathsf{T} & \mathbf{0}
    \end{bmatrix}
    \begin{bmatrix}
        \mathbf{w}\\
        \boldsymbol{\lambda}
    \end{bmatrix}
    =
    \begin{bmatrix}
        \mathbf{c}_*\\
        \mathbf{f}_*
    \end{bmatrix}.
\end{equation}

The BLUP and its mean-squared prediction error are
\begin{equation}
    y_\text{Co-kri}(\mathbf{x}^*) = \mathbf{w}^\mathsf{T} \mathbf{Y},
\end{equation}
\vspace{-0.5em}
\begin{equation}
    \sigma^2_{\text{Co-kri}}(\mathbf{x}^*) =
    \begin{cases}
        C_{\text{tt}}(\mathbf{x}^*,\mathbf{x}^*) - \mathbf{c}_*^\mathsf{T} \mathbf{R}^{-1}\mathbf{c}_* & \text{(SK)},\\[6pt]
        C_{\text{tt}}(\mathbf{x}^*,\mathbf{x}^*) - 
        \begin{bmatrix}\mathbf{c}_*^\mathsf{T} & \mathbf{f}_*^\mathsf{T}\end{bmatrix}
        \begin{bmatrix}
            \mathbf{R} & \mathbf{F}\\
            \mathbf{F}^\mathsf{T} & \mathbf{0}
        \end{bmatrix}^{-1}
        \begin{bmatrix}\mathbf{c}_*\\ \mathbf{f}_*\end{bmatrix}
        & \text{(OK/UK)}.
    \end{cases}
\end{equation}

\begin{figure}[hbt!]
\centering
\includegraphics[width=0.6\linewidth]{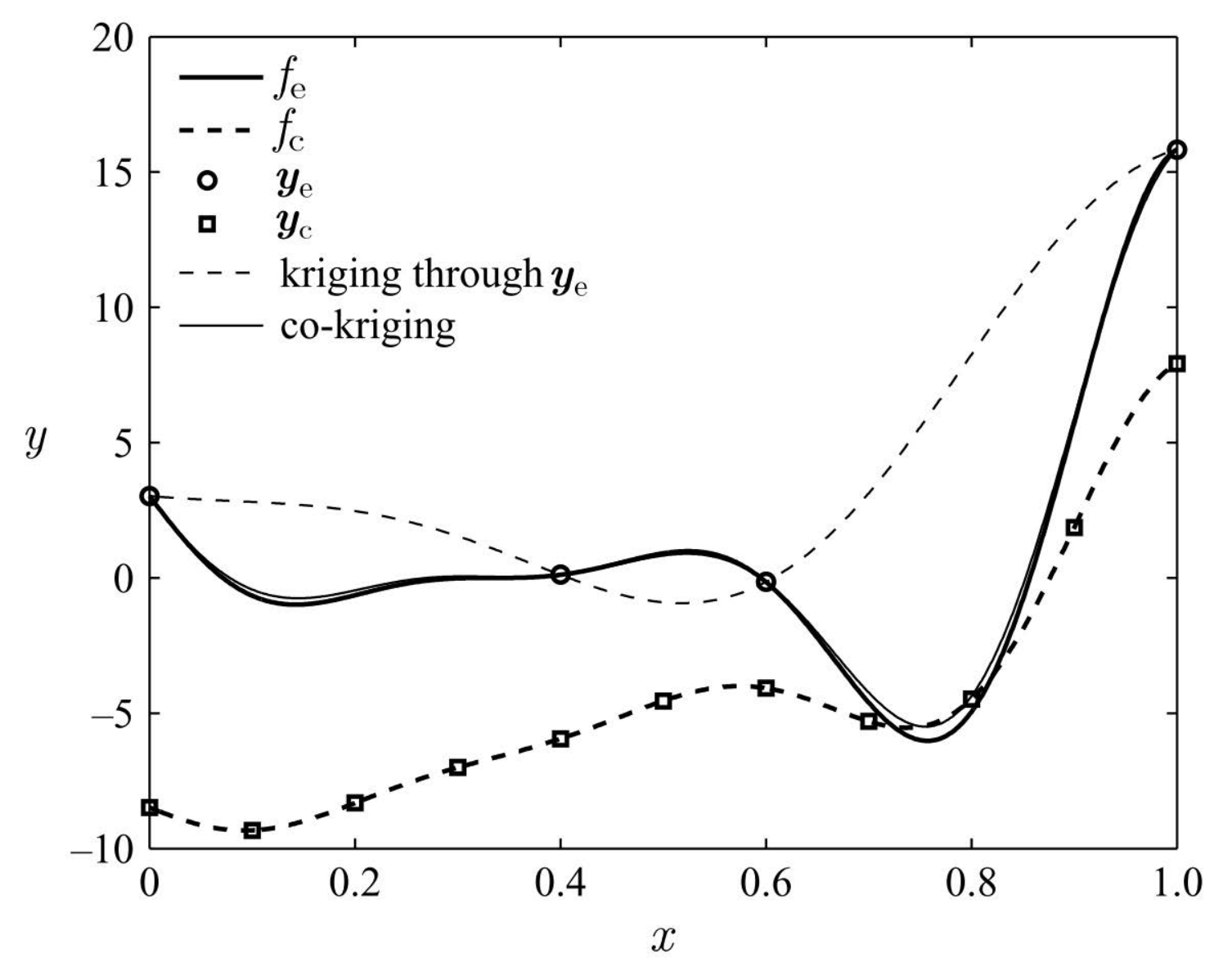}
\caption{Example of one-dimensional input case using Kriging and Co-kriging of Forrester~\emph{et al.}~\cite{doi:10.1098/rspa.2007.1900}. The $f_\text{e}=(\text{6}x-\text{2})^\text{2}\sin(\text{12}x-\text{4})$ is the expensive observation data representing HF data, and the $f_\text{c}=Af_\text{e}+B(x-\text{0.5})-C$ with $A=\text{0.5}$, $B=\text{10}$, and $C=\text{-5}$ is the cheap observation data representing LF data. The kriging approximation using four expensive data points ($y_\text{e}$) has been significantly improved using extensive sampling from the cheap function ($y_\text{c}$).}
\label{fig:kriging_demo}
\end{figure}

The hyperparameters $\zeta=\{\mathbf{B},\sigma^\text{2}_{\varepsilon,\text{q}},\theta_h\}$ can be estimated by MLE or restricted maximum likelihood (REML)~\cite{Corbeil01021976}. Fig.~\ref{fig:kriging_demo} illustrates a simple example of BLUP prediction using Kriging and Co-kriging~\cite{doi:10.1098/rspa.2007.1900}.

\subsection{Multi-fidelity Modeling Approaches}
As it is indicated, what distinguishes MOGP from SOGP or Co-kriging from Kriging is the manipulation of cross-output correlation among the different tasks or dimension variables in the output, which offers a large degree of freedom in the modeling the correlation among varying fidelity levels of data, thereby entitling MOGP or Co-kriging natural advantages in handling multi-fidelity problems. In addition to the single demonstrated ICM modeling, several classic modeling approaches in MOGP or Co-kriging, including the linear model of coregionalization (LMC), the auto-regressive (AR) model, the non-linear auto-regressive multi-fidelity Gaussian process (NARGP), and multi-fidelity deep Gaussian process (MF-DGP), will be introduced in this section.

\subsubsection{Linear Model of Coregionalization}
The introduced ICM uses a unified kernel function and a single covariance matrix $\mathbf{B}$ for cross-output modeling as shown in Eqn.~\ref{mogp-icmkernel}. The scalar coefficient $B_{\text{q}\text{q}^\prime}$ denotes the weight of a specific entry block $(\text{q},\text{q}^\prime)$, and the corresponding block is expressed as
\begin{equation}
    K_{\text{q}\text{q}^\prime}(\mathbf{x},\mathbf{x}^\prime)
    =
    B_{\text{q}\text{q}^\prime}\,k(\mathbf{x},\mathbf{x}^\prime).
    \label{lmc_entry}
\end{equation}

This formulation assumes that all task correlations share an identical spatial structure described by a single kernel function $k(\mathbf{x},\mathbf{x}^\prime)$. While this assumption leads to a simple and computationally efficient model, it can be restrictive when different tasks exhibit distinct correlation patterns across the input space.

To overcome this limitation, the linear model of coregionalization (LMC)~\cite{goulard1992linear,MAL-036} generalizes ICM by introducing multiple kernel functions through a linear combination. The $(\text{q},\text{q}^\prime)$-th entry is given by
\begin{equation}
\begin{aligned}
    K_{\text{q}\text{q}^\prime}(\mathbf{x},\mathbf{x}^\prime)
    &=
    \sum_{\text{r}=1}^{\text{R}} B_{\text{q}\text{q}^\prime}^{(\text{r})}\,k_{\text{r}}(\mathbf{x},\mathbf{x}^\prime),
\end{aligned}
\label{lmc-general}
\end{equation}
where $\{k_{\text{r}}\}_{\text{r}=1}^{\text{R}}$ are kernel functions capturing different spatial correlation structures, and $B_{\text{q}\text{q}^\prime}^{(\text{r})}$ are the corresponding coregionalization weights. The ICM is therefore a special case of LMC when $\text{R}=1$.

This decomposition provides a more flexible representation of multi-output covariance: each kernel $k_{\text{r}}$ models a distinct mode of variation in the input space, while the associated matrix $\mathbf{B}^{(\text{r})}$ determines how this mode is shared across different tasks. As a result, LMC can capture heterogeneous relationships between outputs that cannot be represented by a single shared kernel.

From a generative perspective, LMC defines each output component as a linear combination of latent Gaussian processes,
\begin{equation}
    f_\text{q}(\mathbf{x})=\sum_{\text{r}=1}^{\text{R}} a_{\text{q},\text{r}}\,u_{\text{r}}(\mathbf{x}),
    \label{lmc-rank1}
\end{equation}
where $\{u_{\text{r}}(\mathbf{x})\}_{\text{r}=1}^{\text{R}}$ are independent latent functions following zero-mean GPs with covariance
\[
\text{Cov}(u_{\text{r}}(\mathbf{x}),u_{\text{r}^\prime}(\mathbf{x}^\prime))
=
\delta_{\text{r}\text{r}^\prime}\,k_{\text{r}}(\mathbf{x},\mathbf{x}^\prime),
\]
and $a_{\text{q},\text{r}}$ are scalar mixing coefficients.

This construction provides a clear interpretation: each latent process $u_{\text{r}}$ represents a shared hidden factor influencing all outputs, and the coefficients $a_{\text{q},\text{r}}$ determine how strongly each task depends on that factor. Consequently, correlations between tasks arise naturally from shared latent structures.

Since each latent GP contributes a rank-1 covariance structure across outputs, Eqn.~\ref{lmc-rank1} yields a low-rank representation of the coregionalization matrix. While this can be efficient, it may limit expressiveness when modeling complex inter-task relationships~\cite{teh2005semiparametric,NIPS2007_66368270}. To address this, multiple independent latent processes are introduced for each kernel:
\begin{equation}
    f_\text{q}(\mathbf{x})=\sum_{\text{r}=1}^{\text{R}}\sum_{\text{c}=1}^{C_\text{r}} a^{\text{c}}_{\text{q},\text{r}}\,u_{\text{r}}^{\text{c}}(\mathbf{x}),
    \label{lmc-highrank}
\end{equation}
where $\{u_{\text{r}}^{\text{c}}(\mathbf{x})\}$ are independent realizations sharing the same kernel $k_{\text{r}}$.

The resulting cross-covariance is
\begin{equation}
    \text{Cov}(f_\text{q}(\mathbf{x}),f_{\text{q}^\prime}(\mathbf{x}^\prime))
    =
    \sum_{\text{r}=1}^{\text{R}}\sum_{\text{c}=1}^{C_\text{r}}
    a^{\text{c}}_{\text{q},\text{r}}\,a^{\text{c}}_{\text{q}^\prime,\text{r}}\,k_{\text{r}}(\mathbf{x},\mathbf{x}^\prime).
\end{equation}

Thus, the coregionalization matrix entries can be expressed as
\begin{equation}
    B^{(\text{r})}_{\text{q}\text{q}^\prime}
    =
    \sum_{\text{c}=1}^{C_\text{r}} a^{\text{c}}_{\text{q},\text{r}}\,a^{\text{c}}_{\text{q}^\prime,\text{r}},
\end{equation}
which guarantees positive semidefiniteness by construction.

Accordingly, the full covariance matrix can be written in a compact Kronecker form
\begin{equation}
    \mathbf{K}_{\mathbf{X}\mathbf{X}}
    =
    \sum_{\text{r}=1}^{\text{R}} \mathbf{B}^{(\text{r})} \otimes \mathbf{K}_{\text{r}},
    \label{lmc-geq}
\end{equation}
where $\mathbf{B}^{(\text{r})}\in\mathbb{R}^{\text{Q}\times\text{Q}}$ captures inter-task correlations and $\mathbf{K}_{\text{r}}$ is the kernel matrix with entries
\[
[\mathbf{K}_{\text{r}}]_{\text{i}\text{j}} = k_{\text{r}}(\mathbf{x}^\text{i},\mathbf{x}^\text{j}).
\]

If all tasks share identical input locations ($\mathbf{X}^{(\text{1})}=\cdots=\mathbf{X}^{(\text{Q})}$), this reduces to a sum of separable kernels,
\begin{equation}
    \mathbf{K}_{\mathbf{X}\mathbf{X}}
    =
    \sum_{\text{r}=1}^{\text{R}} \mathbf{B}^{(\text{r})} \otimes \mathbf{K}_{\text{r}}.
    \label{lmc-seq}
\end{equation}

Overall, LMC provides a flexible and interpretable framework for modeling multi-output correlations by decomposing them into shared latent processes. Compared to ICM, it significantly enhances modeling capacity by allowing multiple correlation structures, while still maintaining a principled Gaussian process formulation.

\subsubsection{Auto-Regressive}
The AR method by Kennedy and O’Hagan~\cite{10.1093/biomet/87.1.1} is one of the most classic and widely used approaches in MFSM. It introduces a linear dependency between adjacent fidelities in a hierarchical autoregressive fusion scheme. The AR model assumes that output tasks are ordered by fidelity level (from the lowest $\text{1}$ to the highest $\text{Q}$), where a higher-fidelity GP prior $f_\text{q}$ is expressed in terms of the adjacent lower-fidelity one:
\begin{equation}
    f_\text{q}(\mathbf{x})
    =
    \rho_{\text{q}-1}(\mathbf{x})\,f_{\text{q}-1}(\mathbf{x})
    +
    \delta_\text{q}(\mathbf{x}),
    \quad \text{q}\in[2,Q].
    \label{AR1}
\end{equation}

This formulation provides an intuitive interpretation of multi-fidelity modeling: the higher-fidelity response is constructed by scaling the lower-fidelity prediction and correcting it with a discrepancy term. The scaling factor $\rho_{\text{q}-1}$ captures the global linear relationship between adjacent fidelity levels, while $\delta_\text{q}$ accounts for the local bias that cannot be explained by simple scaling.

It is worth noting that Eqn.~\ref{AR1} implies a conditional independence structure: the distribution of $f_\text{q}$ depends only on $f_{\text{q}-1}$, even though all lower-fidelity models $f_\text{1},\dots,f_{\text{q}-1}$ may be available. In other words, once $f_{\text{q}-1}$ is known, no additional information from cheaper fidelities contributes to $f_\text{q}$. This first-order Markov structure is fundamental to the AR model, as it yields a tractable covariance formulation and enables efficient recursive training strategies~\cite{10.1093/biomet/87.1.1,LeGratiet_2014}.

Assume the lowest fidelity follows
\[
f_\text{1}(\mathbf{x})\sim\mathcal{GP}(0,k_\text{1}(\mathbf{x},\mathbf{x}^\prime)),
\]
and each discrepancy term follows
\[
\delta_\text{q}(\mathbf{x})\sim\mathcal{GP}(0,k_\text{q}(\mathbf{x},\mathbf{x}^\prime)).
\]
For constant scaling $\rho_{\text{q}-1}$, the covariance structure becomes
\begin{equation}
\begin{aligned}
    \text{Cov}(f_\text{1}(\mathbf{x}),f_\text{1}(\mathbf{x}^\prime))
    &=k_\text{1}(\mathbf{x},\mathbf{x}^\prime),\\
    \text{Cov}(f_\text{2}(\mathbf{x}),f_\text{1}(\mathbf{x}^\prime))
    &=\rho_\text{1}\,k_\text{1}(\mathbf{x},\mathbf{x}^\prime),\\
    \text{Cov}(f_\text{2}(\mathbf{x}),f_\text{2}(\mathbf{x}^\prime))
    &=\rho_\text{1}^2 k_\text{1}(\mathbf{x},\mathbf{x}^\prime)+k_\text{2}(\mathbf{x},\mathbf{x}^\prime).
\end{aligned}
\end{equation}

This recursive structure shows how uncertainty propagates through fidelity levels: lower-fidelity correlations are inherited and scaled, while each level introduces additional variability through its discrepancy kernel. In general,
\begin{equation}
\text{Cov}(f_\text{q}(\mathbf{x}),f_{\text{q}^\prime}(\mathbf{x}^\prime))
=
\sum_{\text{j}=1}^{\min(\text{q},\text{q}^\prime)}
\left(\prod_{\text{p}=\text{j}}^{\text{q}-1}\rho_\text{p}\right)
\left(\prod_{\text{p}=\text{j}}^{\text{q}^\prime-1}\rho_\text{p}\right)
k_\text{j}(\mathbf{x},\mathbf{x}^\prime).
\end{equation}

From another perspective, the AR model can be interpreted within the LMC framework. Define latent Gaussian processes
\begin{equation}
\begin{aligned}
    u_\text{1}(\mathbf{x})&=f_\text{1}(\mathbf{x}),\\
    u_\text{q}(\mathbf{x})&=\delta_\text{q}(\mathbf{x}),\quad \text{q}\in[2,Q],
\end{aligned}
\end{equation}
then each output can be expressed as
\begin{equation}
    f_\text{q}(\mathbf{x})
    =
    \sum_{\text{r}=1}^{\text{q}} \alpha_{\text{q},\text{r}}\,u_\text{r}(\mathbf{x}),
\end{equation}
with coefficients
\begin{equation}
\alpha_{\text{q},\text{r}} =
\begin{cases}
\prod_{\text{p}=\text{r}}^{\text{q}-1}\rho_\text{p}, & \text{r}<\text{q},\\
1, & \text{r}=\text{q},\\
0, & \text{r}>\text{q}.
\end{cases}
\end{equation}

This formulation reveals that AR corresponds to a structured LMC model, where each fidelity level accumulates contributions from latent processes associated with all lower levels. The resulting coregionalization matrix is
\begin{equation}
    B^{(\text{r})}_{\text{q}\text{q}^\prime}
    =
    \alpha_{\text{q},\text{r}}\,\alpha_{\text{q}^\prime,\text{r}},
    \quad \text{r}\in[1,Q].
    \label{Bmat_AR1}
\end{equation}

Therefore, AR can be viewed as a special case of LMC with a highly structured coregionalization matrix that encodes the hierarchical dependency among fidelities. This perspective also clarifies the limitation of AR: the assumption of a linear scaling relationship across fidelity levels may be overly restrictive in problems where the correlation between fidelities is strongly nonlinear, such as complex composite materials with intricate mechanical responses. This motivates the development of more flexible nonlinear multi-fidelity models, which will be introduced in the next section.

\subsubsection{Non-linear Auto-Regressive multi-fidelity Gaussian Process}
NARGP can be seen as a generalized version of the AR fusion approach by Perdikaris~\emph{et al.}~\cite{10.1098/rspa.2016.0751}, which replaces the linear mapping between hierarchical fidelities with a nonlinear one. Similar to the linear formulation in Eqn.~\ref{AR1}, the nonlinear relation is expressed as
\begin{equation}
    f_\text{q}(\mathbf{x}) = z_\text{q-1}\left(  f_\text{q-1}(\mathbf{x}) \right)+\delta_\text{q}(\mathbf{x}), \quad \text{q}\in\text{[2,Q]}.
\label{NARGP}
\end{equation}
where $z_\text{q-1}$ is an unknown nonlinear function mapping the lower-fidelity output to the higher-fidelity one.

This formulation significantly enhances modeling flexibility: instead of assuming a global linear relationship as in AR, NARGP allows complex nonlinear transformations between fidelities. Such flexibility is particularly important in applications where discrepancies between fidelities cannot be captured by simple scaling, such as highly nonlinear material behaviors.

If the mapping $z_\text{q-1}$ is assigned a GP prior, then the term $z_\text{q-1}\big(f_\text{q-1}(\mathbf{x})\big)$ becomes a composition of two Gaussian processes. Since $f_\text{q-1}$ is itself a random function, this leads to a deep Gaussian process (DGP) structure~\cite{damianou2013deep}. Consequently, the resulting posterior of $f_\text{q}$ is no longer Gaussian and becomes analytically intractable.

To retain nonlinear expressiveness while avoiding the computational complexity of DGPs, Perdikaris~\emph{et al.}~\cite{10.1098/rspa.2016.0751} adopt the recursive inference strategy proposed by Le Gratiet and Garnier~\cite{LeGratiet_2014}, where the random variable $f_\text{q-1}(\mathbf{x})$ is replaced by a deterministic estimate $f^*_\text{q-1}(\mathbf{x})$. Under this approximation, the model becomes
\begin{equation}
    f_\text{q}(\mathbf{x})=g_\text{q} \left(\mathbf{x},f^*_\text{q-1}(\mathbf{x})\right) \coloneqq z_\text{q-1}\left(f^*_\text{q-1}(\mathbf{x}) \right)+\delta_\text{q}(\mathbf{x}),
\label{NARGP_update}
\end{equation}
where
\[
g_\text{q}\sim\mathcal{GP}\left(0,\,k_\text{q}\big([\mathbf{x},f^*_\text{q-1}(\mathbf{x})],[\mathbf{x}^\prime,f^*_\text{q-1}(\mathbf{x}^\prime)]\big)\right),
\]
and $z_\text{q-1}$ and $\delta_\text{q}$ are assumed independent, following the same assumption as in AR~\cite{10.1093/biomet/87.1.1}.

This approximation transforms the original hierarchical stochastic model into a sequence of standard GP regressions defined on augmented input spaces. Each fidelity level is therefore trained independently using the predictions from the previous level as additional inputs.

The GP priors are assigned as
\begin{equation}
    \begin{aligned}
        z_\text{q-1}(f^*)&\sim\mathcal{GP}(0,k_\text{z,q-1}(f^*,{f^*}^\prime)),\\
        \delta_\text{q}(\mathbf{x})&\sim\mathcal{GP}(0,k_{\delta\text{,q}}(\mathbf{x},\mathbf{x}^\prime)).
    \end{aligned}
\end{equation}

An important consequence of the recursive approximation is that cross-covariances between different fidelity levels vanish. Since $f_\text{q}$ depends on the deterministic quantity $f^*_{\text{q}-1}$ rather than the random process $f_{\text{q}-1}$, the stochastic coupling between fidelity levels is removed. As a result, the covariance structure becomes block-diagonal in the fidelity dimension.

Accordingly, the covariance at each fidelity level is
\begin{equation}
    \begin{aligned}
        \text{Cov}\left(f_\text{1}(\mathbf{x}),f_\text{1}(\mathbf{x}^\prime)\right)&=k_\text{1}(\mathbf{x},\mathbf{x}^\prime),\\
        \text{Cov}\left(f_\text{q}(\mathbf{x}),f_\text{q}(\mathbf{x}^\prime)\right)&=k_\text{z,q-1}\left(f^*_\text{q-1}(\mathbf{x}),f^*_\text{q-1}(\mathbf{x}^\prime)\right)+k_{\delta\text{,q}}(\mathbf{x},\mathbf{x}^\prime), \quad (\text{q}\in\text{[2,Q]}).
    \end{aligned}
\end{equation}

Given local observations $\mathbf{X}^\text{(q)}=\{\mathbf{x}^\text{1},\mathbf{x}^\text{2},\dots,\mathbf{x}^{\text{N}_\text{q}}\}$ at fidelity level $\text{q}$, the covariance matrix for this level is
\begin{equation}
    \mathbf{K}_\text{q}=\Big[
    k_\text{z,q-1}\big(f^*_\text{q-1}(\mathbf{x}^\text{i}),f^*_\text{q-1}(\mathbf{x}^\text{j})\big)
    +
    k_{\delta\text{,q}}(\mathbf{x}^\text{i},\mathbf{x}^\text{j})
    \Big]_{\text{i}\text{j}}.
\end{equation}

Therefore, the overall covariance matrix across all fidelity levels is
\begin{equation}
    \mathbf{K}=\mathrm{blkdiag}(\mathbf{K}_\text{1},\mathbf{K}_\text{2},\dots,\mathbf{K}_\text{Q}).
\end{equation}

While NARGP provides a flexible nonlinear mapping across fidelities, this recursive decoupling introduces an important limitation. Since each fidelity level is trained independently, no information flows backward from higher to lower fidelities. Consequently, lower-fidelity models remain fixed after their initial training and cannot benefit from higher-fidelity observations. This limitation motivates the development of more advanced models, such as MF-DGP, which enable bidirectional information flow across fidelity levels.

\subsubsection{Multi-Fidelity Deep Gaussian Process}
Similar to NARGP, MF-DGP assumes a nonlinear mapping between fidelities in a hierarchical structure, expressed as
\begin{equation}
    f_\text{q}(\mathbf{x})=g_\text{q}\left(\mathbf{x},f_{\text{q-1}}(\mathbf{x})\right).
\end{equation}

As discussed in NARGP, replacing the random variable $f_{\text{q-1}}(\mathbf{x})$ with a deterministic estimate $f^*_{\text{q-1}}(\mathbf{x})$ leads to a tractable model but cuts off uncertainty propagation across fidelity levels. In contrast, MF-DGP retains the stochastic nature of the hierarchical mapping by propagating sampled latent variables $\tilde{f}_{\text{q-1}}(\mathbf{x})$ through layers using variational inference~\cite{damianou2013deep,NIPS2017_82089746}.

This key difference enables MF-DGP to maintain both nonlinear expressiveness and coherent uncertainty propagation across fidelities, avoiding the decoupled structure of NARGP.

Under this formulation, each mapping function is assigned a GP prior on an augmented input space,
\begin{equation}
    \begin{aligned}
        &g_{\text{q}}(h)\sim\mathcal{GP}\!\left(0,\,k_{\text{q}}(h,h^\prime)\right),\\
        &\text{where}\quad
        h_{\text{q}}(\mathbf{x})=
        \begin{cases}
            \mathbf{x}, & \text{q}=1,\\[3pt]
            [\,\mathbf{x},\,\tilde{f}_{\text{q-1}}(\mathbf{x})\,], & \text{q}\ge 2.
        \end{cases}
    \end{aligned}
\end{equation}
Here, the input to each GP is augmented by the latent output from the previous fidelity level, allowing the model to learn complex nonlinear transformations across fidelities.

To enable scalable inference, MF-DGP adopts a sparse variational GP framework. A set of inducing inputs $\mathbf{Z}_{\text{q}} = \{\mathbf{z}_\text{m}\}_{\text{m}=1}^{\text{M}}$ and corresponding inducing variables $\mathbf{u}_{\text{q}} = g_{\text{q}}(\mathbf{Z}_{\text{q}})$ are introduced, following the prior
\begin{equation}
    \mathbf{u}_{\text{q}}
    \sim
    \mathcal{N}\!\left(\mathbf{0},\,K^{(\text{q})}_{\mathbf{Z}_{\text{q}}\mathbf{Z}_{\text{q}}}\right),
\end{equation}
where $K^{(\text{q})}_{\mathbf{Z}_{\text{q}}\mathbf{Z}_{\text{q}}}\in\mathbb{R}^{\text{M}\times\text{M}}$ is the covariance matrix.

Conditioned on $\mathbf{u}_{\text{q}}$, the latent function admits the GP conditional form
\begin{equation}
\begin{aligned}
     f_{\text{q}}(\mathbf{x}) \mid \mathbf{u}_{\text{q}}
    &= A_{\text{q}}(\mathbf{x})\,\mathbf{u}_{\text{q}}
      + \xi_{\text{q}}(\mathbf{x}),\\
    \text{where } 
    A_{\text{q}}(\mathbf{x})
    &= K^{(\text{q})}_{\mathbf{x}\mathbf{Z}_{\text{q}}}
      \big(K^{(\text{q})}_{\mathbf{Z}_{\text{q}}\mathbf{Z}_{\text{q}}}\big)^{-1},\\
    \xi_{\text{q}}(\mathbf{x})
    &\sim 
    \mathcal{N}\!\bigg(
        0,\;
        K^{(\text{q})}_{\mathbf{x}\mathbf{x}}
        - K^{(\text{q})}_{\mathbf{x}\mathbf{Z}_{\text{q}}}
          \big(K^{(\text{q})}_{\mathbf{Z}_{\text{q}}\mathbf{Z}_{\text{q}}}\big)^{-1}
          K^{(\text{q})}_{\mathbf{Z}_{\text{q}}\mathbf{x}}
    \bigg).
\end{aligned}
\end{equation}

Since the true posterior $p(\mathbf{u}_{\text{q}} \mid \mathcal{D}_{\text{q}})$ is intractable, MF-DGP employs a variational approximation
\begin{equation}
    \mathcal{Q}(\mathbf{u}_{\text{q}})
    = \mathcal{N}(\mathbf{m}_{\text{q}},\,\mathbf{S}_{\text{q}}),
\end{equation}
where $\mathbf{m}_{\text{q}}$ and $\mathbf{S}_{\text{q}}$ are learned jointly across all fidelity levels.

This induces a variational posterior over latent function values,
\begin{equation}
    \mathcal{Q}\!\left(f_{\text{q}}(\mathbf{x})\right)
    = \int 
        p\!\left(f_{\text{q}}(\mathbf{x}) \mid \mathbf{u}_{\text{q}}\right)\,
        \mathcal{Q}(\mathbf{u}_{\text{q}})\,
        d\mathbf{u}_{\text{q}}.
\end{equation}

A stochastic realization is obtained by sampling
\begin{equation}
    \tilde{f}_{\text{q}}(\mathbf{x})
    \sim 
    \mathcal{Q}\!\left(f_{\text{q}}(\mathbf{x})\right)
    =
    \mathcal{N}\!\Big(
        A_{\text{q}}(\mathbf{x})\,\mathbf{m}_{\text{q}},\;
        A_{\text{q}}(\mathbf{x})\,\mathbf{S}_{\text{q}}A_{\text{q}}^\top(\mathbf{x})
        + \mathrm{Var}[\xi_{\text{q}}(\mathbf{x})]
    \Big),
\end{equation}
which is then propagated as input to the next fidelity level. Through this recursive stochastic mapping, MF-DGP enables uncertainty to flow across layers while preserving nonlinear transformations between fidelities. Compared to NARGP, this results in a fully coupled probabilistic model, where information can be refined across fidelity levels rather than being fixed in a one-way hierarchical structure.

In summary, this section briefly reviewed several of the most classic fusion approaches in MOGP or Co-kriging, specifically, the generic LMC providing the interpretable formulation, the AR model incorporating an explicit linear autoregressive dependence between fidelity levels, the NARGP introducing the nonlinear mapping, and the MF-DGP propagating uncertainty in a full hierarchical architecture. Those employ varying manipulation in correlation among fidelities of observation data, making the MOGP-based methodology useful in the optimization and UQ in studying composite mechanics. As multi-fidelity datasets grow in dimension, nonlinearity, and heterogeneity, the scalability and expressiveness limits of MOGP-based models have driven increasing interest in the alternative multi-fidelity neural networks, which will be introduced in the following section.

\section{Neural-Networks-based Multi-Fidelity Surrogate Modeling} \label{Section NN-based MFSM}
As it is mentioned, the purpose of MFSM is to replace a certain inefficient physical or engineering process with an efficient surrogate, which uses a fusion technique to manipulate data from multiple fidelity sources and utilizes them to train an instant mapping directly from input to output regardless of the complexity of the original process. Building upon the basis of statistics, GP- and Kriging-based fusion families are widely applied across various scientific and engineering areas for their efficiency and interpretability. In recent decades, the NN-based deep learning surrogate modeling has gradually become the most preferred approach, and it has formed the basis of many key advancements in diverse fields. This review section will first introduce the fundamentals of neural networks and some of the neural network variants that are typically employed in the surrogate modeling of composite mechanics, and then extend the NN-based surrogate to the multi-fidelity framework, where two typical MF fusion architectures that are widely applied in composite mechanics will be introduced.

\subsection{Neural Networks}
\begin{figure}[hbt!]
\centering
\includegraphics[width=0.6\linewidth]{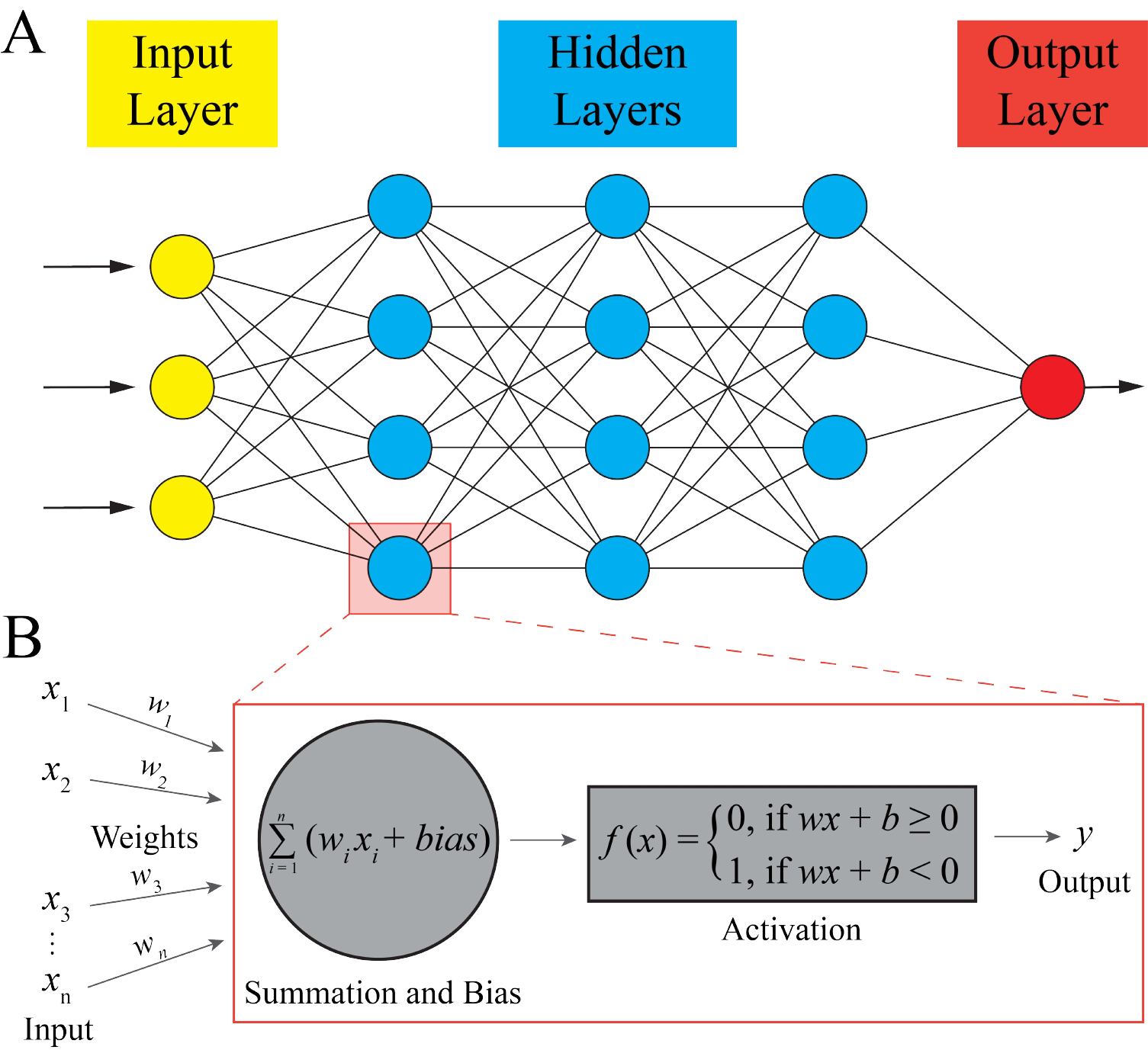}
\caption{Sketch of neural networks: (A) basic architecture of ANN, and (B) typical calculation inside one neuron.}
\label{fig:nn1}
\end{figure}

Neural network is a mathematical tool inspired by the biological nervous system and is used to solve a wide range of engineering and scientific problems by recognizing the underlying relationships from the available data~\cite{anderson1992artificial}. The first layer that receives the given data is defined as the input layer, while the last layer that gives prediction data is defined as the output layer. Between these two layers, there is at least one layer defined as the hidden layer, which contains many neural nodes and is responsible for all the computations~\cite{685648}. A simple and generic artificial neural network (ANN) architecture is shown in Fig.~\ref{fig:nn1}~A. There are 
estimation algorithms within the network that assign synaptic weights and bias to the input parameters and then calculate the output. This ANN architecture can be mathematically defined as 
\begin{equation}
    \mathbf{y}=F(\mathbf{x})=A(\mathbf{W}\times \mathbf{x}+\mathbf{b}_\text{in})+\mathbf{b}_\text{out},
\end{equation}
where $\mathbf{y}$ is the output vector, $\mathbf{x}$ is the input vector, $A$ is the activation function, $\mathbf{W}$ is the matrix that contains the synaptic weights, $\mathbf{b}_\text{in}$ is the column vector of biases from the input layer to the hidden layer and $\mathbf{b}_\text{out}$ is the column vector of biases from the hidden layer to the output layer (Fig.~\ref{fig:nn1}~B). The activation function is nonlinear and has many specific forms (sigmoid, Tanh, softmax, softplus etc.), and it enables the neural network to perform well in approximating complex mapping from inputs to outputs.

\subsubsection{Simple Feedforward Neural Network}

\begin{figure}[hbt!]
\centering
\includegraphics[width=0.6\linewidth]{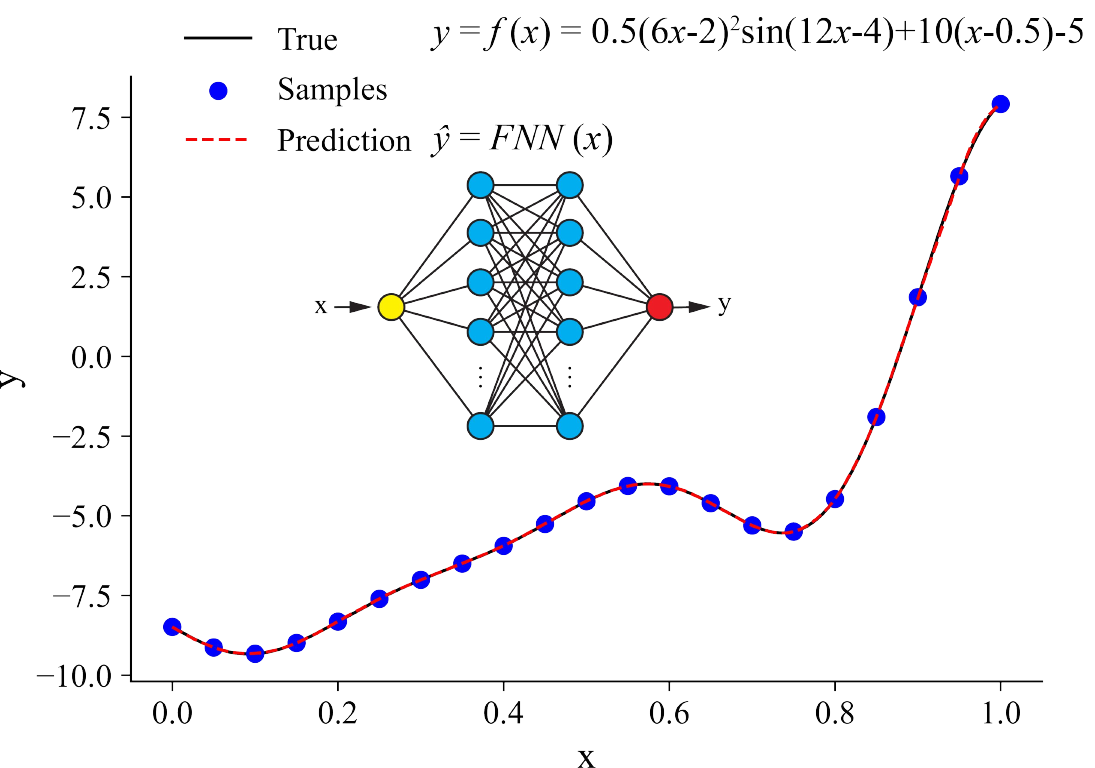}
\caption{Predicting latent function $f(x)=\text{0.5}(\text{6}x-\text{2})^\text{2}\text{sin}(\text{12}x-\text{4})+\text{10}(x-\text{0.5})-\text{5}$ with FNN in architecture of [1]+[20]*2+[1].}
\label{fig:nn2}
\end{figure}

A feedforward neural network (FNN)~\cite{329294} is the most basic form of ANN and is widely used to approximate a direct mapping between a finite-dimensional input vector and output vector. In an FNN, information flows only in the forward direction, meaning the output of each layer is determined by the activations of the previous layer, without feedback connections. For an $L$-layer FNN, the forward propagation can be written as
\begin{equation}
\mathbf{h}^{(0)}=\mathbf{x}, \qquad
\mathbf{h}^{(\ell)} = A^{(\ell)}\left(\mathbf{W}^{(\ell)}\mathbf{h}^{(\ell-\text{1})}+\mathbf{b}^{(\ell)}\right), \ \ \ell=\text{1},\text{2},\dots,L-\text{1},
\end{equation}
and the output layer for regression is commonly expressed as
\begin{equation}
\mathbf{y}=\mathbf{W}^{(L)}\mathbf{h}^{(L-\text{1})}+\mathbf{b}^{(L)},
\end{equation}
where $\mathbf{h}^{(\ell)}$ denotes the hidden-state vector at layer $\ell$, $\mathbf{W}^{(\ell)}$ and $\mathbf{b}^{(\ell)}$ are the weights and biases, and $A^{(\ell)}$ is the activation function.

The trainable parameters ${\mathbf{W}^{(\ell)},\mathbf{b}^{(\ell)}}$ are optimized by minimizing a loss function that measures the mismatch between predictions and training targets~\cite{wang2022comprehensive}. For supervised regression, a typical choice is the mean squared error (MSE),
\begin{equation}
\mathcal{L}(\Theta)=\frac{1}{N}\sum_\text{i=1}^{N}\left|\mathbf{y}_\text{i}-\hat{\mathbf{y}}(\mathbf{x}_\text{i},\Theta)\right|_\text{2}^\text{2},
\end{equation}
where $\Theta$ collects all weights and biases, $N$ is the number of training samples, $\mathbf{y}_\text{i}$ is the ground-truth output, and $\hat{\mathbf{y}}(\mathbf{x}_\text{i},\Theta)$ is the network prediction. The optimization is performed through gradient-based updates, where the gradients $\nabla_{\Theta}\mathcal{L}$ are efficiently computed by backpropagation. Backpropagation applies the chain rule layer by layer to propagate the output error backward through the network, producing partial derivatives of the loss with respect to each weight and bias. With these gradients, parameters are updated using algorithms such as stochastic gradient descent (SGD)~\cite{AMARI1993185} and its adaptive variant like Adam~\cite{kingma2017adammethodstochasticoptimization},
\begin{equation}
\Theta^{(k+\text{1})}=\Theta^{(k)}-l_r\cdot\nabla_{\Theta}\mathcal{L}(\Theta^{(k)}),
\end{equation}
where $k$ is the iteration index and $l_r$ is the learning rate. In practice, mini-batch training is commonly used, where the gradients are estimated from a subset of samples at each iteration, which improves scalability and often enhances generalization. An example using a simple FNN of two hidden layers to predict a simple latent function $f(x)=\text{0.5}(\text{6}x-\text{2})^\text{2}\text{sin}(\text{12}x-\text{4})+\text{10}(x-\text{0.5})-\text{5}$ is shown in Fig.~\ref{fig:nn2}.

A neural network surrogate and a GP surrogate share the same goal, namely learning a mathematical mapping from inputs to outputs to replace repeated high-cost simulations or experiments. Both can be trained in a supervised manner from paired data, and both aim to generalize beyond the training set. Despite the same goal, the neural network and GP start from different modeling philosophies. A GP begins with a probabilistic prior over functions, specified by a mean function and a covariance kernel. Learning is framed as conditioning the prior on observed data to obtain a posterior distribution over functions, and prediction follows from this posterior conditioning. The inductive bias is mainly expressed through the kernel choice and its hyperparameters, which encode statistical assumptions. In contrast, a neural network begins with a parametric function family determined by the architecture. Learning is posed as loss minimization, where weights and biases are optimized to reduce prediction error using gradient-based methods and backpropagation. Here, inductive bias comes primarily from architectural design, activation choices, and regularization. Although GP surrogate can naturally provide an evaluation on uncertainty, the neural network can also be incorporated with indeterministic features by extending the baseline network with additional inference or approximation schemes (for example Bayesian neural networks~\cite{9756596}).

\subsubsection{Physics-Informed Neural network}

\begin{figure}[hbt!]
\centering
\includegraphics[width=1.0\linewidth]{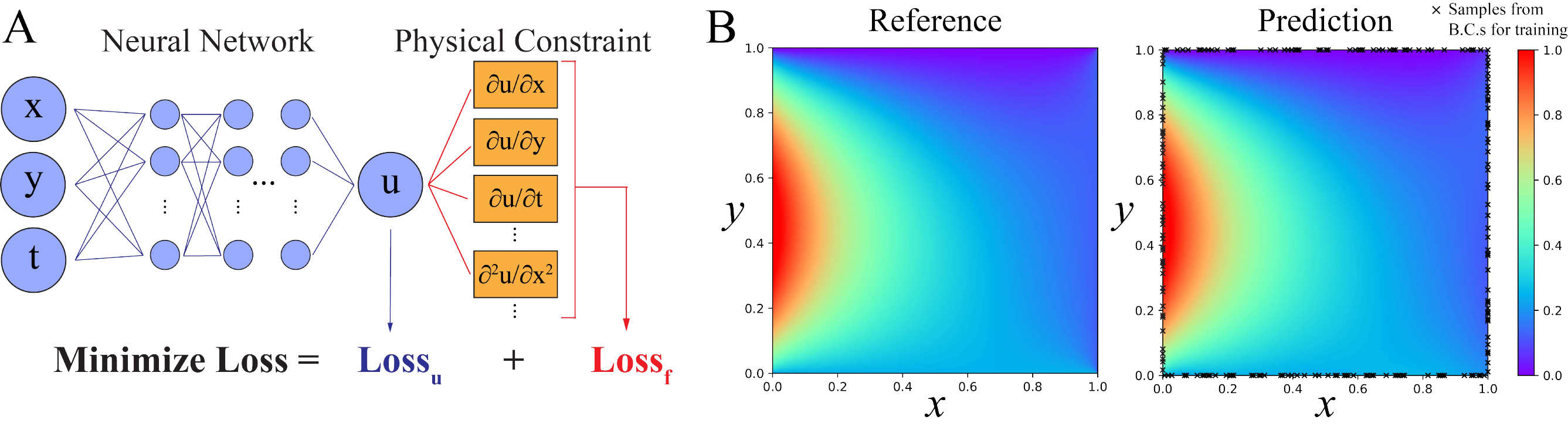}
\caption{Demonstration of PINN: (A) The architecture of PINN and (B) predicted solution $T(x,y)$ to a simple Laplace equation $\partial^2T/ \partial {x}^2 + \partial^2 T/ \partial {y}^2=0$ from PINN and its reference.}
\label{fig:nn5}
\end{figure}

In many physical and engineering processes, there are principled physical laws or empirically validated rules that govern the dynamics of the system. These laws and rules provide prior knowledge, which can be utilized to inform the training of the corresponding modeled surrogate. Therefore, the core concept of PINN~\cite{RAISSI2019686,dreisbach2024pinns4drops,kiyani2025optimizing,luo2025physics} is to use instead of massive data samples but prior governing equations that are usually written as differential equations to train a surrogate model. In the original surrogate formulation, a neural network is used to represent the unknown solution field $u(t,\mathbf{x})$, and the corresponding physics residual is constructed by applying the differential operator to this network through automatic differentiation. For a nonlinear partial differential equation (PDE), its general form can be expressed as 
\begin{equation}
u_t+\mathcal{N}[u;\boldsymbol{\lambda_u}]=0,\qquad \mathbf{x}\in\Omega,\ \ t\in[0,T].
\end{equation}
The physics residual of the PDE is defined as
\begin{equation}
f(t,\mathbf{x}) := u_t+\mathcal{N}[u;\boldsymbol{\lambda_u}],
\end{equation}
where $f(t,\mathbf{x})$ shares the same network parameters as $u(t,\mathbf{x})$ but evaluates the PDE constraint instead of the field value. Training then minimizes a composite loss that combines data mismatch terms, such as initial and boundary conditions, and a residual term that penalizes violations of the PDE at a set of collocation points (Fig.~\ref{fig:nn5}~A), leading to a standard continuous-time PINN objective of the form
\begin{equation}
\mathrm{MSE}=\mathrm{MSE}_u+\mathrm{MSE}_f,
\end{equation}
with $\mathrm{MSE}_u$ enforcing agreement with observed or imposed data, and $\mathrm{MSE}_f$ enforcing the governing equation in the domain. This construction turns the physical law into a regularization mechanism that constrains the admissible function space of the neural surrogate, and it enables data-efficient learning in regimes where measurements are sparse or expensive. An example of using PINN to solve a simple Laplace equation $\partial^2T/ \partial {x}^2 + \partial^2 T/ \partial {y}^2=0$ representing a simple heat conduction problem in steady state is shown in Fig.~\ref{fig:nn5}~B.

Beyond forward solution approximation, PINNs are also designed for inverse problems~\cite{invers2023framework,kim2024review,mohammad2026bayesian,characterization}. In that setting, unknown coefficients or parameters in the operator, collected in $\boldsymbol{\lambda_u}$, are treated as trainable variables together with the network weights, and are identified by minimizing the same physics-informed loss so that the learned field and the inferred parameters jointly explain the observations while remaining consistent with the PDE. More broadly, the concept of "physics-informed" can also incorporate empirically validated rules, conservation constraints, symmetries, process constants, or other domain knowledge as additional loss terms, not only PDE operators. In general, this physics-informed training paradigm provides a general recipe for building structured surrogates by embedding known governing constraints into the learning objective, which helps the network generalize beyond the observed samples.

\subsubsection{Recurrent Neural network}
\begin{figure}[hbt!]
\centering
\includegraphics[width=1.0\linewidth]{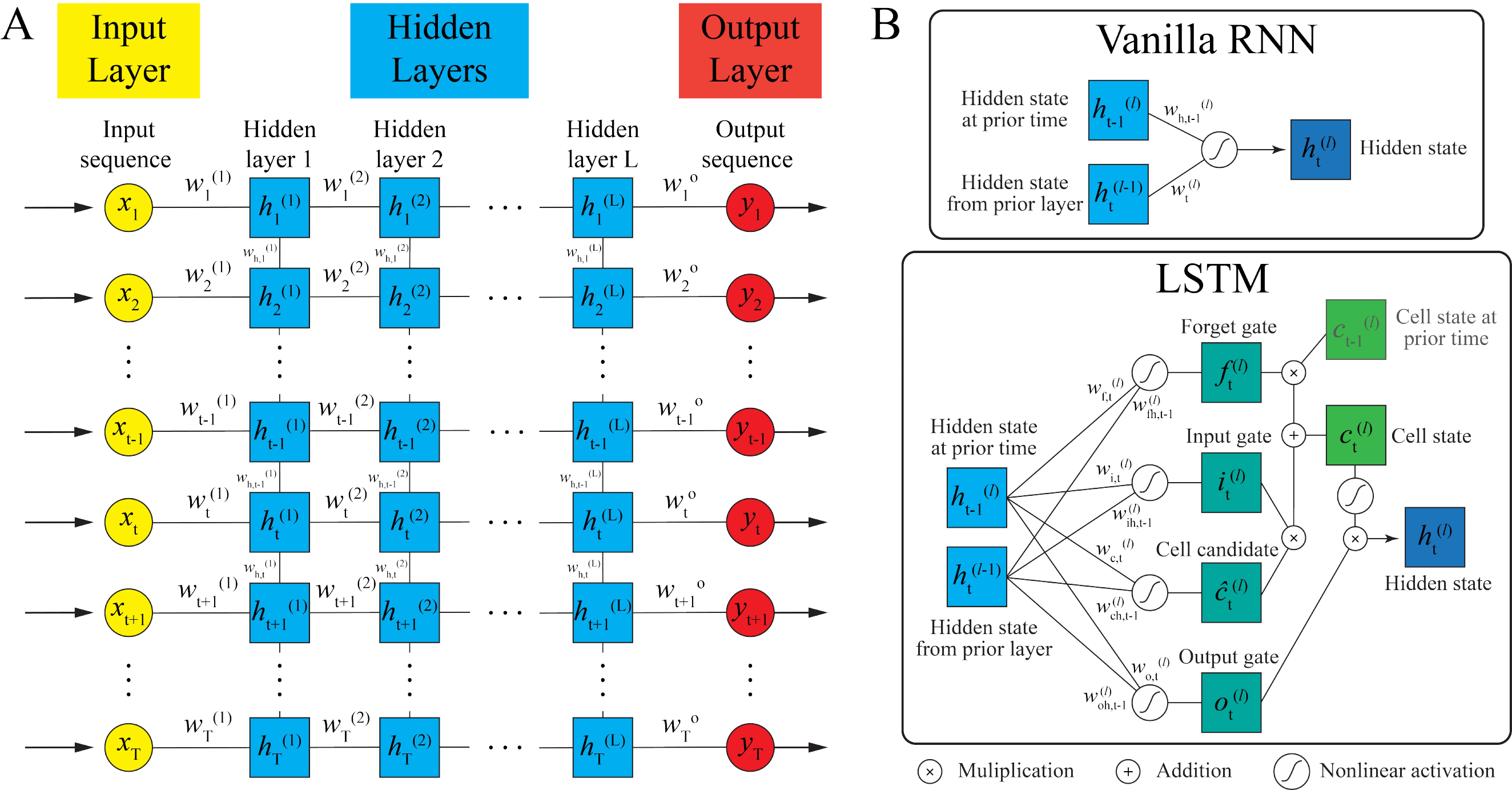}
\caption{Demonstration of RNN: (A) the general RNN architecture and (B) computation design of hidden state inside a RNN neuron in Vanilla RNNs and LSTM.}
\label{fig:nn3}
\end{figure}

In composite mechanics and manufacturing, the quantities of interest are often naturally sequential rather than single-point responses. Typical examples include history-dependent stress-strain behavior of composite from mechanical tests~\cite{CHEUNG2024110359,GHANE2025111163}, load-displacement histories of joints and coupons during loading experiments~\cite{doi:10.1177/002199839102500303}, and crack evaluation of composite in the specific spatio-temporal domain in the damage localization test~\cite{10379118}. In these cases, treating the output as an unordered vector can ignore the temporal or incremental dependence that governs the evolution of the response. To model such sequence dependence, recurrent neural networks (RNNs)~\cite{SHERSTINSKY2020132306} introduce a hidden state that carries information from earlier steps to later steps. Typical RNN architecture is shown in Fig.~\ref{fig:nn3}~A. Given an input sequence $\{\mathbf{x}_t\}_{t=1}^{T}$, a standard vanilla RNN updates the hidden state as
\begin{equation}
\begin{aligned}
    h_1^{(\ell)} &= A\!\left(w_{1}^{(\ell)}h_1^{(\ell-1)} + b_1^{(\ell)}\right),\quad l=1,2,3,\cdots,L,\\
    h_t^{(\ell)} &= A\!\left(w_{t}^{(\ell)}h_t^{(\ell-1)} + w_{h,t-1}^{(\ell)}h_t^{(\ell)} + b_t^{(\ell)}\right),\quad t=2,3,\cdots,T \text{ and } l=1,2,3,\cdots,L,\\
    h_t^{(0)} &= x_t,\quad t=1,2,3,\cdots,T,
\end{aligned}
\end{equation}
where $h_t^{(\ell)}$ is the hidden state at step $t$ at $\ell$-th hidden layer, $A$ is a nonlinear activation function and is chosen according to the task, often an identity map for regression. $w_t^{(\ell)}$, $w_{h,t}^{(\ell)}$, and $b_{t}^{(\ell)}$ are the trainable parameters. Training is commonly performed by minimizing a sequence loss and computing gradients through backpropagation through time, which unfolds the recurrence across the time horizon. For any non-initial hidden state $h_t^{(\ell)}$, it is determined by both the hidden state $h_t^{(\ell-1)}$ at the same time of the former layer and the hidden state $h_{t-1}^{(\ell)}$ at the prior time. Therefore, the RNN propagates temporal information through this unique architecture.

Vanilla RNNs can capture short-range dependence, but learning long-range history effects is often difficult because gradients may vanish or explode when backpropagated through many time steps. Long short-term memory (LSTM) networks~\cite{SHERSTINSKY2020132306} mitigate this issue by introducing a cell state that provides an explicit memory pathway, together with multiplicative gates that control how information is written, forgotten, and exposed. Given an input sequence $\{x_t\}_{t=1}^{T}$, an LSTM maintains, at each layer $\ell$, both a hidden state $h_t^{(\ell)}$ and a cell state $c_t^{(\ell)}$ (Fig.~\ref{fig:nn3}~B). Using $h_t^{(0)} = x_t$, the scalar LSTM updates can be written as
\begin{equation}
\begin{aligned}
c_0^{(\ell)} &= 0,\quad h_0^{(\ell)} = 0,\quad \ell=1,2,\cdots,L,\\[2pt]
i_1^{(\ell)} &= \sigma\!\left(w_{i,1}^{(\ell)} h_1^{(\ell-1)} + w_{ih,0}^{(\ell)} h_{0}^{(\ell)} + b_{i,1}^{(\ell)}\right),\quad \ell=1,2,\cdots,L,\\
f_1^{(\ell)} &= \sigma\!\left(w_{f,1}^{(\ell)} h_1^{(\ell-1)} + w_{fh,0}^{(\ell)} h_{0}^{(\ell)} + b_{f,1}^{(\ell)}\right),\quad \ell=1,2,\cdots,L,\\
o_1^{(\ell)} &= \sigma\!\left(w_{o,1}^{(\ell)} h_1^{(\ell-1)} + w_{oh,0}^{(\ell)} h_{0}^{(\ell)} + b_{o,1}^{(\ell)}\right),\quad \ell=1,2,\cdots,L,\\
\hat{c}_1^{(\ell)} &= \tanh\!\left(w_{c,1}^{(\ell)} h_1^{(\ell-1)} + w_{ch,0}^{(\ell)} h_{0}^{(\ell)} + b_{c,1}^{(\ell)}\right),\quad \ell=1,2,\cdots,L,\\
c_1^{(\ell)} &= f_1^{(\ell)}\, c_{0}^{(\ell)} + i_1^{(\ell)}\, \hat{c}_1^{(\ell)},\quad \ell=1,2,\cdots,L,\\
h_1^{(\ell)} &= o_1^{(\ell)}\, \tanh\!\left(c_1^{(\ell)}\right),\quad \ell=1,2,\cdots,L,\\[4pt]
i_t^{(\ell)} &= \sigma\!\left(w_{i,t}^{(\ell)} h_t^{(\ell-1)} + w_{ih,t-1}^{(\ell)} h_{t-1}^{(\ell)} + b_{i,t}^{(\ell)}\right),\quad t=2,3,\cdots,T,\ \ell=1,2,\cdots,L,\\
f_t^{(\ell)} &= \sigma\!\left(w_{f,t}^{(\ell)} h_t^{(\ell-1)} + w_{fh,t-1}^{(\ell)} h_{t-1}^{(\ell)} + b_{f,t}^{(\ell)}\right),\quad t=2,3,\cdots,T,\ \ell=1,2,\cdots,L,\\
o_t^{(\ell)} &= \sigma\!\left(w_{o,t}^{(\ell)} h_t^{(\ell-1)} + w_{oh,t-1}^{(\ell)} h_{t-1}^{(\ell)} + b_{o,t}^{(\ell)}\right),\quad t=2,3,\cdots,T,\ \ell=1,2,\cdots,L,\\
\hat{c}_t^{(\ell)} &= \tanh\!\left(w_{c,t}^{(\ell)} h_t^{(\ell-1)} + w_{ch,t-1}^{(\ell)} h_{t-1}^{(\ell)} + b_{c,t}^{(\ell)}\right),\quad t=2,3,\cdots,T,\ \ell=1,2,\cdots,L,\\
c_t^{(\ell)} &= f_t^{(\ell)}\, c_{t-1}^{(\ell)} + i_t^{(\ell)}\, \hat{c}_t^{(\ell)},\quad t=2,3,\cdots,T,\ \ell=1,2,\cdots,L,\\
h_t^{(\ell)} &= o_t^{(\ell)}\, \tanh\!\left(c_t^{(\ell)}\right),\quad t=2,3,\cdots,T,\ \ell=1,2,\cdots,L,\\
h_t^{(0)} &= x_t,\quad t=1,2,\cdots,T,
\end{aligned}
\end{equation}
where $i_t^{(\ell)}$, $f_t^{(\ell)}$, $o_t^{(\ell)}$, $\hat{c}_t^{(\ell)}$ $c_t^{(\ell)}$ are respectively the input gates, forget gates, output gates, candidate state, and cell state. $\sigma(\cdot)$ denotes the logistic sigmoid, $\tanh(\cdot)$ denotes the hyperbolic tangent, which are usually chosen for the activation functions inside LSTM, and can be replaced with other functions depending on specific application.
The additive update for $c_t^{(\ell)}$ provides a more stable pathway for information and gradient flow across time, enabling improved learning of long-range dependencies in sequential composite responses.

\subsection{Multi-Fidelity Framework}
Composite mechanics problems naturally generate multi-fidelity information since composite behavior is governed by hierarchical structures and coupled physics. Typical fidelity pairs include homogenization models versus detailed meso-scale models~\cite{CHEUNG2024110359,GHANE2025111163}, coarse FE models versus refined high-resolution analyses~\cite{10.1007/s00158-020-02684-3}, and simulation outputs versus experimental measurements~\cite{TANG2025113106}. A MF framework aims to leverage this complementarity by learning from multiple sources, capturing the global trend from LF data and correcting discrepancies using limited LF data, so that accurate prediction can be achieved with reduced reliance on expensive samples. In addition to GP- and Kriging-based models, neural networks are reliable surrogate models that can be incorporated into the MF framework. Within NN-based surrogate modeling for composite mechanics, two representative strategies are commonly used and will be introduced in this section. The first constructs an end-to-end architecture that explicitly couples LF and HF information inside a single network, referred to here as a multi-fidelity fused network.  The second reuses representations learned from LF tasks and adapts them to the HF target using limited HF samples, which corresponds to transfer learning.

\subsubsection{Multi-Fidelity Fused Network}
A MF fused network refers to an end-to-end neural architecture in which LF and HF or multiple fidelity information are combined inside a single learning system, rather than training separate surrogates and coupling them externally. The central goal is to learn the cross-fidelity relationship that exploits the broad coverage and low cost of LF data while using limited HF samples to correct LF bias. A generic MF fused-network representation is the multi-fidelity neural networks (MFNNs) proposed by Meng and Karniadakis~\cite{MENG2020109020}. For a LF function $y_\text{L}(x)$ and a HF function $y_\text{H}(x)$, a general correlation between two can be constructed as
\begin{equation}
    y_\text{H}=\mathcal{F}_\text{l}(x,y_\text{L})+\mathcal{F}_\text{nl}(x,y_\text{L}),
\end{equation}
where $\mathcal{F}_\text{l}$ and $\mathcal{F}_\text{nl}$ respectively denote the linear and nonlinear terms in the correlation, and they together form the unknown function that maps the low-fidelity data to the high-fidelity level.

\begin{figure}[hbt!]
\centering
\includegraphics[width=0.7\linewidth]{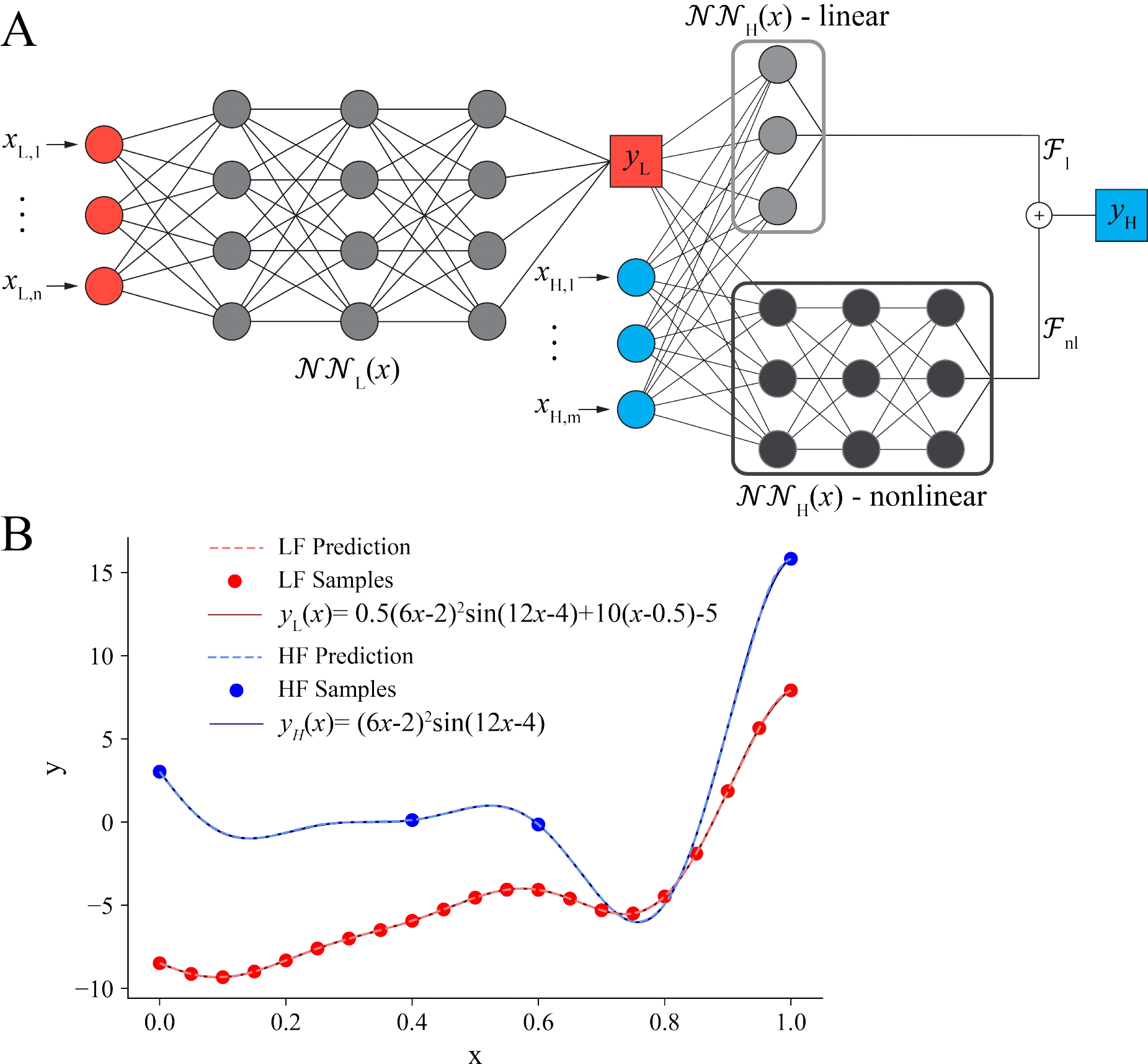}
\caption{Demonstration of MFNN: (A) the general MFNN architecture and (B) an example to predict a HF latent function 
$f_\text{H}(x)=(\text{6}x-\text{2})^\text{2}\text{sin}(\text{12}x-\text{4})$ with few HF samples and many LF samples of a LF function
$f_\text{L}(x)=\text{0.5}(\text{6}x-\text{2})^\text{2}\text{sin}(\text{12}x-\text{4})+\text{10}(x-\text{0.5})-\text{5}$ with MFNN in architecture of LF network [1]+[20]*2+[1], HF network linear component [2]+[1], and HF network nonlinear component [2]+[10]*2+[1].}
\label{fig:nn4}
\end{figure}

The architecture of MFNN is conventionally composed of three essential parts: the LF network $\mathcal{N}\mathcal{N}_\text{L}$ approximating the LF data, the HF network $\mathcal{N}\mathcal{N}_{\text{H}_\text{l}}$ approximating the linear correlation between the LF and HF data, and the high-fidelity network $\mathcal{N}\mathcal{N}_{\text{H}_\text{nl}}$ approximating the nonlinear correlation between the LF and HF data (Fig.~\ref{fig:nn4}~A). The loss function, which evaluates the approximation of MFNN and is aimed to be minimized through the training process, is expressed as

\begin{equation}
    loss_\text{total}=loss_{y_\text{L}}+loss_{y_\text{H}}+\lambda_r \sum \beta_\text{i}^\text{2},
\end{equation}

where $loss_{y_\text{L}}$ and $loss_{y_\text{H}}$ respectively evaluate the approximation of $\mathcal{N}\mathcal{N}_\text{L}$ and $\mathcal{N}\mathcal{N}_\text{H}$, and $\lambda \sum \beta_\text{i}^\text{2}$ denotes the $L_\text{2}$ regularization, which has been widely adopted to prevent overfitting~\cite{ZHANG2019108850}, with any weight $\beta$ in $\mathcal{N}\mathcal{N}_\text{L}$ and $\mathcal{N}\mathcal{N}_\text{H}$ and regularization rate $\lambda_r$. The network component can be varied with different neural network variants depending on the specific approximation task or the studied subject. An example using a simple MFNN to predict a HF latent function 
$f_\text{H}(x)=(\text{6}x-\text{2})^\text{2}\text{sin}(\text{12}x-\text{4})$ with few HF samples and many LF samples of a LF function
$f_\text{L}(x)=\text{0.5}(\text{6}x-\text{2})^\text{2}\text{sin}(\text{12}x-\text{4})+\text{10}(x-\text{0.5})-\text{5}$ is shown in Fig.~\ref{fig:nn4}~B.

\subsubsection{Transfer Learning}
Transfer learning reuses knowledge stored in a trained source model to construct a target model for a related problem, rather than training the target model from random initialization. Conceptually, it transfers parameterized representations learned from the source, using them as an initialization or partially fixed feature extractor for the target, so that the target learning is guided by prior structure instead of being determined solely by scarce target data, and models trained across related settings remain aligned rather than arbitrarily different~\cite{9134370}. This methodology aligns naturally with a MF framework. MF data are commonly available from sources that differ in cost and accuracy. The LF source is usually abundant and inexpensive but biased, while the HF source is limited but trusted. This organization matches the source and target structure required by transfer learning. A LF network can first be trained to learn global trends and feature representations from a large dataset, then a HF network can be initialized from that LF model and adapted using the scarce HF data. In this sense, multi-fidelity learning creates the setting in which transfer learning is most effective. Transfer learning also provides a practical mechanism to reduce overfitting in the HF regime by constraining how the model is updated, and it helps maintain a consistent representation across fidelities when LF and HF surrogates are expected to remain comparable within a unified workflow. 

At the implementation level, transfer learning applied in NN-based MF surrogates is commonly realized through several operations. One operation is weight transfer followed by fine-tuning~\cite{5288526}, where a network trained on LF data provides the initial parameters for the HF model and the parameters are then updated using HF data. Another operation is freezing and partial training~\cite{DAVILA2024105012}, where early layers trained on LF data are fixed to preserve transferable representations while only a small subset of layers is trained to fit the HF data. A more structured strategy is progressive transfer~\cite{liu2019transfer}, where the model is transferred sequentially across increasing hierarchical fidelity levels or across a chain of closely related datasets so that each target model inherits parameters from a closely related predecessor and the family of models remains aligned. When transfer learning is used to maintain cross-fidelity correspondence, structural modifications that would reorder or permute parameters are typically avoided during training so that the inherited parameter structure remains comparable across fidelities. In~\cite{kiyani2025probabilistic}, transfer learning is employed within a Deep Operator Network (DeepONet)~\cite{lu2019deeponet} framework to adapt from a LF simulation-data distribution to the experimental distribution. The DeepONet is first trained on LF simulation data, and the available HF experimental data are then used for fine-tuning. During this transfer-learning stage, the trunk network is kept fixed and only the last layer of the branch network is updated, allowing the model to preserve the operator representation learned from simulation while adapting its final prediction to the experimental domain.

\section{Applications in Composite Mechanics} \label{Section Apps}
Distinguishing from the conventional SM technique creating a surrogate twin for a certain physical or engineering process based on a sole-fidelity source, the MFSM technique requires the existence of multiple fidelity, which should consist of at least a LF source and a HF source. The comparative definition of "low" and "high" (or LF and HF) usually varies in diverse physical topics, such as ideal but abundant CFD simulation versus real but expensive experiment in fluid dynamics~\cite{doi:10.2514/2.456,OBERKAMPF2002209}, a low-order solver or a lower resolution versus a high-order solver or a higher resolution in CFD~\cite{https://doi.org/10.1002/fld.3767}, efficient meso-level coarse-grained particle modeling versus high-cost micro-level molecular modeling in computational biology~\cite{annurev:/content/journals/10.1146/annurev-biophys-083012-130348,Li2016}, and testing on the known sole-component materials versus one on the newly-designed composite consisting of known gradients in evaluating material's mechanical performance~\cite{SABA20191}. Based on that, MFSM offers an effective algorithm for identifying both discrepancy and similarity among different fidelity tiers of data and further fusing them to construct a multi-fidelity SM with higher efficiency in both reducing training time and improving data utilization compared to the single-fidelity SM. Since a composite is formed by two or more distinct constituents, its mechanics is inherently described through a hierarchy of models across multi-scales~\cite{https://doi.org/10.1002/adma.201101683} and multi-physics~\cite{https://doi.org/10.3390/ma17123040}, making it a natural match for multi-fidelity frameworks. In this section, many recent applications utilizing the GP- and NN-based MFSM techniques in studying composite mechanics will be presented (all the introduced applications are organized in Table~\ref{tab:apps_summary_compact}).

\begin{table}[htbp]
\centering
\caption{MFSM applications in composite mechanics reviewed in Section~\ref{Section Apps} (TL: transfer learning).}
{\fontsize{8}{15}\selectfont
\resizebox{\linewidth}{!}{%
\begin{tabular}{p{0.25\linewidth} p{0.2\linewidth} p{0.18\linewidth} c c p{0.22\linewidth}}
\hline
\textbf{Application} & \textbf{LF source} & \textbf{HF source} & \textbf{Fwd} & \textbf{Inv} & \textbf{MF surrogate} \\
\hline

Multi-scale Woven UQ~\cite{BOSTANABAD2018506}
& low-accuracy models
& high-accuracy model
& \yes & \no & MRGP \\

Laminate failure onset~\cite{10.1115/SSDM2025-152303}
& CUF--ESL
& CUF--LW
& \yes & \no & Discrepancy GP \\

\makecell[l]{Progressive damage\\[-1.5ex] in laminate~\cite{CHAHAR2023106647}}
& Abaqus built-in PD
& UMAT 3D-CDM
& \yes & \no & MF-MOGP with AR \\

\makecell[l]{PID in CFRP\\[-1.5ex] thermo-processing~\cite{SCHOENHOLZ2024111499}}
& 2D reduced simulation
& \makecell[l]{3D refined simulation\\[-1.5ex] + few experiments}
& \yes & \no & SWGPR \\

Dispersion-stiffness inversion~\cite{KALIMULLAH2025111872}
& low-cost dispersion
& expansive dispersion
& \no & \yes & Inverse MOGP \\

\makecell[l]{VS fiber steering\\[-1.5ex] buckling design~\cite{10.1007/s00158-020-02684-3}}
& coarse-mesh FE
& refined-mesh FE
& \yes & \yes & Hierarchical Kriging \\

Blast armor optimization~\cite{valladares2020design}
& simplified blast FE
& damage-resolved blast FE
& \yes & \yes & MF-GP in BO \\

Al--Nb--Ti composition design~\cite{10.1063/5.0015672}
& ML interatomic potential
& DFT
& \yes & \yes & MF-GP in BO \\

Thermo-buckling optimization~\cite{YOO2021106655}
& coarse FE buckling
& refined FE buckling
& \yes & \yes & NARGP in RBDO \\

Open-hole stress field~\cite{doi:10.1177/00219983241281073}
& Lekhnitskii field
& Abaqus stress field
& \yes & \no & MF U-net \\

Fiber suspension rheology~\cite{10.1063/5.0087449}
& constitutive fits
& immersed-boundary DNS
& \yes & \no & MFNN \\

Open-hole tensile curve~\cite{doi:10.1177/002199839102500303}
& Abaqus PD curves
& experimental curves
& \yes & \no & MF Triple LSTM \\

Autoclave curing temperature~\cite{9816983}
& composite system 1
& composite system 2
& \yes & \no & MFPINN \\

SPR parameter optimization~\cite{LI2023812}
& CAE simulations
& physical experiments
& \yes & \yes & MORNN with TL \\

AE impact localization~\cite{KALIMULLAH2023110360}
& TOF physics residual
& experimental TOF
& \yes & \no & Probabilistic MFPINN with TL \\

SFRC history-dependent stress~\cite{CHEUNG2024110359}
& mean-field micromechanics
& full-field FE/FFT
& \yes & \no & GRU-based RNN with TL \\

Woven inelastic response~\cite{GHANE2025111163}
& mean-field homogenization
& full-field FFT
& \yes & \no & GRU-based RNN with TL \\

\hline
\end{tabular}%
}}\label{tab:apps_summary_compact}
\end{table}

\subsection{Applications of GP-based MFSM}
This first section organizes GP-based multi-fidelity applications by how MF surrogates are used within an engineering workflow, rather than by specific GP or Kriging variants. The discussion begins with fundamental studies that construct MF-GP surrogates for forward prediction, where the primary goal is to accelerate expensive HF simulations or experiments while maintaining sufficient accuracy. Then it further moves to the inverse optimization tasks built on top of the learned surrogate, which aim to inversely identify the optimal input parameters or state of the process under a target output. A variety of applications will be introduced to show how MF-GP surrogates are embedded into the design and testing pipeline of composite materials in order to support practical decision-making, quality control, and iterative design refinement.

\subsubsection{Efficient Forward Prediction}
\begin{figure}[hbt!]
\centering
\includegraphics[width=0.8\linewidth]{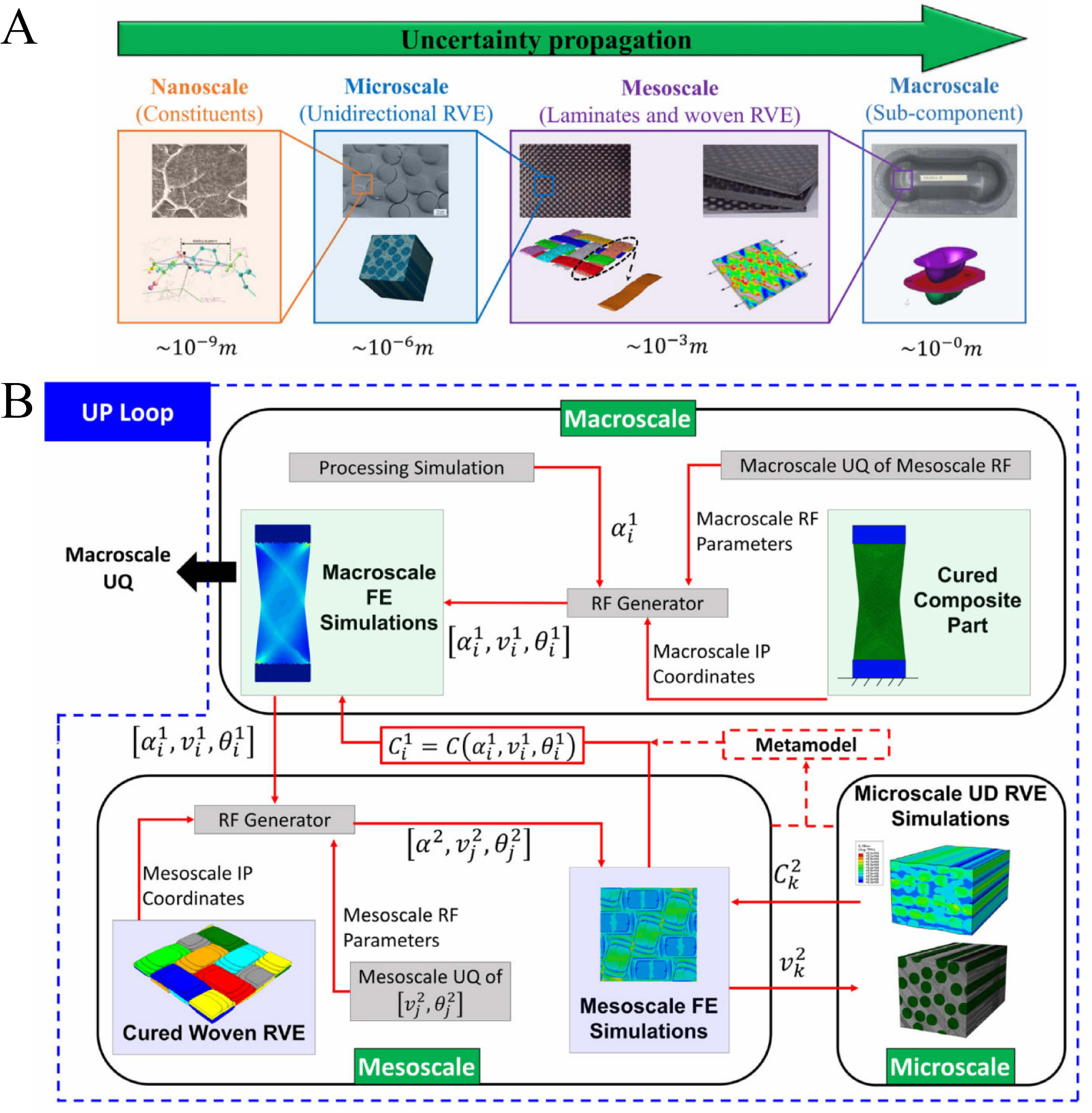}
\caption{Demonstration of (A) a four-scale woven fiber composite with polymer matrix and (B) MF surrogate (metamodel) replaces micro- and meso-simulations in the conventional UQ loop for a woven fiber composite. The images are from the reference~\cite{BOSTANABAD2018506} with authorization.}
\label{fig:app1}
\end{figure}

Woven fiber composites, which have been increasingly used in aerospace, construction, and transportation industries due to their superior mechanical performance~\cite{TABIEI2002149,KOMEILI2012163}, are the typical composite materials that possess a hierarchical structure that spans multiple length-scales~\cite{BOSTANABAD2018506} (Fig.~\ref{fig:app1}~A). The macroscale properties of such are usually uncertain due to multiple uncertainty sources at the lower length-scales, including the varying architecture of the fiber yarn in the mesoscale, the fibers' relative location change inside the yarn in the microscale, and variations in constituent properties and the interaction between them in the nanoscale. To solve this problem of uncertainty, the conventional UQ approach uses a nested multiscale computational homogenization loop through the chain of scales, which causes inevitable repetition in different simulations under different length-scales and thereby a high computational burden. Therefore, an efficient surrogate model based on multi-response Gaussian process (MRGP)~\cite{WANG2015159} is introduced to replace micro- and meso-simulations in the conventional UQ routine. MRGP is a MOGP variant, which additionally models a vector of correlated outputs jointly, rather than fitting independent scalar GPs for each output component. After the MRGP is trained, the nested UQ workflow is modified by replacing the repeated micro- and meso-scale simulations with direct probabilistic predictions. Instead of re-running lower-scale simulations for each uncertainty realization and each location to obtain the corresponding effective stiffness-related quantities, the MRGP takes uncertainty descriptors such as yarn-architecture variations and constituent-property variations as inputs and predicts multiple effective properties jointly, thereby retaining cross-output correlations in a single MF surrogate.
These predicted effective properties of error below 1\% are then passed to the macroscale analysis to propagate uncertainty to structural responses with far fewer expensive lower-scale evaluations.

Evaluating the mechanical failure of newly-designed composite laminates is crucial~\cite{ORIFICI2008194}, especially in areas like aerospace, where composite materials are widely utilized for their high strength-to-weight ratio. The failure onset, representing the earliest stage of the mechanical failure, must be estimated reliably to ensure structural safety. However, since composites are usually anisotropic and sensitive to manufacturing defects, the prediction of failure onset is inherently affected by the unavoidable variability in composite manufacturing and service conditions. Small deviations in constituent properties, ply-level stiffness parameters, and applied loads can shift the laminate stress state and consequently alter which damage mode first becomes critical~\cite{HAO2021107465}. Therefore, practical failure assessment requires instead of a single deterministic estimate but the distribution of failure indices and deflections evaluated under uncertain inputs. Numerical implementation of such assessment consumes massive computational resources, thereby an efficient surrogate model is needed. To address this issue, a multi-fidelity surrogate based on discrepancy GP is introduced~\cite{10.1115/SSDM2025-152303}, which utilizes discrepancy learning to couple LF data from Carrera Unified Formulation (CUF)~\cite{10.1007/BF02736224} using Equivalent Single Layer (ESL) model and HF data from CUF using Layer-Wise (LW) model. Using this MF surrogate model, the evaluation properties, including maximum deflection and four failure indices of tension and compression, are accurately predicted with a small dataset that contains only 20\% LF data (2 HF + 8 LF training samples).

In addition to the failure evaluation in composite laminates, progressive damage in notched composite laminates is also a safety‐critical concern. The damage can initiate and evolve through multiple interacting mechanisms due to the composite's inevitable variability in material properties and ply orientations, causing a nonlinear and complex pattern that is challenging to analyze~\cite{SRIRAMULA20091673,SHARMA2022108724}. Therefore, similar to the evaluation of mechanical failure, the evaluation of progressive damage in composite laminates is also probabilistic, which thereby needs thousands of realizations of a single simulation or experiment process. Although FE progressive-damage numerical simulations inside Monte Carlo–type UQ~\cite{doi:10.1177/002199838401800304} are relatively economical compared to real experiments, massive repetitions in simulations make evaluation prohibitive. To reduce this cost, a surrogate based on MF-MOGP is built to fuse a cheaper LF Abaqus built-in progressive damage model with a more accurate but expensive HF user-defined-material-based (UMAT-based) 3D continuum damage mechanics model~\cite{CHAHAR2023106647}. This MF-MOGP adopts an AR-style structure (Eqn.~\ref{AR1}), thus abundant LF data shapes global trends while fewer HF samples correct the bias, yielding forward predictions of multiple laminate responses without rerunning the full HF model each time. Specifically, the MF-MOGP surrogate costs 3700 mins with 200 LF + 400 HF samples in the 15D setting, whereas the compared single HF model costs 3900 mins with 600 HF samples and achieves essentially the same relative $L_\text{2}$ errors, such as 0.2208\% versus 0.2228\% for reaction force and 2.924\% versus 3.1477\% for matrix damage. Overall, the MF-MOGP surrogate utilizing the AR fusion strategy is proven effective and efficient in predicting progressive damage of composite laminates.

\begin{figure}[hbt!]
\centering
\includegraphics[width=0.99\linewidth]{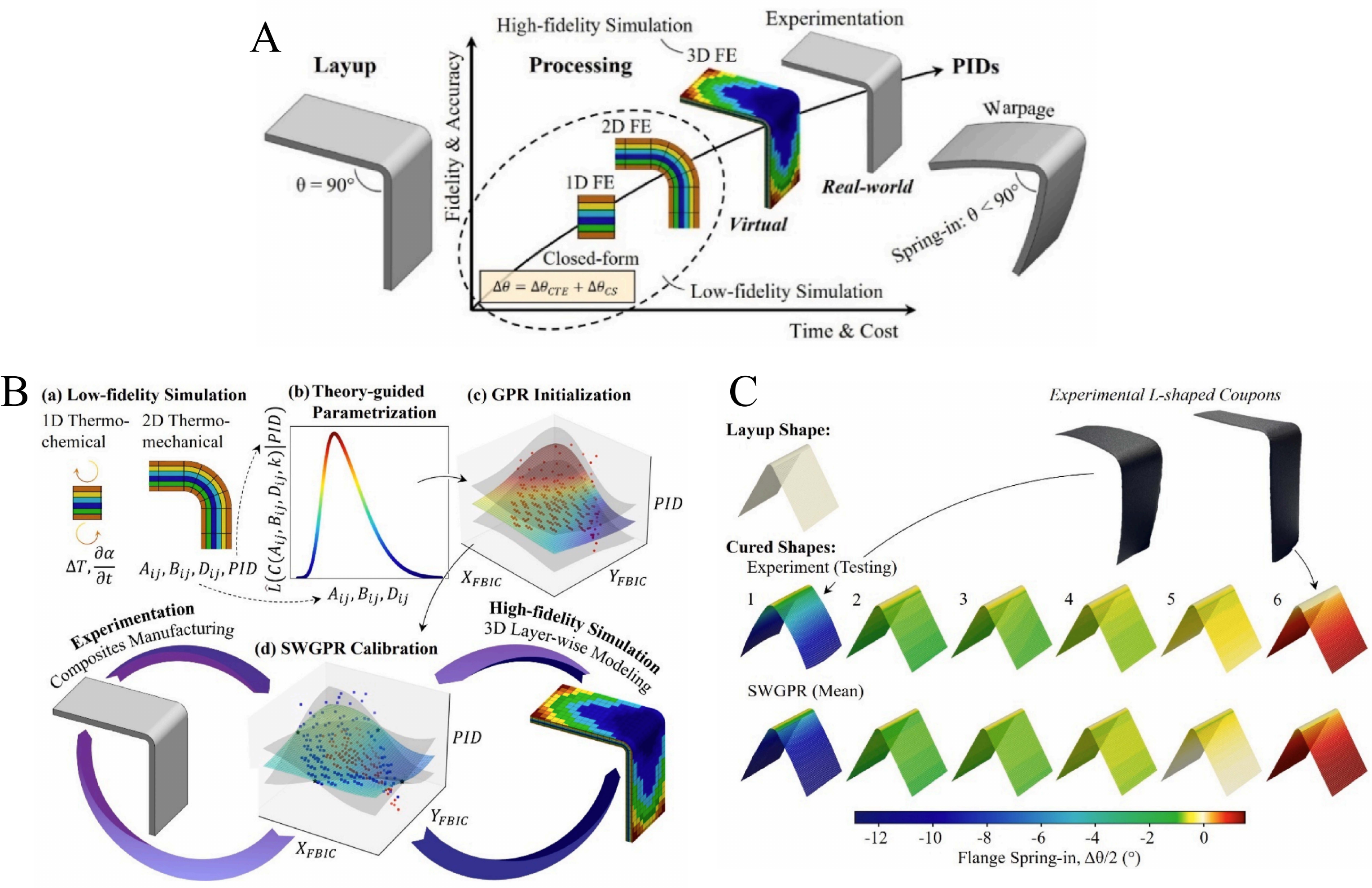}
\caption{Demonstration of PID assessment in a L-shaped CFRP part with SWGPR surrogate: (A) the trade-off between time/cost and fidelity/accuracy in analysis methods, (B) Flowchart of efficient composites manufacturing analysis method using multi-fidelity simulation, and (C) Comparative results of experimental PIDs and predictions using calibrated SWGPR models. The images are from the reference~\cite{SCHOENHOLZ2024111499} with authorization.}
\label{fig:app2}
\end{figure}

Process-induced deformations (PIDs) such as spring-in and warpage are a major manufacturing bottleneck for aerospace parts built with carbon fiber reinforced polymer (CFRP) because they create assembly gaps, increase production costs, and can degrade final mechanical performance~\cite{10.1098/rsta.2015.0278,lee1982heat,limaye2025numerical,shah2018optimal,johnston1997integrated,kiyani2025probabilistic,limaye2024use}. These distortions originate from coupled thermo-chemical and thermo-mechanical phenomena during the high-temperature/pressure processing, such as autoclave cure~\cite{ZOBEIRY201543}. Therefore, the PID assessment of CFRP is an essential step to the quality control of thermo-processing products. Conventional PID assessment typically relies on either fast but biased LF 1D or 2D reduced-order models or expensive HF 3D layer-wise simulations and repeated experiments, and it becomes inefficient when processing settings such as composite layups and structural geometry are complicated. To address this, a study proposes a probabilistic MF-GP surrogate built upon spatially weighted Gaussian process regression (SWGPR)~\cite{5288526} and uses a typical L-shaped component (Fig.~\ref{fig:app2}) as the studied subject~\cite{SCHOENHOLZ2024111499}. The study utilizes three fidelity sources, including a LF 2D thermo-mechanical simulation informed by free-strain data from cheaper models, a HF 3D simulation based on a refined CUF scheme, and a few true but scarce thermal experiments. The surrogate is initially trained with a broad LF dataset, and is later calibrated by targeted HF simulations and limited experiments through data-point–dependent noise weighting strategy. Quantitatively, SWGPR achieves root mean squared error (RMSE) = 0.1° using 114 LF simulations + 138 HF simulations + only 4 experiments, whereas the full 256 LF simulations with regular GP gives large RMSE = 27.7°, demonstrating that MF-GP calibration can deliver near-experimental accuracy with minimal experimental effort.

In this subsection, the main motivation for introducing multi-fidelity GP surrogates in forward prediction is efficiency. Many composite-mechanics tasks, such as multiscale uncertainty propagation, failure onset screening, progressive damage evaluation, and PID assessment, rarely depend on a single high-precision solve. Instead, they require repeated evaluations across different designs, loading cases, and uncertainty realizations. Under this repeated-use setting, conventional single HF simulation pipelines become prohibitively expensive because computational cost increases. Multi-fidelity GP models mitigate this burden by leveraging abundant low-cost information to capture the global response trend and a limited amount of high-cost data to correct systematic bias, enabling fast forward prediction.

\subsubsection{Inverse Optimization}
Accurate stiffness characterization of carbon composite laminates is essential for structural analysis, quality assurance, and structural health monitoring, as the stiffness tensor governs load transfer, vibration response, and damage sensitivity~\cite{SEGERS2020115360}. In many practical settings, it is not realistic to directly obtain the stiffness parameters from in-use components, and destructive testing is undesirable. A common non-destructive method for determining viscoelastic properties and the medium’s symmetry is through the analysis of guided elastic waves by linking stiffness to measurable dispersion behavior or, specifically, dispersion curves~\cite{KUDELA2021114178}. Although Numerous research efforts have focused on analyzing the dispersion curves in composite laminates, a robust inversion algorithm to extract the constitutive properties of fully anisotropic materials remains a challenge. One difficulty arises from the strong coupling among stiffness components and the sensitivity of dispersion features to measurement noise, thus conventional inversion typically relies on repeatedly calling a forward dispersion solver within an iterative search procedure, which becomes inefficient as it demands many forward evaluations for a stable solution. To address this, an inverse surrogate based on MOGP regression is introduced to learn a direct mapping from dispersion-curve information to the nine stiffness parameters jointly~\cite{KALIMULLAH2025111872}, which also enables uncertainty-aware estimation through predictive intervals. Quantitative result shows that the reported maximum estimation bias is about 0.67\% for specimen I and 1.33\% for specimen II without noise, and increases to about 12.11\% under 5\% noise with appropriately widened uncertainty bounds. On experimental datasets, the inferred stiffness parameters remain close to genetic-algorithm-based inversion with an overall deviation below 8.6\%, demonstrating the efficiency advantage of GP-based inverse identification for anisotropic laminate characterization. It is worth clarifying that the GP surrogate is trained as a direct inverse model, mapping THE dispersion curve to the stiffness tensor parameters. This differs from the other following inverse optimization applications.

\begin{figure}[hbt!]
\centering
\includegraphics[width=0.99\linewidth]{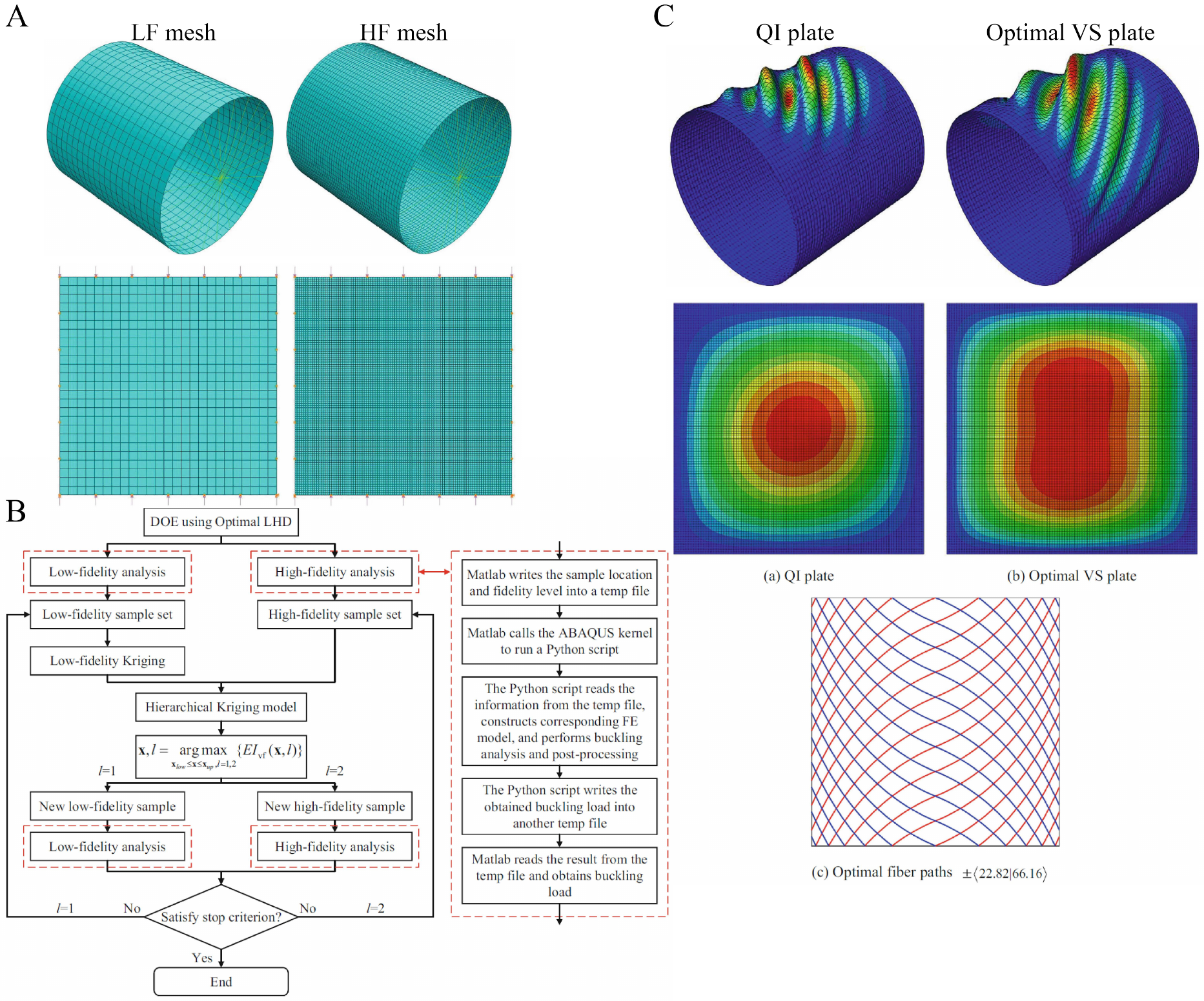}
\caption{Demonstration of fiber arrangement optimization in a VS composite with MF surrogate: (A) LF and HF FE models of the VS cylinder and plate, (B) Flowchart of the optimization procedure, and (C) buckling mode shapes of the both cylinder and plate separately using the QI laminate and optimal VS laminate with the below shown optimal fiber pattern (buckling load in QI plate and optimal VS plate are 14.93 kN and 22.78 kN, respectively; critical bending moments in QI cylinder and optimal VS cylinder are 93.17 kN·m and 116.40 kN·m, respectively). The images are from the reference~\cite{10.1007/s00158-020-02684-3} with authorization.}
\label{fig:app3}
\end{figure}

Variable stiffness (VS) composites enabled by fiber steering have become favorable to designers working on buckling-critical structures since the spatial freedom of fiber angles allows composite stiffness to be tailored to resist instability~\cite{doi:10.2514/3.11613}. On the one hand, this design freedom can deliver substantial performance gains compared to the traditional quasi-isotropic (QI) composite laminates, particularly for plates and shells used in aerospace structures. On the other hand, the high spatial variability results in high sensitivity in the buckling response to the local stiffness patterns, turning identifying an optimal fiber path design into an optimization problem, which further requires the high-resolution numerical models in repeated HF FE analyses~\cite{doi:10.2514/1.42490}. To address this computational bottleneck, the study formulates the design search based on forward FE buckling simulations, where each candidate VS laminate is evaluated by solving a linear eigenvalue buckling problem to obtain the critical response, such as the plate buckling load or the cylinder critical bending moment. In this setting, the expensive step is the repeated HF FE evaluation throughout the optimization loop. Therefore, a MF surrogate based on hierarchical Kriging is introduced to emulate the mapping from VS design variables to buckling capacity~\cite{10.1007/s00158-020-02684-3}, where the inputs are the parameters defining the fiber angle field, such as the fiber angles prescribed at selected locations across the structure, and the output is the corresponding critical buckling indicator. The surrogate fuses abundant LF data generated from coarse mesh FE models with a small number of HF data from refined FE models, so that the global trend is learned efficiently while HF samples correct the bias. Guided by this surrogate, an efficient global optimization (EGO) procedure iteratively updates the design by maximizing a given criterion, thereby moving from an initial set of feasible fiber path candidates to an optimized VS configuration that maximizes buckling capacity with substantially fewer HF FE calls. As shown in Fig.~\ref{fig:app3}, in the plate study, this strategy reaches the reported optimum using only 12 HF evaluations and improves the buckling load from 14.93 kN for a QI baseline to 22.78 kN for the optimized VS design. A similar gain is observed for the cylinder in pure bending, where the critical bending moment increases from 93.17 kN·m to 116.40 kN·m, illustrating the effectiveness of MF surrogate-assisted optimization for VS composites.

Composite sandwich armors have become a practical option for blast protection in ground vehicles, since they can improve resistance to underbody explosions while avoiding the large mass increase associated with purely metallic solutions~\cite{BATRA2008513}. In this setting, the design goal is inherently multi-objective, which is that an armor must simultaneously limit local damage such as penetration, and meanwhile reduce the load transmitted to the vehicle structure. However, blast response is strongly nonlinear and highly sensitive to design choices such as layer thickness distribution and CFRP layup, so identifying a high-performing configuration becomes an optimization problem that would traditionally require many high-fidelity explicit FE blast simulations, leading to prohibitive cost. To address this bottleneck, the study conducts forward explicit FE simulations where several input parameters including the thickness of each composite layer and layup angles are mapped to the performance outputs including maximum penetration and support reaction force~\cite{valladares2020design}, and then uses Bayesian optimization (BO) to estimate the optimal input configuration. BO is a sequential design strategy for global optimization of black-box functions or objectives~\cite{NIPS2012_05311655}. In this specific study, a GP surrogate is first trained on existing simulation results, then BO uses an acquisition function to decide the next design to evaluate by balancing exploitation of promising regions and exploration of uncertain regions. Furthermore, the GP surrogate is upgraded to a MF form using AR co-kriging, fusing a cheap LF blast model with simplified material assumptions and a costly high-fidelity model that includes progressive damage, so that LF data guide global exploration while sparse high-fidelity runs provide correction. In terms of practical performance, the optimized 100 kg sandwich composite armor is reported to achieve penetration comparable to a 195 kg steel armor, indicating substantial potential for weight-efficient blast protection. With the MF strategy applied in the GP surrogate, the need for 150 to 170 HF simulation runs reduces to 95 mixed (LF + HF) simulation runs.

\begin{figure}[hbt!]
\centering
\includegraphics[width=0.8\linewidth]{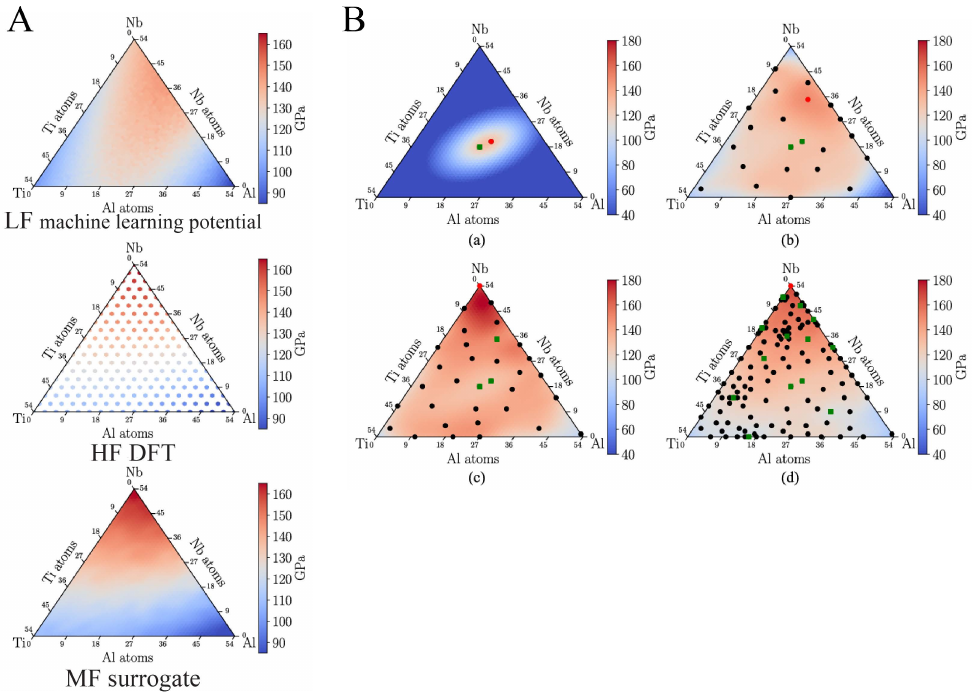}
\caption{Demonstration of composition–modulus triangular diagram: (A) the results from LF source, HF source, and MF-GP surrogate, respectively; (B) the results showing the prediction progression of the MF BO at iterations 4, 24, 35, and 130, respectively (the red dot indicates the best HF data point evaluated so far, while the green squares indicate the other HF data points and the green black dots indicate LF data points). The images are from the reference~\cite{10.1063/5.0015672} with authorization.}
\label{fig:app4}
\end{figure}

Ternary random alloys such as Al–Nb–Ti offer a large compositional design space for tuning mechanical properties and identifying the composition of three components to maximize a target quantity such as the bulk modulus is naturally an inverse materials design problem~\cite{GUBAEV2019148}. On the one hand, first-principles computations such as density functional theory (DFT)~\cite{guerra1998towards} provide high-accuracy property predictions that can guide discovery. On the other hand, the vast ternary composition space makes a brute-force DFT sweep impractical, since the optimization would require many repeated expensive evaluations. To overcome this limitation, the study performs forward atomistic evaluations for the bulk modulus at two fidelity levels, where DFT is used as the HF source and a machine-learning interatomic potential~\cite{PhysRevLett.104.136403} is used as the LF source. A MF-GP surrogate using AR is built to emulate the mapping from ternary composition to bulk modulus~\cite{10.1063/5.0015672}. Then the surrogate is coupled with BO, where a GP-based acquisition function sequentially selects the next composition to evaluate while also deciding whether to query LF or HF to minimize cost for a given uncertainty reduction. Quantitatively, the MF-GP surrogate is able to reconstruct a high-resolution composition–modulus triangular diagram (Fig.~\ref{fig:app4}) with high accuracy, which is nearly close to the original DFT approach, with only 50\% of the HF data. The MF surrogate modeling demonstrates a substantial reduction in expensive DFT calls during inverse search of bulk modulus of Al–Nb–Ti ternary random alloys.

Composite stiffened panels are widely used as lightweight load-carrying components in aircraft structures, yet their stability can be threatened when thermal loads act along with mechanical shortening, potentially triggering premature thermomechanical buckling in some extreme environmental conditions. At the design stage, this motivates reliability-based design optimization (RBDO)~\cite{Meyers01101991}, since geometric tolerances and variability in elastic and thermal properties can shift the critical buckling temperature, which makes safety estimation biased from reality. However, RBDO requires repeated reliability evaluations, and each evaluation calls many FE buckling analyses for different uncertainty realizations, so a direct HF FEM-driven loop becomes prohibitively expensive. To mitigate this computational burden, this study builds a MF surrogate using NARGP with a sampling strategy that efficiently utilizes data of two fidelities~\cite{YOO2021106655}. The data for the surrogate is from the Abaqus FE thermomechanical buckling simulations of a mono-stringer stiffened composite panel, where the inputs include stringer geometry and key material parameters, and the outputs are the critical temperature change and the mass. Fidelity is defined by discretization or mesh quality in the FE simulations. The surrogate is then embedded in an RBDO loop optimized using a multi-objective genetic algorithm, aiming to maximize thermomechanical buckling resistance while minimizing mass under a target reliability index. Quantitatively, the proposed MF surrogate recovers solutions comparable to the HF source using 20 HF and 70 LF simulations, whereas the single HF surrogate requires 150 HF simulations, yielding about 75\% normalized computational time savings while maintaining comparable optimal designs and reliability statistics.

In this subsection, the main motivation for introducing the MF-GP surrogate in inverse design is to make inverse needs in composite mechanics computationally feasible. Many key tasks, such as stiffness identification from sensing data, fiber path tailoring for buckling resistance, blast-resistant armor design, and reliability-based design of stiffened panels, are formulated as searching for inputs that achieve target performance rather than predicting a single response at a fixed condition. The conventional route treats inversion as an optimization loop wrapped around a forward solver, where candidate designs are repeatedly evaluated by HF simulations or experiments until convergence, and it becomes prohibitively expensive when objectives are nonlinear, multi-objective, or uncertainty-aware. GP-based Bayesian optimization offers an efficient alternative by learning a probabilistic surrogate from available evaluations and selecting new candidates sequentially through an acquisition function that balances improvement and uncertainty reduction. Under the MF framework, abundant low-cost data support global exploration while sparse high-cost evaluations provide correction, so the inverse search can reach high-quality solutions with substantially fewer expensive forward calls.

\subsection{Applications of NN-based MFSM}
Differing from classifying GP-based applications into forward prediction and inverse optimization based on the surrogate's major purposes in the last section, this section introduces NN-based applications under the two MF frameworks or strategies to fuse data of different fidelities, which are previously mentioned as the MF fused network and transfer learning. The MF fused network learns cross-fidelity correlation within a single coupled architecture using LF and HF data jointly, thereby reducing the need for expensive HF samples. The transfer learning strategy first pretrains on LF data and then fine-tunes on limited HF data, which is suitable when the two fidelities share aligned inputs and outputs but differ in accuracy and cost. Based on this organization, the following applications are reviewed to illustrate typical network designs and information pathways for MF learning in composite mechanics.

\subsubsection{Applications using MF Fused Network}

\begin{figure}[hbt!]
\centering
\includegraphics[width=0.99\linewidth]{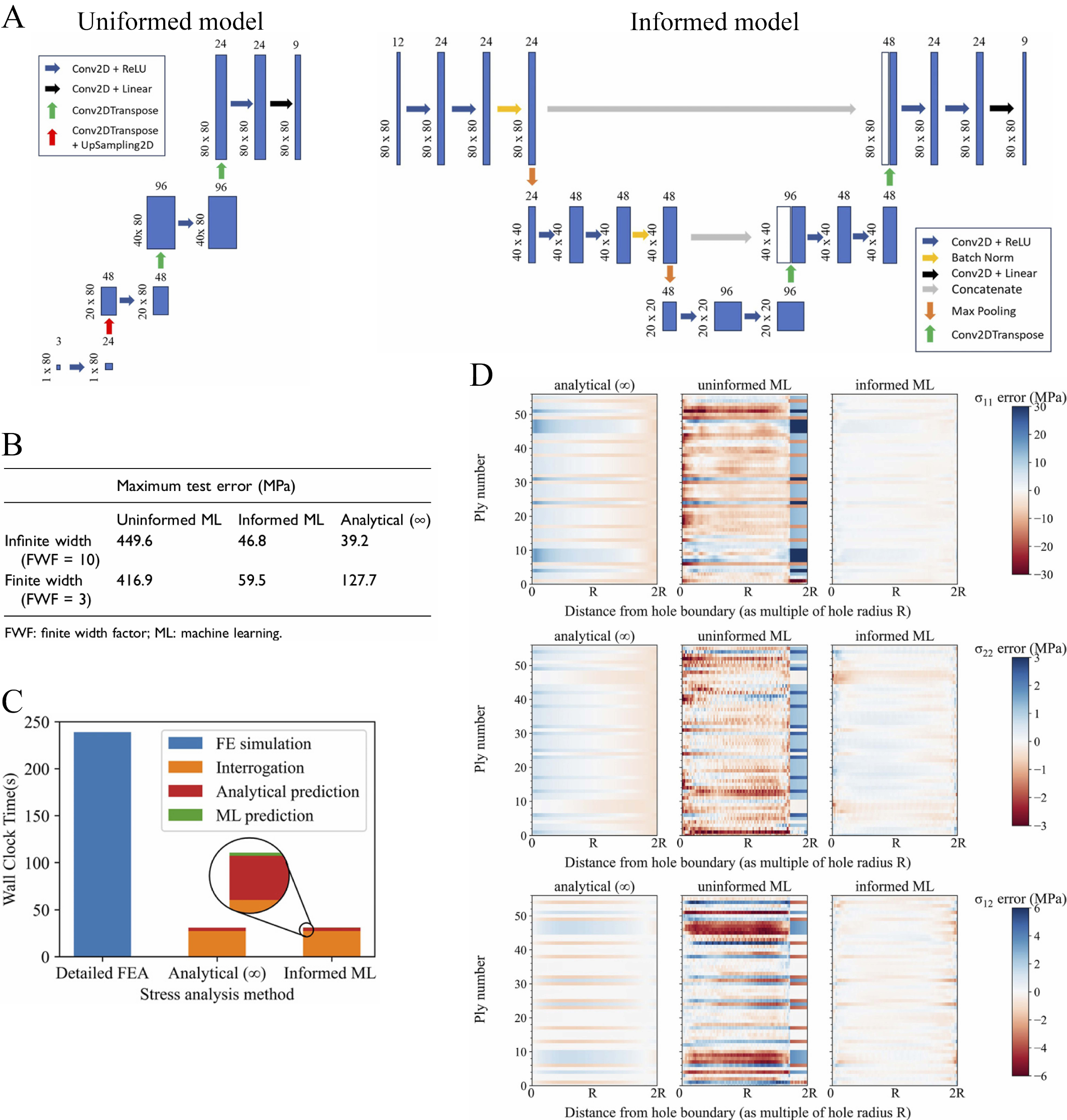}
\caption{Demonstration of two U-net surrogate models and their prediction result: (A) block diagrams of the two U-net neural networks in the uniform model and informed model; (B) table of maximum test error for prediction methods for varying finite width factor; (C) comparison of simulation time for different stress analysis methods; (D) example stress prediction errors for multiple stress components across the net section plane for a finite-width plate. The images are from the reference~\cite{doi:10.1177/00219983241281073} with authorization.}
\label{fig:appn1}
\end{figure}

Open-hole laminates are common design features in composite structures~\cite{doi:https://doi.org/10.1002/9781119013228.ch2}, and fast stress prediction around the hole is essential in global-local modeling, where a coarse global model supplies boundary information and a local model resolves the detailed stress field for failure assessment. Analytical solutions based on Lekhnitskii’s formalism provide efficient stress estimates for infinite-width anisotropic plates~\cite{lekhnitskii1964theory}, but they can not be directly applied to the real finite-width scenario. Meanwhile, FE-based numerical modeling approaches give accurate prediction but they are costly for repetition across many geometries, layups, and loading cases. To address this gap, a neural network surrogate is developed to learn finite-width corrections while retaining a near-analytical efficiency~\cite{doi:10.1177/00219983241281073}. In this study, LF information is defined as the analytical solution of the stress field in infinite-width computed from Lekhnitskii open-hole theory, and HF information is defined as the corresponding stress field images extracted from linear elastic Abaqus simulations with a refined mesh. A surrogate based on U-net is trained to predict the HF stress fields for multiple 2D stress components under uniaxial and biaxial membrane loading. Two neural designs are built and compared (Fig.~\ref{fig:appn1}~A). The first is the uninformed model, which takes only designed variables as input and can be seen as a single fidelity model. The second is the informed model, which has a fused input of the LF analytical stress field channels and design variables and can be seen as the MF model. Furthermore, the informed model utilizes a fusion strategy that embeds low-cost LF fields into the network input, thereby, the network focuses on learning the discrepancy induced by finite-width boundaries rather than learning the entire stress field from scratch. Quantitatively, the fusion strategy substantially reduces the data requirement and improves worst-case accuracy. With the largest dataset setting reported, the maximum test error of the uninformed model is 449.6 MPa for infinite-width and 416.9 MPa for finite-width factor 3, while the informed model reduces these to 46.8 MPa and 59.5 MPa, respectively, which is close to the analytical infinite-width baseline in the infinite-width case and markedly better than the analytical solution in the finite-width case (Fig.~\ref{fig:appn1}~B). The further stress prediction indicates that an informed model trained with even fewer than 50 samples can outperform an uninformed model trained with more than 400 samples, highlighting that analytical field inputs provide a strong inductive bias for stress field learning (Fig.~\ref{fig:appn1}~D). The surrogate reduces the stress analysis process time to nearly 1/10 of the original FEA time, demonstrating high efficiency of the surrogate (Fig.~\ref{fig:appn1}~C). Overall, this application exemplifies an MF fused network in which physics-based analytical information serves as the LF input and FE information serves as the HF target, enabling fast prediction using much fewer expensive FE simulations.

\begin{figure}[hbt!]
\centering
\includegraphics[width=0.99\linewidth]{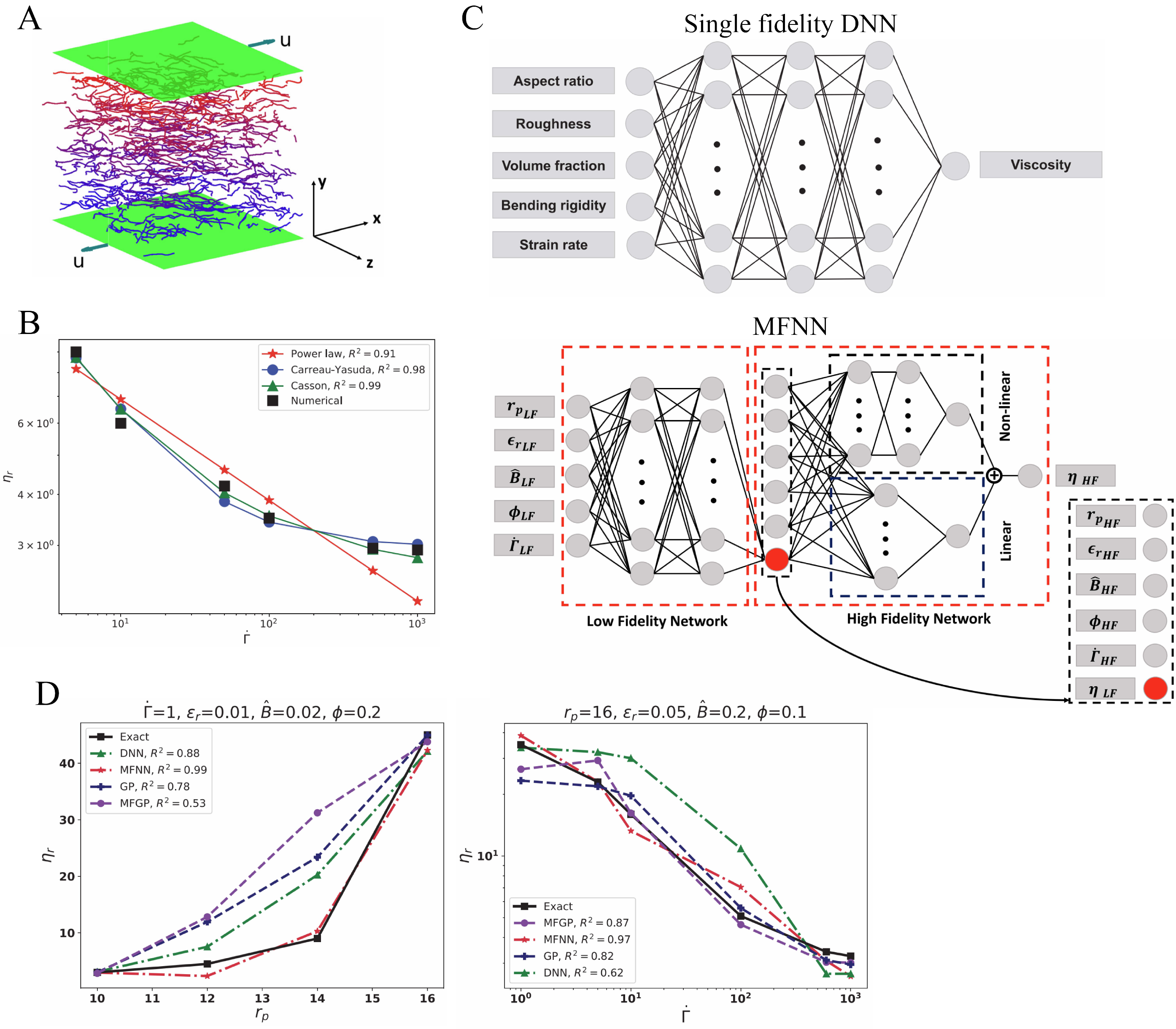}
\caption{Demonstration of MFNN surrogate applied in learning rheological relation of fiber suspension: (A) sketch of the rough fiber in suspension under the microscale modeling; (B) comparison of the curve fitting accuracy for different constitutive equations
for one case with $r_\text{p}=\text{10}$, $\epsilon_\text{r}=\text{0.01}$, $\hat{B}=\text{0.02}$, and $\phi=\text{0.4}$; (C) Schematic view of the single fidelity DNN and MFNN; (D) two cases showing predictions of DNN, MFNN, GP, and MF-GP surrogates in two specific cases. The images are from the reference~\cite{10.1063/5.0087449} with authorization.}
\label{fig:appn2}
\end{figure}

Fiber suspensions are widely encountered in industrial processes involving transportation and mixing, and determining their rheological properties is essential because these suspensions often show strong non-Newtonian behaviors that directly affect processability and performance~\cite{annurev:/content/journals/10.1146/annurev.fluid.36.050802.122132,goto1986flow}. Moreover, rheology is highly sensitive to microstructural and physical parameters such as fiber aspect ratio and volume fraction, so understanding how macroscopic rheological response changes with these properties is a central practical need. Although many empirical constitutive equations are defined to describe rheological relations, they are often restricted to specific parameter ranges and fitting conditions and thereby cannot reliably capture the complex microstructure-driven behavior of fiber suspensions across broad regimes. High-accuracy numerical simulations that resolve microscale fiber–fluid interactions can provide more reliable rheology data. However, they are computationally expensive, especially when evaluating numerous parameter combinations. To address this cost issue, this work builds a MFNN surrogate (Fig.~\ref{fig:appn2}~C) to predict the steady-state apparent relative viscosity $\eta_\text{r}$ with five dimensionless inputs, including fiber aspect ratio $r_\text{p}$, volume fraction $\phi$, roughness height $\epsilon_\text{r}$, dimensionless bending stiffness $\hat{B}$, and dimensionless shear rate $\dot{\Gamma}$~\cite{10.1063/5.0087449}. For the LF source, three constitutive equations, including power-law, Carreau–Yasuda, and Casson, are used to generate LF training data (Fig.~\ref{fig:appn2}~B). For the HF source, HF training data are generated using direct numerical simulations based on an immersed boundary method that couples fluid and solid motion (Fig.~\ref{fig:appn2}~A). The result of this study indicates that the benefit of introducing LF data is most evident in the small-HF regime. When only 53 HF training points are available, the single fidelity deep neural networks (DNNs) achieves Normalized Root Mean Square Error (NRMSE) $=\text{0.10}$ and R-squared metric $R^\text{2}=\text{0.76}$, whereas MFNN improves to $\text{NRMSE }=\text{0.05}$ and R-squared metric $R^\text{2}=\text{0.95}$ with 1000 LF training points. The study also builds GP and MF-GP surrogates to predict the apparent relative viscosity. In the lateral comparison of four surrogates of DNN, MFNN, GP, and MF-GP, the MFNN gives the highest $R^\text{2}$ in nearly all cases (Fig.~\ref{fig:appn2}~D). This study employs a full MFNN fused framework where both the LF network component and the HF network component are learnt, and provides an efficient solution for learning the rheological relation of fiber suspension. 

\begin{figure}[hbt!]
\centering
\includegraphics[width=1.0\linewidth]{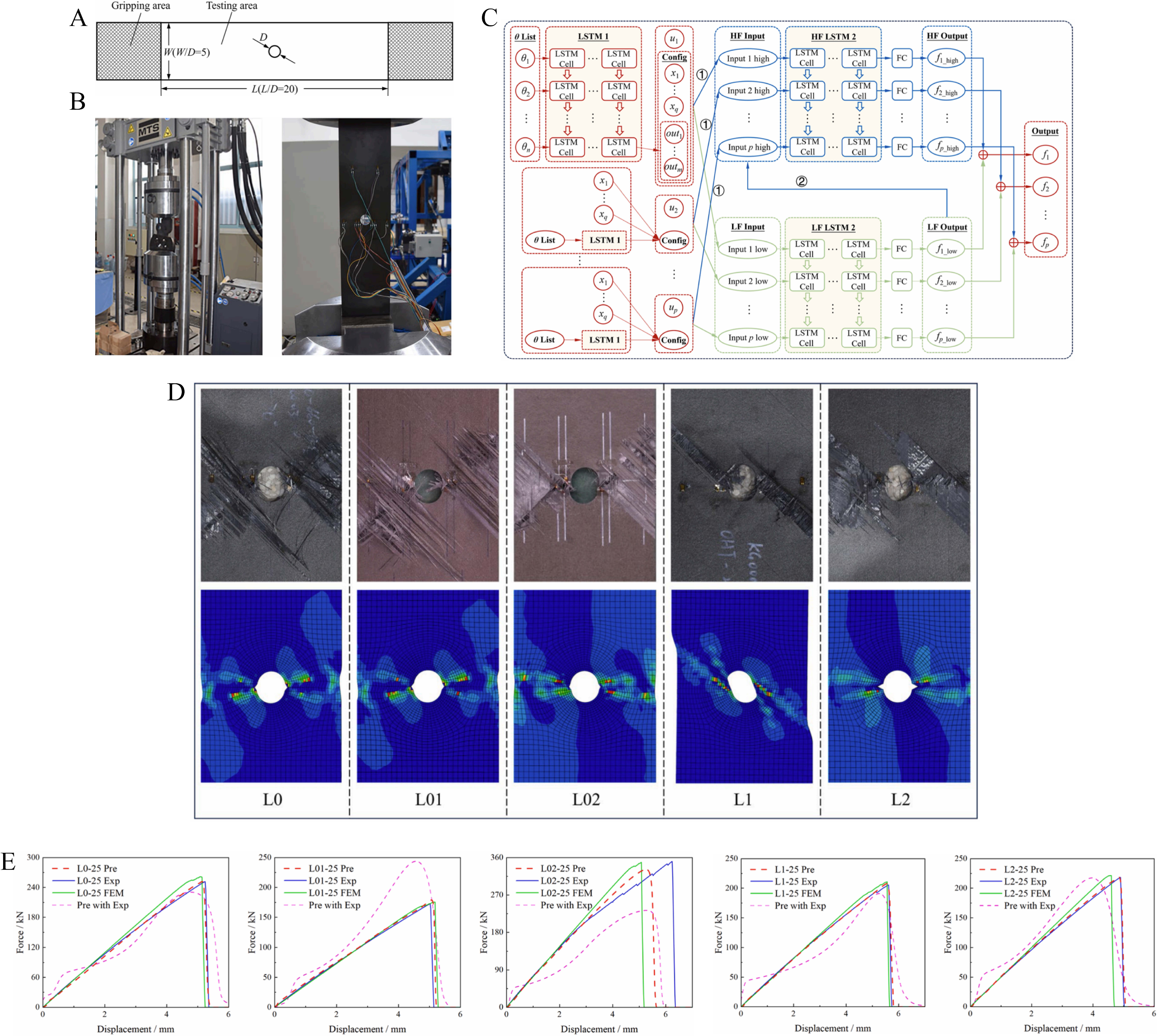}
\caption{Demonstration of MF Triple LSTM surrogate for predicting load–displacement curve of open-hole CFRP laminates: (A) Schematic diagram of open-hole tensile test specimen; (B) the test system and specimen in the tensile experiment; (C) MF Triple LSTM model architecture; (D) comparison of damage morphologies between HF experiment and LF FEA (shown cases use hole diameter of 25mm); (E) the prediction results of the MF Triple LSTM surrogate of five 25mm-hole cases. The images are from the reference~\cite{doi:10.1177/002199839102500303} with authorization.}
\label{fig:appn6}
\end{figure}

Composite open-hole tensile tests are widely used for design allowables and damage assessment~\cite{doi:10.1177/002199839102500303}, yet generating enough experimental load–displacement curves for data-driven surrogates is difficult because specimen fabrication and testing are expensive, and the layup sequence introduces a large and discrete design space. To address this data scarcity, this study proposes a MF Triple LSTM surrogate to predict the full tensile load–displacement curve of open-hole CFRP laminates using limited experiments together with abundant simulations~\cite{TANG2025113106} (Fig.~\ref{fig:appn6}). The HF source is a small set of experimental curves from multiple layup configurations and hole sizes, while the LF source is a large dataset from Abaqus-based progressive damage simulations. The model uses one LSTM to encode the layup sequence, then couples a LF LSTM branch and a HF LSTM branch, where the HF branch takes the LF predicted curve as an additional input and mainly learns the residual correction, so the network does not simply mix two datasets but learns how to correct simulation trends toward experimental behavior. With 3000 LF simulated curves and 11 HF experimental curves, the surrogate model reaches an average $R^2$ of 0.918 on the experimental test set and clearly outperforms training on experiments alone, showing that this MF fusion technique can recover experimental-level curves with very limited testing. In addition, replacing the LF dataset with a less accurate version causes only a 1.4\% average performance drop, indicating that the framework is robust to imperfect LF simulations and mainly needs LF data to provide global curve trends rather than exact agreement.

Autoclave curing requires controlling the transient temperature history inside a composite part, since thermal deviation during the cure cycle can affect final quality and performance. In practice, once the material system changes, the thermal response changes as well, which means the calibration or testing effort for building a reliable temperature predictor must be repeated for the new composite system. To reduce this repeated burden, this study builds a multi-fidelity physics-informed neural network (MFPINN) surrogate for the 1D composites heat-transfer problem with convective boundary conditions~\cite{9816983}. Fidelity is defined by two carbon-fiber epoxy systems, where Composite 1 is treated as the data-abundant source and Composite 2 is the data-scarce target. A coupled MFPINN is trained by first learning a LF PINN on Composite 1, then injecting its predictions into a HF PINN for Composite 2 together with the HF PDE residual and limited labeled data when available. The result indicates that transferring knowledge of physical PDEs across material systems can reduce HF temperature errors, while targeted labeled points are still needed in subdomains where the two materials diverge, especially during cooldown. In the reported comparison, the relative $l_\text{2}$ error decreases from 12.9\% (PINN) to 2.8\% (MFPINN), and further to 1.7\% when labeled HF data are imported to MFPINN.

In this subsection, four applications, including the stress prediction in open-hole laminates, learning the rheological relation of fiber suspension, predicting the load-displacement curve of open-hole CFRP laminates, and temperature prediction during autoclave curing, are presented. The MF fused-network framework is characterized by learning cross-fidelity correlation within a single coupled neural network system, where LF information is injected into the HF predictor as an additional input. Relying on the core concept that LF captures global trends while the HF component learns the remaining discrepancy, MF fused networks typically improve sample efficiency and robustness in small-HF regimes. Furthermore, this framework remains flexible in how fidelity is defined, such as conventional cheap–expensive modeling hierarchies and cross-system transfer across different material sets.

\subsubsection{Applications using Transfer learning MF Network}
\begin{figure}[hbt!]
\centering
\includegraphics[width=0.95\linewidth]{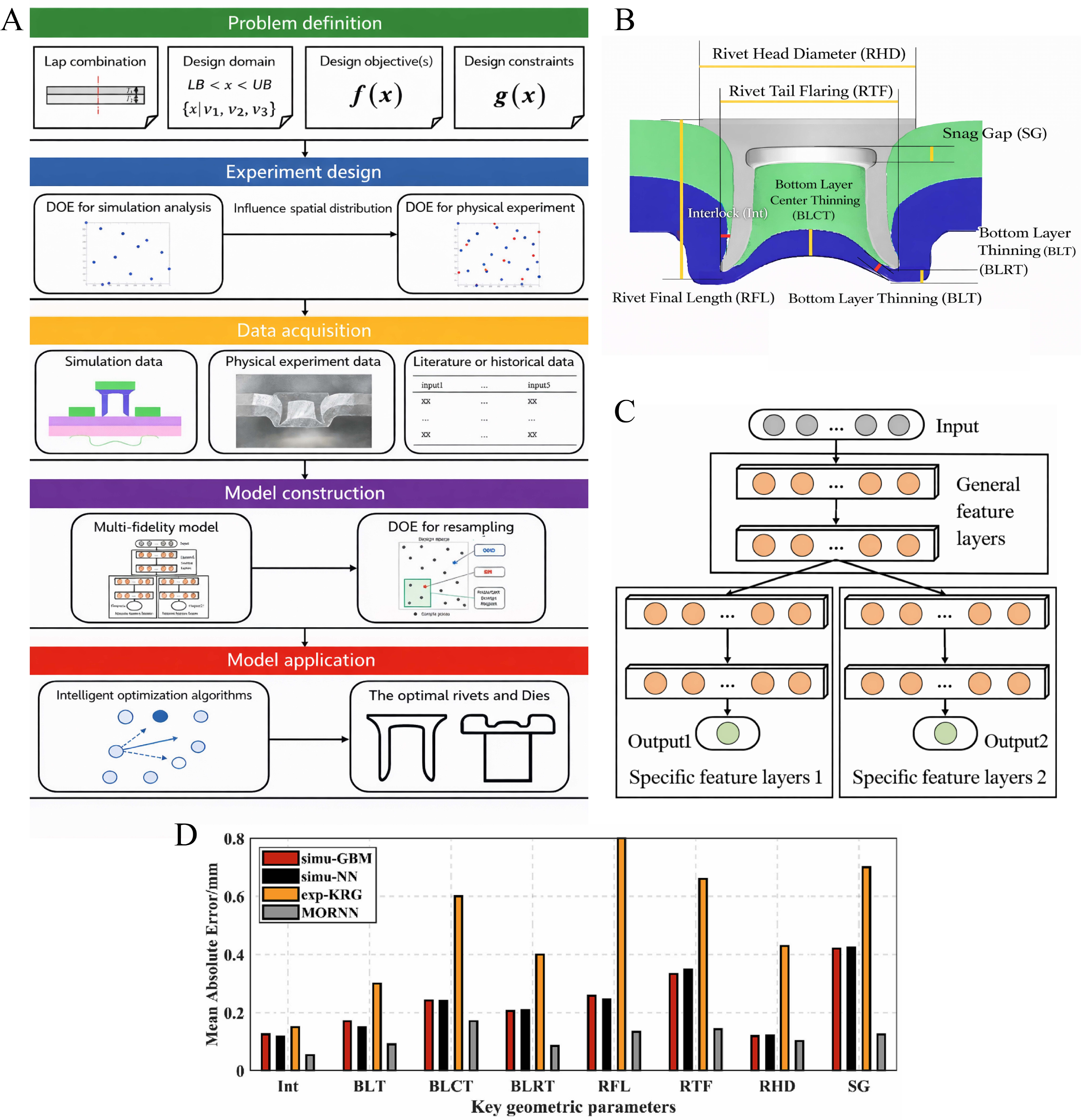}
\caption{Demonstration of MORNN surrogate applied in efficient design of process parameters in SPR: (A) the complete flow schematic of the study framework; (B) definition of 9 geometric parameters in a single SPR cross-section; (C) The network architecture of the MORNN; (D) The accuracy comparison of four surrogates (simu-GBM: gradient boosting machine with LF simulation data; simu-NN: single fidelity neural networks with LF simulation data; exp-KRG: Kriging with experimental data). The images are from the reference~\cite{LI2023812} with authorization, and sub-figures A B C are upscaled to the higher resolution with a generative AI tool.}
\label{fig:appn3}
\end{figure}

Self-piercing riveting (SPR) is widely used in auto body manufacturing, and efficient design of process parameters is increasingly important because joint quality is governed by cross-section geometric indicators such as interlock value (Int) and bottom layer thinning (BLT)~\cite{HAQUE201883,doi:10.1179/174329306X131866}. Traditional trial-and-error design and full physical testing are costly, while simulation can generate large datasets efficiently but still deviates from experiment, thereby neither a single-source dataset alone is sufficient for reliable optimization. To address this, this study proposes a MF optimization framework that fuses LF CAE simulation data with HF physical experiment data, then uses a multi-output regression neural network (MORNN) based on the transfer learning strategy to calibrate simulation-learned features toward experimental truth~\cite{LI2023812} (Fig.~\ref{fig:appn3}~C). MORNN uses a shared trunk and separate specific feature layer branches for LF outputs and HF outputs, which are designed to capture the common relationship and individual fidelity bias, respectively. In the case study, the MF training set contains 30 experiments and 2458 simulations, with 15 input variables and 8 output geometric parameters (Fig.~\ref{fig:appn3}~B), such as Int and BLT, and the resulting model achieves a mean absolute error (MAE) below 0.1 mm on Int and BLT and below 0.2 mm across all predicted parameters (Fig.~\ref{fig:appn3}~D). The MF surrogate is then coupled with an optimization routine to rank feasible process configurations, and physical experiments verify that the recommended process configurations are the optimal solutions within the available option range. Furthermore, this work demonstrates a workflow-oriented use of multiple tools to make the overall SPR design pipeline more efficient and automated (Fig.~\ref{fig:appn3}~A). In addition to the MORNN surrogate, Latin hypercube sampling (LHS) is used to generate a space-filling design that improves coverage of the input space for both LF simulation and HF experimental campaigns, and a deep-learning-based image identification step is integrated to automatically extract cross-section geometric labels from HF experimental images, thereby reducing manual measurement effort and improving data acquisition efficiency. Together with the surrogate-assisted optimization loop, these components form a more intelligent end-to-end framework, where sampling, data extraction, MF calibration, and decision-making are connected into one streamlined process.

\begin{figure}[hbt!]
\centering
\includegraphics[width=1.0\linewidth]{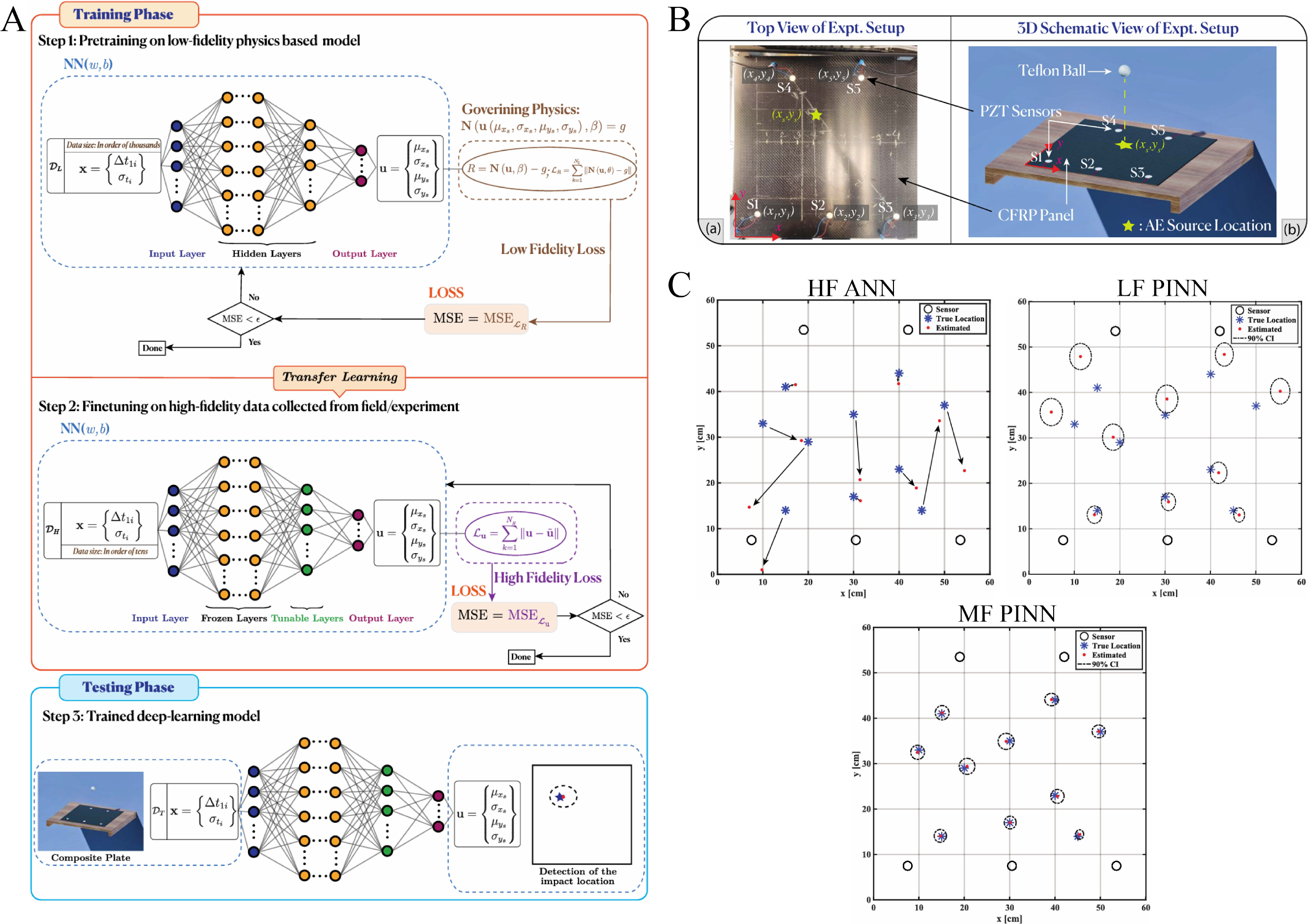}
\caption{Demonstration of probabilistic MFPINN surrogate applied in AE impact localization: (A) schematic representation of proposed mfPINN framework for estimating AE source location and corresponding uncertainty; (B) the top view and 3D schematic view of the experimental setup; (C) True and estimated impact location results using HFANN, LFPINN, and HFPINN. The images are from the reference~\cite{KALIMULLAH2023110360} with authorization.}
\label{fig:appn4}
\end{figure}

Acoustic emission (AE) impact localization is a key step in structural health monitoring of anisotropic CFRP panels, since identifying the source position supports rapid inspection and damage assessment~\cite{staszewski2004health}. In practice, reliable localization is difficult when only a sparse sensor array is available and the measured time of flight (TOF) signals contain noise, while purely data-driven models typically need more clean labeled data than experiments can provide. To address this gap, this work constructs a probabilistic MFPINN surrogate (Fig.~\ref{fig:appn4}~A), where the LF source is the anisotropic guided-wave TOF localization relation imposed as a PINN residual on collocation points, and the HF source is a limited set of measured TOF signals from impact experiments on an actual CFRP panel collected by a sparse sensor array~\cite{KALIMULLAH2023110360} (Fig.~\ref{fig:appn4}~B). The probabilistic component is introduced by explicitly modeling TOF uncertainty as Gaussian noise, so the network outputs both the estimated source location and its uncertainty, enabling confidence bounds rather than a single deterministic point. The physics component is enforced by minimizing the residual of the anisotropic localization relation on collocation points, which stabilizes training and reduces data dependence. It is worth noting that the LF information here is not an explicit labeled dataset. Instead, it is the anisotropic TOF localization relation, which is imposed as a physics residual constraint in the PINN training loss. Under the MF strategy, the LFPINN is trained first and then transferred to the experimental setting by freezing most layers and fine-tuning only the last layers with limited HF impacts, thereby correcting systematic LF–HF mismatch while retaining physics-consistent structure. By comparing three surrogates including HF artificial neural network (ANN), LFPINN and MFPINN, the quantitative result shows the average localization error drops from 10.3217 cm (HFANN) and 3.5529 cm (LFPINN) to 0.5075 cm for MFPINN, while the maximum error drops from 19.9423 cm and 7.8926 cm to 0.8062 cm, respectively (Fig.~\ref{fig:appn4}~C).

\begin{figure}[hbt!]
\centering
\includegraphics[width=1.0\linewidth]{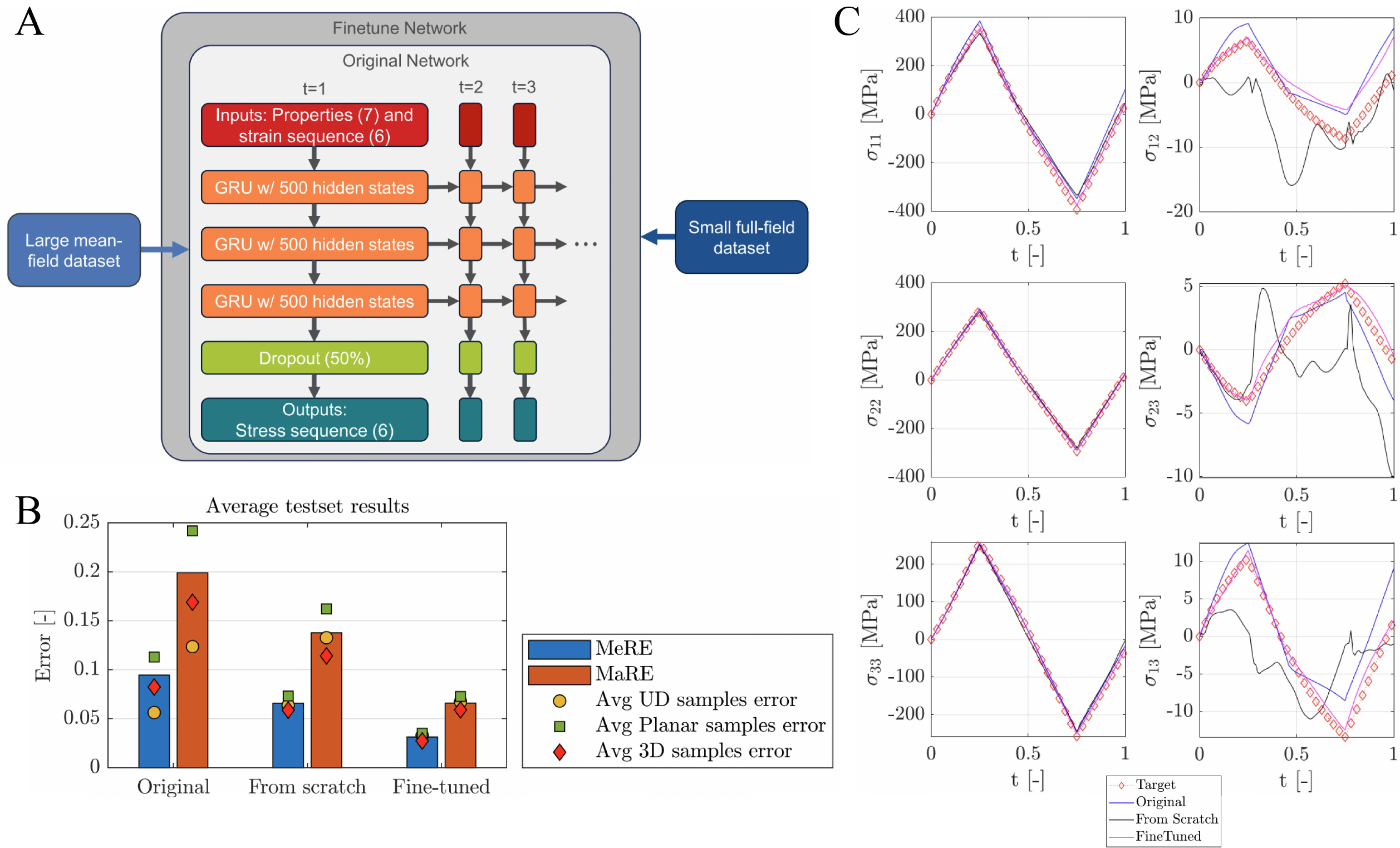}
\caption{Demonstration of GRU-based RNN surrogate with transfer learning applied to predict full stress field of SFRC: (A) Network architecture and the transfer learning approach utilizing a large mean-field and a small full-field data set; (B) Average mean relative error (MeRE) and maximum relative error (MaRE) results for the test data set ("UD", "Planar", and "3D" are referred to three fiber orientation cases); (C) specific 6 stress tensor result (the blue "Original" line represents the single fidelity RNN surrogate using LF data, the black "From Scratch" line represents the single fidelity RNN surrogate using HF data, and the purple "FineTuned" line represents the RNN surrogate with transfer learning using both LF and HF data). The images are from the reference~\cite{CHEUNG2024110359} with authorization.}
\label{fig:appn5}
\end{figure}

Short fiber reinforced composites (SFRCs) are widely used because they are compatible with injection molding, yet predicting their nonlinear elasto-plastic response is difficult since the matrix yielding and microstructure effects make the stress history strongly path-dependent~\cite{MIRKHALAF2022107097}. HF full-field micromechanical simulations, where FEM or fast Fourier transformation (FFT) are involved, can capture these mechanisms, but they are computationally demanding, which limits the amount of training data that can be generated for a purely data-driven surrogate. To resolve this data bottleneck, this work builds a RNN surrogate using gated recurrent unit (GRU) and adopts a transfer learning strategy that utilizes the LF and HF source~\cite{CHEUNG2024110359} (Fig.~\ref{fig:appn5}~A). The LF source is a large mean-field dataset (around 40,000 samples) generated by a mean-field micromechanics model, while the HF source is a much smaller full-field dataset (547 samples) generated by FE- or FFT-based full-field analyses. The GRU-based RNN uses 13 inputs, including 6 independent orientation tensor components, fiber volume fraction, and a sequence of 6 independent strain tensor components, and outputs a sequence of 6 independent stress tensor components. The core of the transfer learning is to fine-tune the pretrained LF RNN using the limited HF full-field data, with interpolation used to reconcile the different time-step resolutions between LF and HF sequences. In the reported result, the fine-tuned model reaches a much lower validation loss (17.92 MPa²) than training from scratch (99.37 MPa²), and both mean and maximum relative errors on test cases are substantially reduced compared with the original LF network and the scratch-trained HF network (Fig.~\ref{fig:appn5}~B \& C).

The surrogate modeling of GRU-based RNN with transfer learning is further extended from SFRCs to inelastic woven composites~\cite{GHANE2025111163}, where the composite physics is different. In SFRCs, short fibers are scattered in the matrix basis, and thereby the target is to predict the strain-stress relation affected by the fiber orientation. While in woven composites, fibers are arranged as woven yarns that cross each other in a repeating pattern, and thereby the target is to capture the stress response affected by complex yarn geometry~\cite{ODDY2022111696}. Similar to the previous study, the LF source is still a larger dataset generated from mean-field homogenization (MFH), and the HF source is a full-field FFT-based homogenization dataset that explicitly incorporates mesoscale modeling of the fiber yarn. To enhance the surrogate's performance, the transfer learning is formulated to handle not only the fidelity gap between MFH and FFT, but also the shift between different loading histories. In addition, an interpolation-based preprocessing step is introduced before training. The study’s results indicate that these additions improve robustness of the HF prediction under irregular loading paths. After excluding one outlier case, the mean error and its standard deviation drop by 30\% and 32\%, and MFH pretraining further reduces the mean error by 21\% and the standard deviation by 40\%.

In this subsection, four applications, including SPR process parameter optimization, estimating AE impact localization in CFRP panels, GRU-based RNN surrogate modeling for predicting stress response of SFRCs, and its extension to inelastic woven composites, are presented. Transfer learning MF network applications are characterized by a different logic from MF fused networks. Instead of injecting LF outputs into the HF predictor, transfer learning first uses the LF setting to pretrain a backbone that learns reusable representations of the input–output relation, and then adapts this backbone to the HF target through fine-tuning with HF information. Under this mechanism, LF mainly contributes a strong initialization and stabilizes learning when HF data are scarce, while HF fine-tuning corrects systematic mismatch caused by measurement noise, model bias, or physics simplifications.

\section{Open Challenges and Research Opportunities} \label{Section OCRO}
Despite the increasing number of successful demonstrations reviewed in the previous section, MFSM in composite mechanics is still far from a routine and standardized tool for engineering analysis, design, and manufacturing decision-making. This gap does not mainly come from the lack of modeling techniques, but from several composite-specific difficulties that repeatedly appear across applications, including how fidelity should be defined under multi-scale and multi-physics hierarchies, how LF and HF data can be aligned under scarcity and heterogeneity, and how surrogate reliability can be ensured when models are used for inverse design and workflow deployment. Accordingly, this section summarizes open challenges and research opportunities by focusing on fidelity definition, data and alignment, method suitability between GP- and NN-based MFSM, and the requirements introduced by inverse optimization and real-world adoption.

First, the reviewed composite-mechanics applications indicate that fidelity is defined in diverse and problem-dependent ways, rather than by a single universal LF-to-HF ordering. In other words, the meaning of “LF” and “HF” is not fixed in composites, but is instead determined by which physical mechanisms, kinematic descriptions, numerical approximations, or data sources are emphasized in a given task. In progressive-damage prediction, fidelity is defined through constitutive richness, where a cheaper built-in Abaqus progressive damage model is fused with a more accurate but expensive UMAT-based 3D continuum damage mechanics model ~\cite{CHAHAR2023106647}. In failure onset screening of composite laminates, fidelity is tied to the structural kinematics and through-thickness representation, since the LF and HF tiers are CUF–ESL and CUF–LW models that differ in how ply-level effects are resolved~\cite{10.1115/SSDM2025-152303}. In process-induced deformation analysis, fidelity becomes explicitly multi-tier and mixed-source, because a LF 2D thermo-mechanical simulation, a HF 3D simulation, and scarce experiments are combined in one calibration framework~\cite{5288526}, implying that HF is not only referred to a refined solver but also the closeness to physical reality anchored by measurements. In NN-based full-field learning, fidelity can be defined by boundary realism and geometric idealization, where Lekhnitskii infinite-width analytical stress fields serve as LF information and refined Abaqus finite-width stress field images provide HF targets~\cite{doi:10.1177/00219983241281073}, so the MF gain comes from learning finite-width corrections rather than relearning the entire field. Fidelity can also be defined as a cross-system transfer problem, where one composite material system is treated as data-abundant and another as data-scarce in heat-transfer modeling, and the MF coupling is realized through physics-informed transfer between material systems~\cite{9816983}. Taken together, these cases show that composites naturally generate multi-axis fidelity, where each axis reflects a different source of mismatch, such as kinematic idealization, constitutive simplification, boundary condition implementation, numerical resolution, or experimental noise. Under this setting, the naive assumption that LF is globally inferior and can be corrected by a single smooth discrepancy term can break, because the LF–HF discrepancy may be mechanism- or regime-dependent, especially when the response transitions across elastic behavior, damage initiation and propagation, or cure stage changes. Consequently, an important research opportunity is to move from fidelity as a label to fidelity as a structured description, namely to develop fidelity-aware MF formulations that explicitly encode what changes across tiers, together with regime-aware fusion strategies that can determine when LF is globally informative, when it is only locally valid, and when MF coupling risks negative transfer due to shifting mismatch mechanisms.

Second, beyond the conceptual difficulty of defining fidelity, the dominant practical barrier across composite MF applications is the data itself, which is typically scarce at HF, heterogeneous across sources, and imperfectly aligned between tiers. In calibration-oriented problems, HF data are not only limited but also noisy and costly to obtain, and the measured quantities often differ from what simulations naturally output. This is visible in PID assessment, where only a few experiments are available and must anchor a MF simulation ensemble through calibration~\cite{SCHOENHOLZ2024111499}, and it is even more explicit in AE impact localization, where the HF source is sparse TOF measurements from a limited sensor array, rather than a dense field label~\cite{KALIMULLAH2023110360}. In history-dependent prediction tasks, HF information is a full trajectory instead of a single scalar, but the number of available trajectories remains small, as in the MF Triple LSTM that learns experimental load–displacement curves from very limited tests while relying on abundant simulated curves to supply global trends~\cite{doi:10.1177/002199839102500303}. Meanwhile, even when LF and HF are both generated numerically, alignment is not guaranteed, because the two tiers may differ in their governing assumptions, discretization choices, internal variables, and postprocessing definitions, which creates systematic mismatch that cannot be removed by simply pooling data. This issue is implied by discrepancy learning in failure onset assessment~\cite{10.1115/SSDM2025-152303} and by fusing a built-in progressive damage model with a UMAT-based continuum damage model~\cite{CHAHAR2023106647}, where the outputs may share names but the underlying physics content and error structures differ. For full-field learning, alignment becomes more stringent, since HF labels are spatial maps whose definition depends on meshing, interpolation, and the selected observation plane, which directly affects training stability and error interpretation, as reflected in the open-hole stress field learning problem~\cite{doi:10.1177/00219983241281073}. Overall, these applications suggest that HF data in composites are usually limited and noisy, while LF and HF are often not aligned in inputs, outputs, or postprocessing definitions. A key further direction is to develop MF learning strategies that tolerate this mismatch, for example by learning residual corrections and introducing noise-aware training~\cite{MENG2021110361}.

In addition, the reviewed studies suggest that the choice between GP-based and NN-based MFSM in composite mechanics is mainly governed by the task scenario. GP-based MFSM is most natural when the outputs are scalar indicators or a small vector of correlated responses, HF data are extremely limited, and uncertainty is required to support screening, optimization, or reliability assessment. This setting appears in multiscale UQ for woven composites using MRGP~\cite{WANG2015159}, failure onset screening using discrepancy GP~\cite{10.1115/SSDM2025-152303}, progressive damage evaluation using AR-based MF-MOGP~\cite{CHAHAR2023106647}, and PID calibration where sparse experiments anchor multi-fidelity simulations through SWGPR~\cite{SCHOENHOLZ2024111499}. The same GP advantage becomes more explicit once the surrogate is placed inside inverse loops, because BO and RBDO rely on uncertainty-aware exploration and feasibility control, as shown in VS laminate fiber steering optimization using hierarchical Kriging and EGO~\cite{10.1007/s00158-020-02684-3}, blast-resistant sandwich armor design using MF Co-kriging inside BO~\cite{valladares2020design}, ternary alloy discovery using fidelity-aware BO~\cite{10.1063/5.0015672}, and stiffened panel RBDO using NARGP~\cite{YOO2021106655}. By contrast, NN-based MFSM is more suitable when the quantities of interest are spatial fields or history-dependent trajectories, since these outputs demand representation learning that is difficult for conventional GP surrogates to scale to. This is exemplified by the informed U-net that injects analytical stress fields to learn finite-width corrections~\cite{doi:10.1177/00219983241281073}, the MFNN surrogate for fiber-suspension rheology that exploits LF constitutive trends~\cite{10.1063/5.0087449}, and the MF Triple LSTM that predicts experimental load–displacement curves by learning simulation-to-experiment correction within a coupled sequence model~\cite{doi:10.1177/002199839102500303}. Transfer learning MF networks further reinforce this pattern in LF-abundant and HF-scarce regimes, including MORNN calibration for self-piercing riveting~\cite{LI2023812} and GRU-based RNN surrogates for SFRCs and woven composites~\cite{CHEUNG2024110359,GHANE2025111163}. With the fast iteration of modern computing hardware and software, the computational cost of NN training is becoming less of a limiting factor in practice, which partially explains why NN-based MFSM is increasingly adopted when field and sequence outputs dominate. Under this trend, the main remaining bottleneck shifts toward trust and explanation. NN-based MFSM is often harder to interpret, and its uncertainty treatment is less standardized than GP-based models, which becomes critical in composite mechanics where understanding of the mechanism is needed for failure, damage, and manufacturing-defect diagnosis. Consequently, a promising direction is interpretable~\cite{pmlr-v119-koh20a} and physics-guided NN-based MFSM, where physics or mechanics insights are either provided by surrogate structures or explicitly incorporated into the surrogate learning process to boost predictive performance.

Meanwhile, the reviewed studies suggest that inverse design and engineering workflow deployment are the scenarios that also strongly reshape the requirements of MFSM, because the surrogate is not used for a single forward prediction but is repeatedly queried inside a decision loop and is connected to other pipeline components. On the inverse side, MF surrogates are embedded into search procedures where constraints, safety margins, and uncertainty propagation become part of the problem definition. This is reflected by VS laminate fiber-steering optimization, where hierarchical Kriging is coupled with EGO to navigate a high-dimensional fiber angle design space with limited HF evaluations~\cite{10.1007/s00158-020-02684-3}, and by blast-resistant sandwich armor design, where MF Co-kriging is placed inside BO to reduce expensive explicit FE calls while maintaining design quality~\cite{10.1063/5.0015672}. The ternary alloy study further shows that the inverse loop can be fidelity-selective, since MF BO decides both the next candidate composition and which fidelity level to query so that accuracy gain and evaluation cost are balanced~\cite{YOO2021106655}. In stiffened panel RBDO, MF surrogates are placed inside reliability evaluation and multi-objective search, where meeting a target reliability requirement is as important as improving nominal performance~\cite{YOO2021106655}. On the workflow side, the surrogate is also expected to remain reliable when it is integrated with sampling, data extraction, calibration, and validation procedures. PID assessment links MF simulations with scarce experiments for manufacturing quality control, so surrogate credibility depends on calibration quality and robustness under processing variability rather than test error alone~\cite{SCHOENHOLZ2024111499}. A stronger illustration of workflow integration is the SPR framework, where multiple machine learning techniques are combined in one pipeline, including space-filling sampling, deep-learning-based image identification for extracting geometric labels, transfer learning MF calibration from simulation to experiments using MORNN, and automatic ranking of process configurations with experimental verification~\cite{LI2023812}. AE impact localization highlights another deployment setting under sparse sensing and noisy signals, where uncertainty output becomes necessary for inspection decisions and practical adoption~\cite{KALIMULLAH2023110360}. Overall, these applications suggest two potential research directions. The first is MF inverse design methods that explicitly handle constraints and reliability targets, instead of optimizing only nominal performance. The second is workflow-level verification and validation that tests whether surrogate-based decisions remain reliable when conditions change in practice, such as shifts in layup, geometry, loading history, and manufacturing variability.

Finally, the reviewed applications show that MFSM can accelerate forward prediction, reduce the cost of inverse search, and support composite engineering workflows. However, this review also indicates that broader and more routine use still requires substantial effort in MFSM, and the following challenges represent only key directions rather than an exhaustive list. Fidelity in composites is often multi-axis and non-nested, which motivates structured fidelity descriptors and regime-dependent LF–HF coupling. LF and HF data are frequently scarce, noisy, and misaligned in definitions and outputs, which motivates MF learning that tolerates imperfect pairing, incorporates noise awareness, and supports partial supervision across heterogeneous outputs. GP-based and NN-based MFSM fit different scenarios, which motivates hybrid designs that retain scalable representation learning while improving interpretability through physics-guided structures and providing decision-relevant uncertainty. Inverse design and workflow deployment introduce constraint and reliability requirements and demand validation under realistic shifts, which motivates decision-aware surrogate-assisted optimization and workflow-level verification under changes in layup, geometry, loading history, and manufacturing variability.

\section{Discussions and Summary} \label{Section summary}
MFSM has matured from a niche acceleration technique into a general modeling paradigm for composite mechanics, where the cost–accuracy trade-off is amplified by multiscale architectures, strong anisotropy, evolving damage, and manufacturing-induced variability. Across the studies reviewed in this paper, MFSM is rarely used as an isolated regression tool. Instead, it acts as a coupling mechanism that connects heterogeneous evidence, including simplified solvers, reduced-order physics, high-resolution simulations, and sparse experiments. This role is particularly consequential in composites because the dominant uncertainty is often not only parametric, but also structural and regime-dependent. It arises from constitutive idealizations, boundary condition realism, discretization choices, and measurement limitations.

From a methodological perspective, the classical GP and Kriging family provides a principled baseline for multi-fidelity fusion because it encodes cross-fidelity dependence through explicit correlation assumptions and yields uncertainty in a native manner. The reviewed GP-based formulations span autoregressive constructions, discrepancy-based corrections, hierarchical extensions, and multi-output variants that exploit correlations among multiple responses. In composite applications, these models are most effective when the quantity of interest is a scalar indicator or a low-dimensional vector, when HF evaluations are extremely limited, and when uncertainty quantification must directly support screening, Bayesian optimization, or reliability analysis. At the same time, the composite setting repeatedly exposes the limits of globally smooth discrepancy assumptions. Fidelity gaps may vary across elastic and inelastic regimes, across damage initiation and propagation stages, or across cure and forming phases. This motivates fusion strategies that are regime-aware and robust to non-nested or multi-axis fidelity definitions.

NN-based MFSM extends the modeling frontier by offering scalable representation learning for outputs that are difficult to handle with conventional GP surrogates, particularly spatial fields and history-dependent trajectories. Two dominant constructions appear across composite mechanics demonstrations. Multi-fidelity fused networks inject LF information into the HF predictor through architectural coupling, while transfer learning frameworks adapt a LF-pretrained backbone to HF targets via fine-tuning under HF scarcity. In addition, physics-guided learning emerges as a unifying theme that improves data efficiency and stabilizes training by constraining admissible function classes through conservation laws, constitutive structure, or analytically derived trends. These developments collectively indicate a shift in MFSM practice. When field outputs and history-dependent sequence outputs dominate, the primary bottleneck moves away from raw training cost and toward reliability, interpretability, and decision-relevant uncertainty.

At the application level, the reviewed composite studies naturally cluster by how the surrogate is used in the engineering loop. In forward prediction, MFSM mainly serves efficiency and coverage, enabling fast evaluation of response surfaces, full-field maps, or temporal histories under broad parameter exploration. In inverse optimization, MFSM becomes a query engine inside design search, where uncertainty-guided exploration, constraint handling, and cost-aware fidelity selection strongly influence final design quality. In workflow integration, MF surrogates are embedded into end-to-end pipelines that connect sampling, simulation, data extraction, calibration, and experimental verification. Performance is then judged not only by test error, but also by robustness under operational variability and by the credibility of surrogate-assisted decisions. In this sense, the practical value of MFSM in composites is best measured at the workflow level, where the surrogate changes how design iterations, manufacturing tuning, and qualification decisions are executed.

Several cross-cutting lessons emerge from these applications. Fidelity in composites is typically problem-defined rather than universally ordered, and the same nominal output can correspond to different physics content across tiers due to differing internal variables, postprocessing definitions, or boundary realizations. Data limitations remain the dominant practical barrier. HF measurements are scarce and noisy, LF and HF are frequently misaligned in inputs and outputs, and full-field supervision introduces additional sensitivity to interpolation and meshing choices. These characteristics make negative transfer a realistic risk and emphasize the need for MF learning mechanisms that tolerate imperfect pairing, incorporate noise awareness, and admit partial supervision across heterogeneous labels. In parallel, broader adoption requires uncertainty that is not only present but also actionable. This includes calibrated predictive intervals, sensitivity to distribution shifts, and uncertainty propagation compatible with constraint satisfaction and reliability targets.

Overall, MFSM provides a coherent bridge from early multi-source GP fusion methods to modern multi-fidelity neural systems for composite mechanics, and the reviewed literature demonstrates clear gains in computational efficiency, data utilization, and workflow acceleration. Routine use, however, will likely depend on advances that treat fidelity as a structured descriptor rather than a label, formalize alignment under heterogeneous and partially observed data, and deliver trustworthy prediction through interpretable and physics-guided surrogates with decision-relevant uncertainty. Progress along these directions, together with reproducible benchmarks and workflow-level verification and validation under realistic shifts in layup, geometry, loading history, and manufacturing variability, will determine whether MFSM becomes a standard component of composite analysis, design, and manufacturing practice.

\section*{Acknowledgments}
This work was supported as part of the AIM for Composites, an Energy Frontier Research Center funded by the U.S. Department of Energy, Office of Science, Basic Energy Sciences at Clemson University under award \#DE-SC0023389.

\bibliographystyle{unsrt}
\bibliography{references}

\end{document}